\documentclass[twocolumn,english]{revtex4-1}
\usepackage[T1]{fontenc}
\usepackage[latin9]{inputenc}
\usepackage{bm}
\usepackage{amsmath}
\usepackage{amssymb}
\usepackage{stmaryrd}
\usepackage{graphicx}
\usepackage{esint}

\makeatletter

\providecommand{\tabularnewline}{\\}

\@ifundefined{textcolor}{}
{%
 \definecolor{BLACK}{gray}{0}
 \definecolor{WHITE}{gray}{1}
 \definecolor{RED}{rgb}{1,0,0}
 \definecolor{GREEN}{rgb}{0,1,0}
 \definecolor{BLUE}{rgb}{0,0,1}
 \definecolor{CYAN}{cmyk}{1,0,0,0}
 \definecolor{MAGENTA}{cmyk}{0,1,0,0}
 \definecolor{YELLOW}{cmyk}{0,0,1,0}
}

\newcommand{\mbf}[1]{\mathbf{#1}}
\usepackage{babel}

\usepackage{babel}

\usepackage{babel}

\usepackage{babel}

\usepackage{babel}

\usepackage{babel}

\usepackage{babel}

\makeatother

\usepackage{babel}
\begin{document}

\title{Low-energy microscopic models for iron-based superconductors: a review}

\author{Rafael M. Fernandes}

\affiliation{School of Physics and Astronomy, University of Minnesota, Minneapolis
55455, USA}

\author{Andrey V. Chubukov}

\affiliation{School of Physics and Astronomy, University of Minnesota, Minneapolis
55455, USA}
\begin{abstract}
The development of sensible microscopic models is essential to elucidate
the normal-state and superconducting properties of the iron-based
superconductors. Because these materials are mostly metallic, a good
starting point is an effective low-energy model that captures the
electronic states near the Fermi level and their interactions. However,
in contrast to cuprates, iron-based high-$T_{c}$ compounds are multi-orbital
systems with Hubbard and Hund interactions, resulting in a rather
involved 10-orbital lattice model. Here we review different minimal
models that have been proposed to unveil the universal features of
these systems. We first review minimal models defined solely in the
orbital basis, which focus on a particular subspace of orbitals, or
solely in the band basis, which rely only on the geometry of the Fermi
surface. The former, while providing important qualitative insight
into the role of the orbital degrees of freedom, do not distinguish
between high-energy and low-energy sectors and, for this reason, generally
do not go beyond mean-field. The latter allow one to go beyond mean-field
and investigate the interplay between superconducting and magnetic
orders as well as Ising-nematic order. However, they cannot capture
orbital-dependent features like spontaneous orbital order. We then
review recent proposals for a minimal model that operates in the band
basis but fully incorporates the orbital composition and symmetries
of the low-energy excitations. We discuss the results of the renormalization
group study of such a model, particularly of the interplay between
superconductivity, magnetism, and spontaneous orbital order, and compare
theoretical predictions with experiments on iron pnictides and chalcogenides.
We also discuss the impact of the glide-plane symmetry on the low-energy
models, highlighting the key role played by the spin-orbit coupling. 
\end{abstract}
\maketitle

\tableofcontents

\makeatletter
\let\toc@pre\relax
\let\toc@post\relax
\makeatother 

\section{Introduction \label{sec:Introduction}}

The discovery of a rich family of iron-based superconductors (FeSC)
with a variety of different chemical compositions \cite{Kamihara08,Rotter08},
such as LaFeAsO (1111 material), BaFe$_{2}$As$_{2}$ (122 material),
NaFeAs (111 material), and FeSe (11 material), opened a new route
to study high-temperature superconductivity. Similarly to high-$T_{c}$
cuprates, which are made of coupled CuO$_{2}$ layers, FeSC are also
layered systems made of coupled FeAs layers. In both cases, the Cu
and Fe atoms form a simple square lattice.

\begin{figure}
\begin{centering}
\includegraphics[width=0.95\columnwidth]{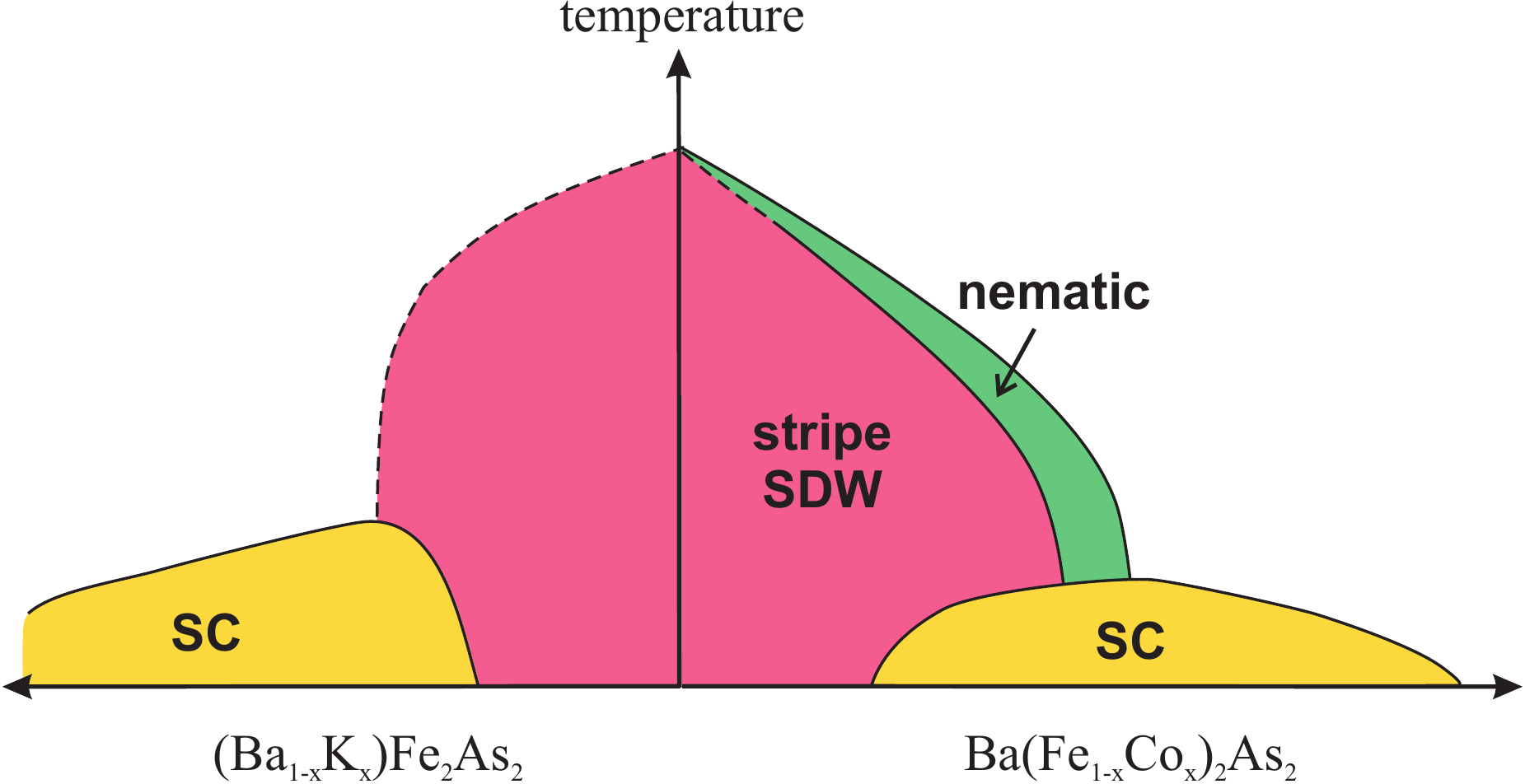} 
\par\end{centering}

\protect\protect\protect\protect\protect\protect\caption{Schematic phase diagram of electron-doped (Co-doped) and hole-doped
(K-doped) BaFe$_{2}$As$_{2}$, displaying stripe spin-density wave
(SDW) order, nematic order, and superconductivity (SC). \label{fig:_phase_diagram}}
\end{figure}

The phase diagrams of FeSC are also quite similar to those of the
cuprates. Although details of the phase diagrams vary between different
families of FeSC, most materials display the key features shown in
Fig. \ref{fig:_phase_diagram} (for reviews, see \cite{Ishida_review,Johnston10,Paglione10,Stewart11}).
Specifically, the parent compounds of most (but not all) FeSC are
magnetically ordered metals. In most cases, the magnetic order is
of a stripe type -- i.e. spins are ferromagnetically aligned in one
direction in the Fe plane and antiferromagnetically aligned in the
other. This is usually known as the $(0,\pi)/(\pi,0)$ spin-density
wave (SDW) state. Upon hole or electron doping, or upon substitution
of one pnictide atom by another, magnetic order goes away and a dome
of superconductivity emerges. In addition, there is a region on the
phase diagram where the system displays nematic order, in which the
$C_{4}$ lattice rotation symmetry is spontaneously broken ($C_{4}$
is the point group symmetry associated with a square, whereas $C_{2}$
is the point group symmetry associated with a rectangle). The nematic
order naturally coexists with the stripe magnetic order and in some
systems also coexists with superconductivity \cite{RMF14}.

Despite the similarities in their phase diagrams, there are important
differences between the cuprates and FeSC. The most pronounced difference
is that the low-energy electronic states of the cuprates arise from
Cu$^{2+}$, which is in a $3d^{9}$ electronic configuration, while
in the FeSC the low-energy states arise from Fe$^{2+}$, which is
in a $3d^{6}$ configuration. One immediate consequence of this difference
is that parent compounds of the cuprates are Mott insulators, while
parent compounds of FeSC are metals. The relevance of metallicity
of FeSC has been discussed in earlier reviews and we will not dwell
on this \cite{Mazin_review,Chubukov_review,Chubukov_book}. In this
review we focus on another immediate consequence of the difference
between $3d^{9}$ and $3d^{6}$ electronic configurations, namely
the fact that the $3d^{6}$ configuration involves five $3d$ orbitals
-- $d_{xz}$, $d_{yz}$, $d_{xy}$, $d_{x^{2}-y^{2}}$, and $d_{3z^{2}-r^{2}}$,
while $3d^{9}$ configuration contains a single $d_{x^{2}-y^{2}}$
orbital. This brings important consequences for microscopic models
constructed to describe $3d^{9}$ and $3d^{6}$ systems.

\begin{figure}
\begin{centering}
\includegraphics[width=0.8\columnwidth]{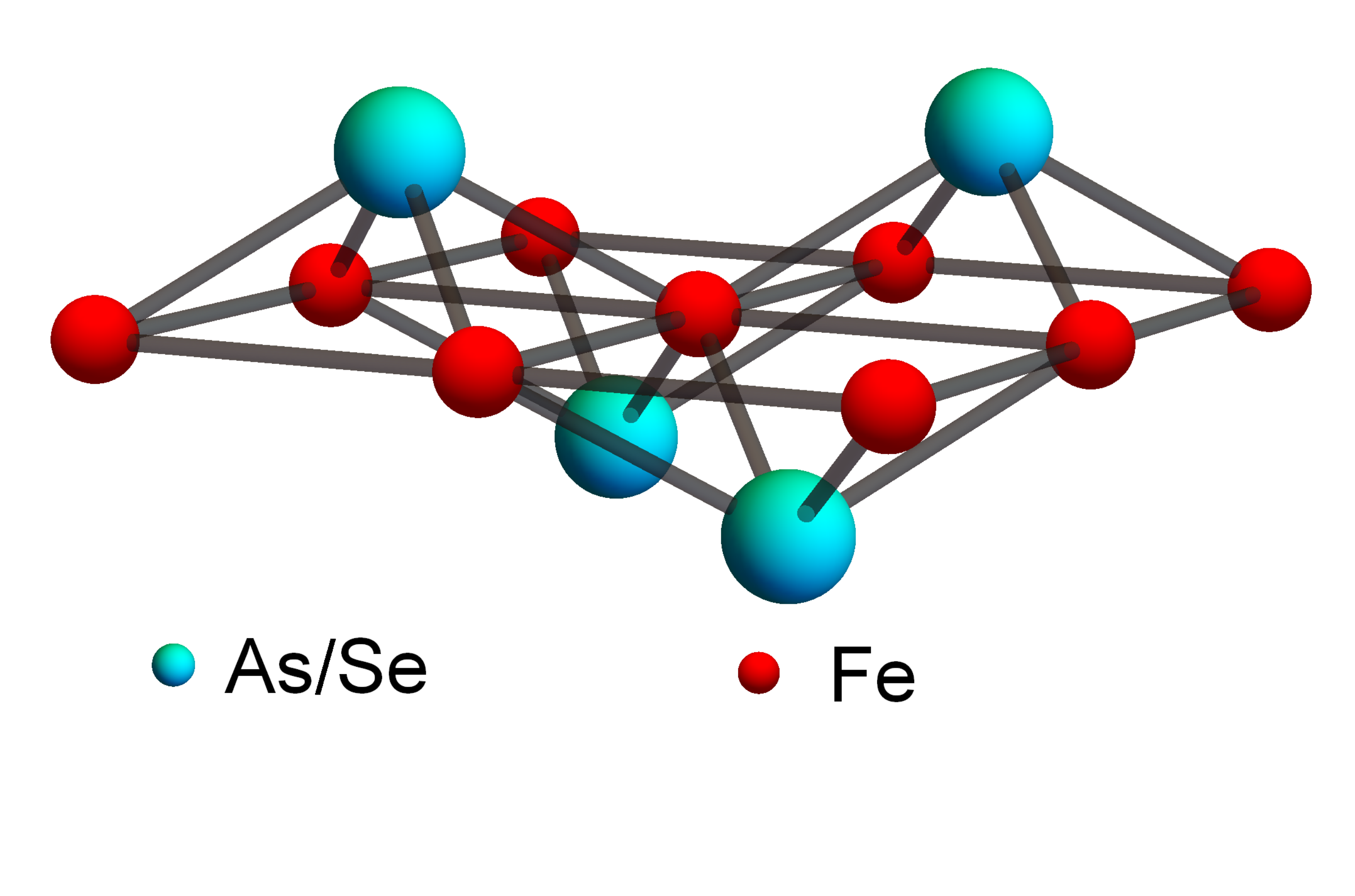} 
\par\end{centering}

\medskip{}

\begin{centering}
\includegraphics[width=0.7\columnwidth]{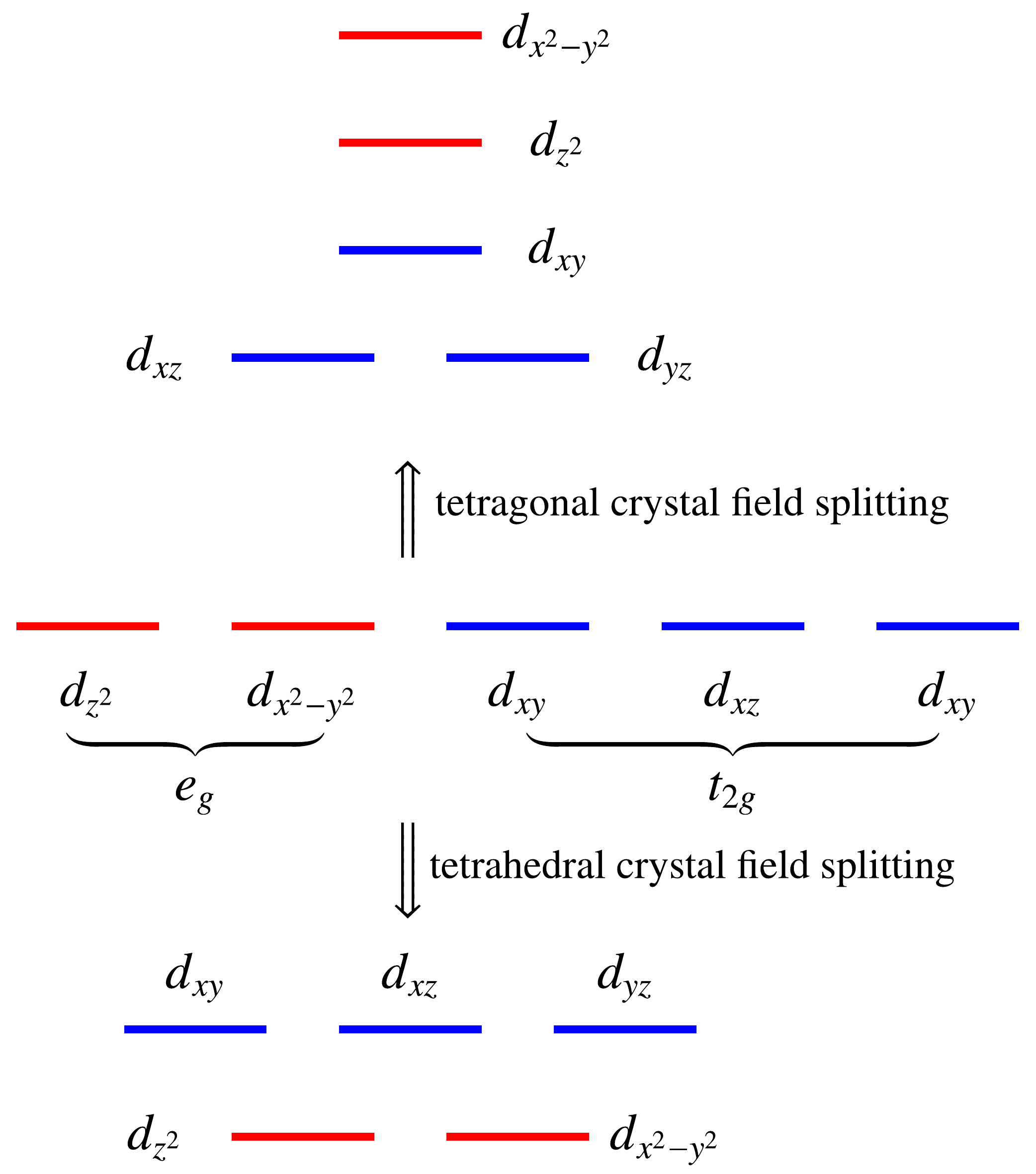} 
\par\end{centering}

\protect\protect\protect\protect\protect\protect\protect\caption{(upper panel) Schematic crystal structure of an FeAs or FeSe plane,
displaying the puckering of the As/Se atoms above and below the square
Fe plane. (lower panel) The crystal field splittings of the $3d$
$e_{g}$ (red) and $t_{2g}$ (blue) orbitals from a tetragonal and
a tetrahedral environment (see also Ref. \cite{Tesanovic09}). \label{fig_crystal_field}}
\end{figure}

In a free space, the five $3d$ orbitals are all degenerate. In a
crystalline environment the degeneracy is lifted, and the energy levels
are split into two subsets, $t_{2g}$ and $e_{g}$, with three and
two orbitals, respectively: $d_{xz}$, $d_{yz}$, and $d_{xy}$ for
$t_{2g}$ and $d_{x^{2}-y^{2}}$ and $d_{3z^{2}-r^{2}}$ for $e_{g}$
(the subscript $g$ implies that the states are symmetric under inversion).
In some multi-orbital systems, such as the manganites ($3d^{5}$)
and the cobaltates ($3d^{7}$), the crystal-field splitting is large,
and this allows one to focus on only one subset. In FeSC the situation
is more subtle because the As/Se positions alternate between the ones
above and below the center of the Fe plaquettes, as shown in Fig.
\ref{fig_crystal_field}. Because of such puckering of the As/Se atoms,
the crystalline environment experienced by Fe atoms is somewhat in
between a tetrahedral one, in which the energy of the $t_{2g}$ orbitals
is higher than that of the $e_{g}$ orbitals, and a tetragonal one,
in which the energy of the $t_{2g}$ orbitals is lower (see Fig. \ref{fig_crystal_field}
and Ref. \cite{Tesanovic09}). As a result, the crystal splitting
between the orbitals is weakened in FeSC and, consequently, all five
$d$-orbitals must be kept in the kinetic energy Hamiltonian: 
\begin{equation}
\mathcal{H}_{0}=\sum_{ij,\mu\nu}\sum_{\sigma}t_{\mu i,\nu j}d_{\mu,i\sigma}^{\dagger}d_{\nu,j\sigma}\label{H0_orb}
\end{equation}

Here $d_{\mu,i\sigma}^{\dagger}$ creates an electron at site $i$
and orbital $\mu$ ($\mu=1,...,5$) with spin $\sigma$, and $t_{\mu i,\nu j}$
are hopping amplitudes. The diagonal terms describe the dispersions
of electrons from separate orbitals, whereas the non-diagonal terms
account for the hopping from one orbital to the other. The latter
give rise to hybridization of the eigenstates from different orbitals.
The hopping parameters $t_{\mu i,\nu j}$ can either be directly fit
to the band dispersions obtained in first-principle calculations \cite{Graser09,Sadovskii08}
or calculated in a perturbative Slater-Koster approach as functions
of the distance between Fe and As \cite{Bascones09}. In the former
case, one usually needs several-neighbors hopping parameters to achieve
a good fit, which makes the fitting procedure itself involved. In
the latter, one has to rely on first principle calculations to get
several parameters which are inputs for the Slater-Koster approach.

Both diagonal and non-diagonal $t_{\mu i,\nu j}$ between different
sites $i$ and $j$ result from either a direct hopping from one Fe
site to the other, or indirect hopping via As/Se. Because of the two
non-equivalent position of the As/Se atoms with respect to the Fe
plane, the fundamental period in the Fe plane is the distance between
next-nearest-neighbor Fe atoms, i.e. the crystallographic unit cell
must contain two Fe atoms. Thus, to respect all symmetries of the
lattice, the kinetic energy must include ten Fe orbitals \cite{Eschrig09,Boeri11}.

Because $\mathcal{H}_{0}$ is not diagonal in the orbital basis, one
invariably needs to diagonalize $10\times10$ matrices in the orbital
space to obtain quasiparticle dispersions. The diagonalization yields
a $10$-band non-interacting Hamiltonian

\begin{equation}
\mathcal{H}_{0}=\sum_{m=1}^{N}\varepsilon_{m}\left(\mathbf{k}\right)c_{m,\mathbf{k}\sigma}^{\dagger}c_{m,\mathbf{k}\sigma}\label{H0_band}
\end{equation}
where $c_{m,\mathbf{k}\sigma}^{\dagger}$ creates an electron in band
$m$ with momentum $\mathbf{k}$ and spin $\sigma$. The band and
orbital operators are related by the matrix elements associated with
the diagonalization of $\mathcal{H}_{0}$, $a_{m\mu}\left(\mathbf{k}\right)\equiv\left\langle m\mathbf{k}\left|\right.\mu\right\rangle $

\begin{equation}
c_{m,\mathbf{k}\sigma}=\sum_{\mu}a_{m\mu}\left(\mathbf{k}\right)d_{\mu,\mathbf{k}\sigma}\label{band_orbital}
\end{equation}

Although diagonalizing $10\times10$ matrices is numerically straightforward,
it becomes difficult to gain qualitative understanding and insights
into the problem once interactions are included, even if the Coulomb
interaction is heavily screened and can be approximated as a local
one. In the cuprates, the interaction between electrons from a single
orbital is fully described by the Hubbard repulsion $U$. In FeSC,
there are at least four onsite interaction terms involving $3d$ electrons
\cite{Graser09,Kuroki1,Ikeda10}: 
\begin{align}
\mathcal{H}_{\mathrm{int}} & =U\sum_{i,\mu}n_{\mu,i\uparrow}n_{\mu,i\downarrow}+J\sum_{i,\mu<\nu}\sum_{\sigma,\sigma'}d_{\mu,i\sigma}^{\dagger}d_{\nu,i\sigma'}^{\dagger}d_{\mu,i\sigma'}d_{\nu,i\sigma}\nonumber \\
 & +U'\sum_{i,\mu<\nu}n_{\mu,i}n_{\nu,i}+J'\sum_{i\mu\neq\nu}d_{\mu,i\uparrow}^{\dagger}d_{\mu,i\downarrow}^{\dagger}d_{\nu,i\downarrow}d_{\nu,i\uparrow}\label{H_int_orb}
\end{align}

Here $U$ is the usual Hubbard repulsion between electrons on the
same orbitals, $U'$ is the onsite repulsion between electrons on
different orbitals, $J$ is the Hund's exchange that tends to align
spins at different orbitals, and $J'$ is another exchange term, often
called the pair-hoping term. The presence of four different interactions
enlarges the parameter space and makes calculations much more involved.

Several works attempted to simplify the $U,U',J,J'$ model by invoking
rotational invariance to argue that the interaction must be expressed
in terms of the squares of the total number and the total spin of
$3d$-electrons on a given site, $\sum_{\mu,\alpha}d_{\mu,i\alpha}^{\dagger}d_{\mu,i\alpha}$
and $\sum_{\mu,\alpha}d_{\mu,i\alpha}^{\dagger}{\vec{\sigma}}_{\alpha\beta}d_{\mu,i\beta}$,
respectively. This would reduce the number of independent interaction
terms in the Hamiltonian to two via the relationships $U'=U-2J$ and
$J'=J$. However, this would be true if As/Se states were irrelevant.
This is not the case in FeSC because the hopping from one Fe site
to the other partly goes through As/Se atoms. These As/Se states must
then be included also in the interaction term. They are high-energy
states (around $5$ eV away from the Fermi level) and one can integrate
them out for studies of the physics at much smaller scales, related
to magnetism, superconductivity, and electronic nematic order. But
by integrating out As/Se states, one breaks spin rotational invariance
of the $3d$ orbitals, and, as a result, breaks the relations $U'=U-2J$
and $J'=J$. Besides, by integrating high-energy parts of the spectra
of the Fe $3d$ orbitals, one necessarily generates interactions between
neighboring Fe sites. This additionally breaks the relations between
$U'$ and $U-2J$ and between $J'$ and $J$.

All these complications raise the important question of whether one
can construct a sensible and simpler minimal microscopic model to
capture the low-energy physics of the FeSC without the need for $10\times10$
(or $5\times5$) matrices and a large number of interaction terms.
In this review, we discuss microscopic models that have been proposed
and solved to understand distinct aspects of the FeSC. We will highlight
the advantages of these models and their drawbacks.

In Section \ref{sec:Orbital-basis-models} we discuss approximate
orbital models with a smaller number of $3d$ Fe orbitals and review
the computations done solely in terms of orbital operators. In Section
\ref{sec:Band-basis-models} we discuss the models which use the experimental
knowledge of the location of the Fermi surfaces as an input and analyze
the effects of the interactions in the band basis, without referring
to the orbital content of the excitations. In Section \ref{sec:Hybrid-band-orbital-models}
we discuss works in which the analysis of the instabilities is done
in the band basis, but the interactions in all channels are constructed
from the orbital basis and retain the full memory about the orbital
content of the low-energy states. We review RPA studies of magnetically-mediated
pairing interaction and discuss recent works on the interplay between
superconductivity, magnetism, and a spontaneous orbital order. We
discuss the minimal model for the analysis of the competing orders
and show the results of the renormalization group (RG) study of such
a model.

The models in Sections II-IV are constructed in the 1-Fe unit cell
and as such neglect the Fe-As/Se hybridization. In Section \ref{sec:1-Fe-versus-2-Fe}
we analyze the consequences of this approximation and discuss extensions
of these models to the 2-Fe unit cell. We first show how the dispersions
change if we just convert from 1-Fe to 2-Fe basis, then briefly discuss
the effect of additional terms with momentum transfer $(\pi,\pi)$
in $\mathcal{H}_{0}$ and $\mathcal{H}_{\mathrm{int}}$, which originate
from the actual non-equivalence of neighboring Fe cells in 1-Fe basis,
and then discuss the role of spin-orbit coupling. We present concluding
remarks in Section \ref{sec:Concluding-remarks}.

The main points of this comparative analysis are the following: 
\begin{itemize}
\item Approximate orbital models (hereafter called orbital-basis models)
with two and three orbitals are attractively simple and offer interesting
insights into the orbital physics of FeSCs. However, because the analysis
in the orbital basis does not rely on the presence of the Fermi surface,
it necessarily involves excitations with all momenta. It turns out
that the three-band model correctly captures the low-energy sector
of the full five-orbital model, but \textit{cannot} correctly describe
how the excitations evolve from one low-energy sector to the other.
The minimum model which correctly describes both the low-energy sectors
and the evolution of excitations between them must involve at least
four orbitals. 
\item Multi-band models (hereafter called band-basis models) with phenomenologically-derived
interactions between low-energy electronic states offer an appealing
and simple framework to study superconductivity and magnetism, the
interplay between the two, and vestigial Ising-nematic order caused
by magnetic fluctuations. They ignore, however, the orbital content
of the low-energy states, and as such they are generally blind to
phenomena involving orbital physics. 
\item The models which operate in the band basis but use the full knowledge
of the orbital content of the low-energy excitations (hereafter called
orbital-projected band models) seem to be the most promising ones.
These models include three orbitals ($d_{xz},d_{yz}$, and $d_{xy}$),
from which the low-energy excitations are constructed, and the interactions
between low-energy states contain angle-dependent prefactors that
reflect the orbital composition of the Fermi surfaces. The full model
of this kind still contains too many coupling constants, but most
of the physics is captured already by simplified models with a smaller
number of couplings. 
\item The phenomena associated with the sizable spin-orbit coupling of the
FeSC can only be captured in the 2-Fe unit cell. The orbital-projected
band models can naturally be extended to this case without the need
to double the number of terms in the kinetic part of the Hamiltonian. 
\end{itemize}
Throughout this review we assume that none of the low-energy electronic
states is localized by interactions. We believe this is a sensible
starting point, as most of the FeSC are metals, with a pronounced
Drude peak in the AC conductivity (see, for instance, \cite{Degiorgi15}).
This does not imply that we consider weak coupling. Rather, in the
analysis of band models in Sections \ref{sec:Band-basis-models} and
\ref{sec:Hybrid-band-orbital-models} we assume that the renormalizations
by high-energy electronic states change the ``band masses'' and
the offset energies of low-energy excitations, and modify the residues
$Z_{i}$ of low-energy states, while keeping these excitations coherent.
The renormalized dispersion parameters can be extracted from the experimental
data on the electronic dispersion, and the residues $Z_{i}$ can be
incorporated into the interactions. This indeed changes the values
of the bare interaction terms, but we will see that the interactions
flow under RG (renormalization group) towards universal values, independent
on the bare ones. The actual (measured) electronic excitations do
indeed have a finite lifetime $1/\tau$. Our assumptions imply that
the dominant contribution to $1/\tau$ for each low-energy fermion
comes from the processes involving only low-energy states, i.e., $1/\tau$
is not an input but rather has to be determined within the low-energy
analysis.

Alternative low-energy models have been proposed based on Heisenberg
or Kugel-Khomskii type Hamiltonians \cite{JPHu08,Si08,Kruger09,Applegate12},
which effectively assume that the system is an insulator. The argument
here is that, while FeSC do display the metallic behavior at low temperatures,
some orbitals may be either localized or near localization \cite{WKu10,Dagotto_review,Si12,Ding_review}\textbf{.}
Because of space constraints, we will not discuss these models further
in the present review. We also will not discuss here an interesting
concept that the Hund's interaction $J$ plays an important role in
promoting bad metallic (but still metallic) behavior up to large values
of the Hubbard $U$ \cite{Bierman_correlations,Kotliar11,Georges13,Medici14,Valenti_correlations,Bascones_review,Medici_book}.
As we said, in the next three sections we discuss the electronic structure
and the interplay between superconductivity, magnetism, and nematic
order within the 1-Fe unit cell, i.e., we restrict ourselves to the
five-orbital model ($N=5$). Physically, this assumption implies that
we neglect terms in the Hamiltonian with momentum transfer $(\pi,\pi)$
and also neglect the spin-orbit interaction.

\section{Orbital-basis models \label{sec:Orbital-basis-models}}

In this section we focus on approximate models defined and analyzed
in the orbital basis. We first discuss the non-interacting part of
the Hamiltonian $\mathcal{H}_{0}$, and investigate whether it is
possible to restrict the number of orbitals to $2$ or $3$ and keep
the symmetry constraints intact. We then briefly review the studies
of interactions in the orbital basis.

\subsection{Non-interacting Hamiltonian}

We discuss the full 5-orbital model in subsection \ref{sub:Five-orbital-model}
and discuss the models that restrict the number of orbitals to $2$
and $3$ in Subsections \ref{sub:Two-orbital-model} and \ref{sub:Three-orbital-model},
respectively. We remind that the goal to analyze models with smaller
number of orbitals is to simplify the analysis in order to gain qualitative
understanding and insights into the issue of competing instabilities,
once interactions are included.

\subsubsection{Five-orbital model \label{sub:Five-orbital-model}}

The non-interacting part of the Hamiltonian of the five-orbital model
is given by $\mathcal{H}_{0}$ in Eq. (\ref{H0_orb}). Taking its
Fourier transform gives:

\begin{equation}
\mathcal{H}_{0}=\sum_{\mu\nu}\left[\epsilon_{\mu\nu}\left(\mathbf{k}\right)-\bar{\mu}\delta_{\mu\nu}\right]d_{\mu,\mathbf{k}\sigma}^{\dagger}d_{\nu,\mathbf{k}\sigma} \label{matrix_5orb}
\end{equation}
where $\bar{\mu}$ is the chemical potential. The explicit expressions for the tight-binding dispersions $\epsilon_{\mu\nu}\left(\mathbf{k}\right)$
with hopping up to fourth-neighbors are given in Appendix A, together with the values of the tight-binding parameters of Ref. \cite{Graser09} (see Table \ref{tab_5orb}). 

\begin{figure*}
\begin{centering}
\includegraphics[width=0.9\textwidth]{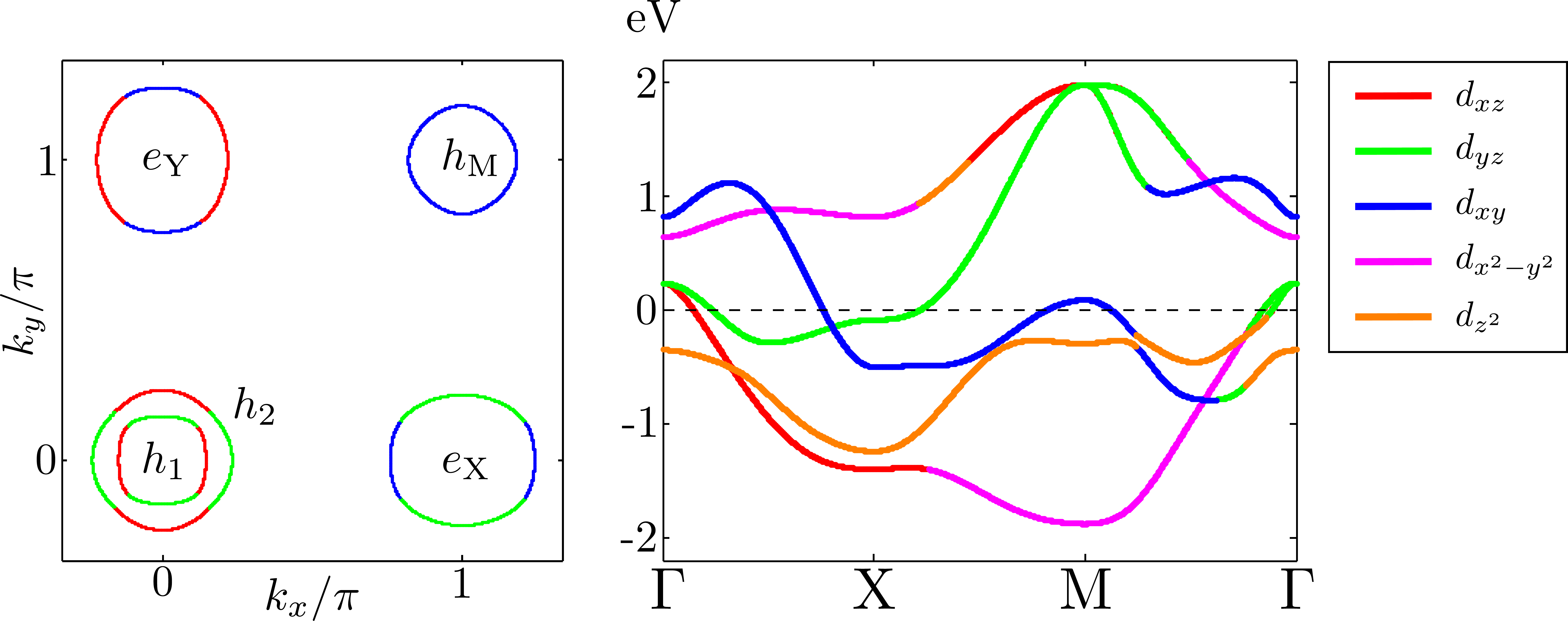} 
\par\end{centering}

\protect\protect\protect\protect\protect\protect\protect\caption{Tight-binding dispersion of Ref. \cite{Ikeda10} and the resulting
Fermi surface in the 1-Fe BZ. The bands are colored according to the
orbital that contributes the largest spectral weight. \label{fig_5orbital_band}}
\end{figure*}

In Fig. \ref{fig_5orbital_band},
we show the band dispersion and the Fermi surfaces corresponding to
the parameters for LaFeAsO from Ref. \cite{Ikeda10}. The Fermi surfaces
are colored according to which orbital gives the largest spectral
weight, the latter being defined by the matrix element $\left|a_{m\mu}\left(\mathbf{k}\right)\right|^{2}$
in Eq. (\ref{band_orbital}). Along the Fermi surface, we have ${\mathbf{k}}={\mathbf{k}}_{F}$
and $\left|a_{m\mu}\left(\mathbf{k}\right)\right|=\left|a_{m\mu}\left(\theta\right)\right|$,
where $\theta$ is the angle with respect to $k_{x}$. The Fermi surface
is composed of small pockets centered at high-symmetry points of the
Brillouin zone (BZ), namely $\Gamma=\left(0,0\right)$, $X=\left(\pi,0\right)$,
$Y=\left(0,\pi\right)$, and $M=\left(\pi,\pi\right)$ (all momenta
hereafter are given in units of $1/a$, where $a$ is the length of
the corresponding unit cell).

\begin{figure*}
\begin{centering}
\includegraphics[width=1\textwidth]{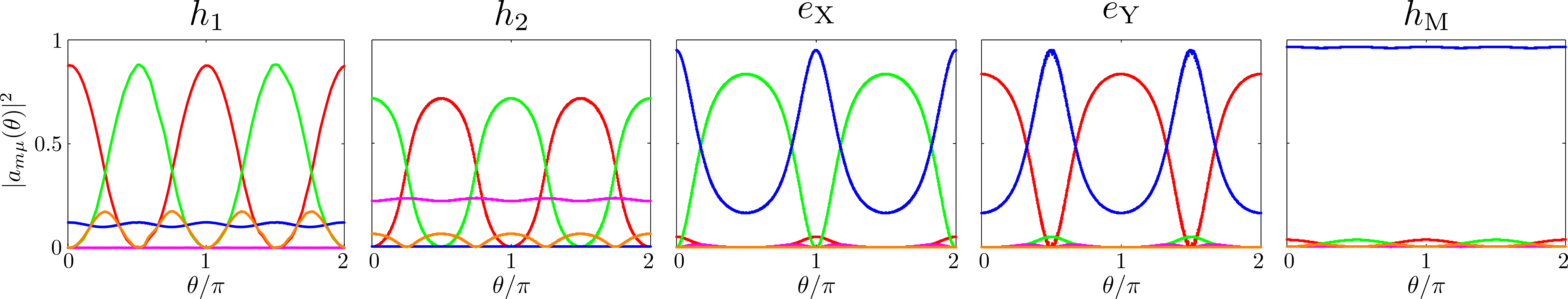} 
\par\end{centering}

\protect\protect\protect\protect\protect\protect\protect\caption{Orbital spectral weight $\left|a_{m\mu}\left(\theta\right)\right|^{2}$
of each Fermi surface as function of the angle $\theta$ measured
relative to the $k_{x}$ axis. The color code is the same as in Fig.
\ref{fig_5orbital_band}. \label{fig_5orbital_spectrum}}
\end{figure*}

There are two hole-like bands that cross the Fermi level near the
$\Gamma$ point, giving rise to two hole pockets $h_{1}$ and $h_{2}$.
As shown in Fig. \ref{fig_5orbital_spectrum}, the angle-dependent
spectral weights $\left|a_{h_{i}\mu}\left(\theta\right)\right|^{2}$
on these Fermi pockets mostly come from the $d_{xz}$ and $d_{yz}$
orbitals. Similarly, two electron-like bands cross the Fermi level
near the $X$ and $Y$ points, giving rise to two electron pockets
$e_{X}$ and $e_{Y}$. The spectral weight on the pocket $e_{X}$
is dominated by the $d_{yz}$ and $d_{xy}$ orbitals, whereas the
spectral weight on the pocket $e_{Y}$ is dominated by $d_{xz}$ and
$d_{xy}$. Tetragonal symmetry enforces the following conditions,
which can be readily observed in the figure:

\begin{align}
\left|a_{e_{X}d_{yz}}\left(\theta\right)\right|^{2} & =\left|a_{e_{Y}d_{xz}}\left(\theta+\pi/2\right)\right|^{2}\nonumber \\
\left|a_{e_{X}d_{xy}}\left(\theta\right)\right|^{2} & =\left|a_{e_{Y}d_{xy}}\left(\theta+\pi/2\right)\right|^{2}
\end{align}

An additional hole-pocket $h_{M}$ crosses the Fermi level near the
$M$ point. Its spectral weight is almost entirely due to the $d_{xy}$
orbital, as shown in the figure. Inspection of the band dispersion
reveals that the top of this hole-like band is very close to the Fermi
level, and that small changes in the crystal lattice parameters or
in the chemical potential may make it sink below the Fermi level,
effectively erasing the corresponding Fermi pocket \cite{Kuroki2}.
Thus, the presence of this third hole pocket is rather material dependent.
It is absent in NaFeAs and FeSe, but present in BaFe$_{2}$As$_{2}$
and LiFeAs. Note that, while all hole pockets must have $C_{4}$-symmetric
shapes, the electron pockets have $C_{2}$ symmetric shapes, which
are related to each other by a $\pi/2$ rotation.

This generic Fermi surface can be tuned by changes in the chemical
potential, which is achieved via electron doping (such as Co-doped
NaFeAs) or hole doping (such as Na-doped BaFe$_{2}$As$_{2}$), see
Refs. \cite{Canfield_review,HHWen_review}. For sufficiently electron-doped
systems, such as K$_{1-y}$Fe$_{2-x}$Se$_{2}$ and electrostatically
gated FeSe, the hole pockets disappear and only electron pockets remain.
Analogously, for systems with strong hole doping, such as K-doped
BaFe$_{2}$As$_{2}$, the electron pockets disappear and only hole
pockets are left. Isovalent substitution, achieved e.g. via gradual
replacement of As by P or Fe by Ru in 122 systems, alters the Fermi
surface due to the changes in the crystal lattice parameters (more
prominently the Fe-As distance) and also by the disorder potential
which isovalent substitution introduces to the system.

\subsubsection{Two-orbital model \label{sub:Two-orbital-model}}

It is clear from Fig. \ref{fig_5orbital_band} that not all five orbitals
contribute equally to the low-energy states near the Fermi energy.
In fact, the Fermi surface states are made almost exclusively from
$d_{xz}$, $d_{yz}$, and $d_{xy}$ orbitals (see Figs. \ref{fig_5orbital_band}
and \ref{fig_5orbital_spectrum}). One can then conjecture that at
least some of the physics of FeSC can be understood within a simplified
model with only this subset of orbitals. Raghu \emph{et al.} assumed,
on top of this, that the hopping via the $d_{xy}$ orbital could be
integrated out and absorbed into next-nearest-neighbor hopping terms
involving $d_{xz}$ and $d_{yz}$ orbitals \cite{Raghu08}. They proposed
the effective two-orbital model: 
\begin{equation}
\mathcal{H}_{0}=\sum_{\mu\nu=xz,yz}\left[\epsilon_{\mu\nu}\left(\mathbf{k}\right)-\bar{\mu}\delta_{\mu\nu}\right]d_{\mu,\mathbf{k}\sigma}^{\dagger}d_{\nu,\mathbf{k}\sigma}\label{2_orbitals}
\end{equation}
with tight-binding parameters (see Table \ref{tab_2orb} in Appendix
A): 
\begin{align}
\epsilon_{xx,xz}\left(\mathbf{k}\right) & =-2t_{1}\cos k_{x}-2t_{2}\cos k_{y}-4t_{3}\cos k_{x}\cos k_{y}\nonumber \\
\epsilon_{yz,yz}\left(\mathbf{k}\right) & =-2t_{2}\cos k_{x}-2t_{1}\cos k_{y}-4t_{3}\cos k_{x}\cos k_{y}\nonumber \\
\epsilon_{xz,yz}\left(\mathbf{k}\right) & =\epsilon_{yz,xz}\left(\mathbf{k}\right)=-4t_{4}\sin k_{x}\sin k_{y}\label{aux_2_orbitals}
\end{align}

\begin{figure}
\begin{centering}
\includegraphics[width=0.7\columnwidth]{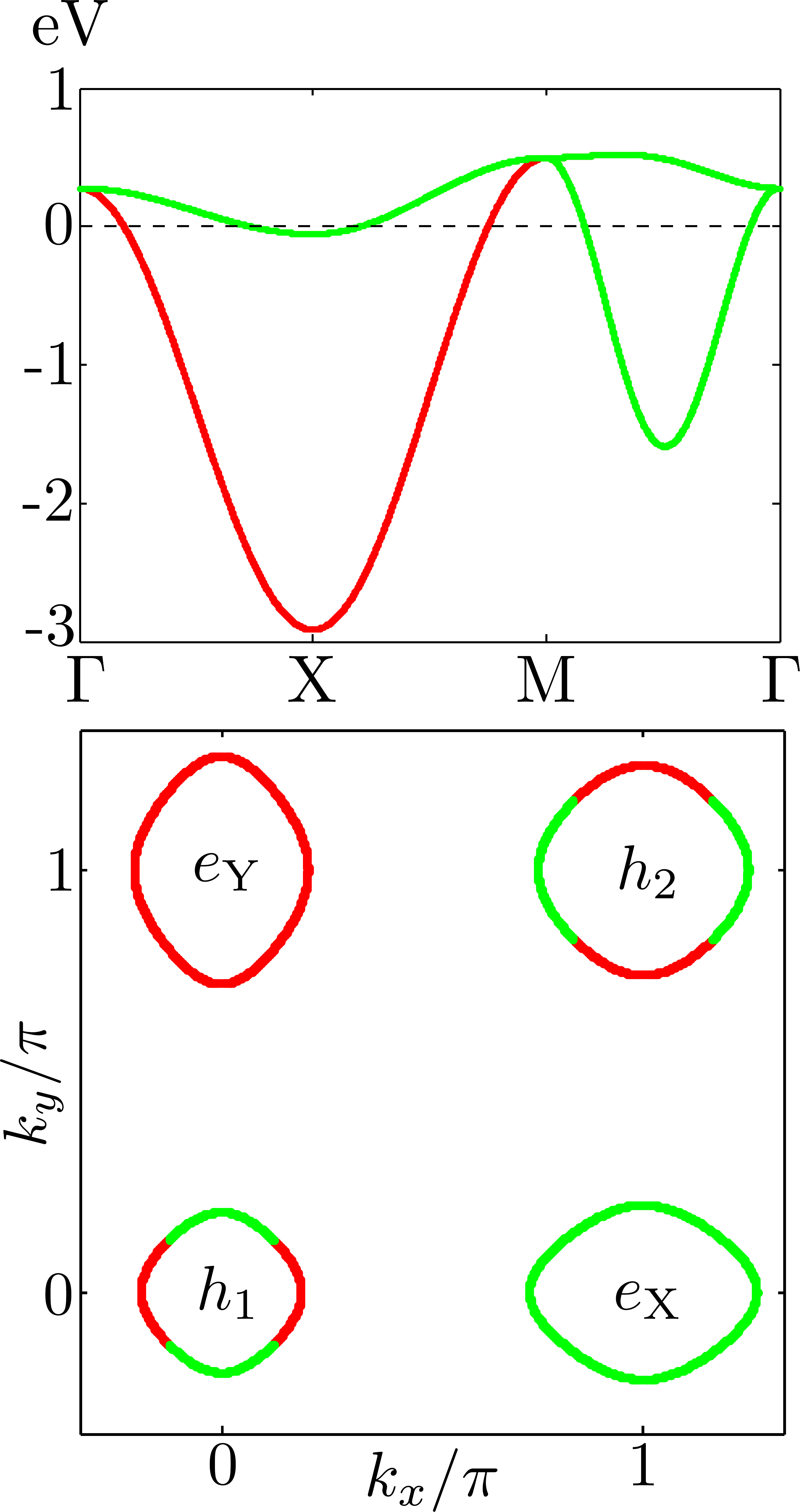} 
\par\end{centering}

\protect\protect\protect\protect\protect\protect\protect\caption{Two-orbital model of Ref. \cite{Raghu08}: band dispersion (upper
panel) and the Fermi surface (lower panel). In the latter, the Fermi
surface is colored according to the orbital that contributes the largest
spectral weight (red for $d_{xz}$ and green for $d_{yz}$). The tight-binding
parameter used here are those from Ref. \cite{Sknepnek09}. \label{fig_2orbital}}
\end{figure}

Fig. \ref{fig_2orbital} shows the corresponding band dispersion and
the Fermi surfaces. In contrast to the 5-orbital model, one of the
two $d_{xz}/d_{yz}$ hole pockets is centered at the $M$ point instead
of the $\Gamma$ point. Such an artifact of the 2-orbital model was
not originally considered to be problematic because in the true crystallographic
unit cell, containing two Fe atoms, the $M$ point is folded onto
the $\Gamma$ point, restoring the existence of two hole pockets at
the center of the BZ.

The simplicity of the 2-orbital model, which can be conveniently written
in terms of Pauli matrices in the orbital space, led to many studies
about the electronic properties and instabilities of this model \cite{Moreo_2orbital,DHLee_Dirac,Ghaemi11,CSTing13,Yamase13,Nevidomskyy,Coleman16,Dumitrescu15,Chubukov_Xing}.
Despite its appeal, there are issues with this model that go beyond
the incorrect position of one of the hole pockets. Most importantly,
the 2-orbital model does not respect all the symmetries of the FeAs
plane -- in particular, the symmetry related to a translation by $\left(\frac{1}{2},\frac{1}{2}\right)$
followed by a mirror reflection with respect to the $xy$ plane. As
explained in Ref. \cite{Vafek13}, the two hole pockets formed by
the $d_{xz}$ and $d_{yz}$ orbitals must be odd under this symmetry,
whereas in the 2-orbital model only one of the pockets is odd. The
absence of the $d_{xy}$ orbital is also a potential issue, as it
has been argued to play an important role in certain FeSC \cite{Kotliar11,Valenti_LiFeAs,Schmalian_Kotliar}.

\subsubsection{Three-orbital model \label{sub:Three-orbital-model}}

A possible way to remedy the issues of the 2-orbital model is to include
the third orbital that contributes significantly to the spectral weight
of the low-energy states, namely the $d_{xy}$ orbital. The corresponding
3-orbital model is described by \cite{Lee_Wen08,Daghofer10} 
\begin{equation}
\mathcal{H}_{0}=\sum_{\mu\nu=xz,yz,xy}\left[\epsilon_{\mu\nu}\left(\mathbf{k}\right)-\bar{\mu}\delta_{\mu\nu}\right]d_{\mu,\mathbf{k}\sigma}^{\dagger}d_{\nu,\mathbf{k}\sigma}\label{3_orbitals}
\end{equation}
with the tight-binding dispersions (see Table \ref{tab_3orb} in Appendix
A):

\begin{align}
\epsilon_{xz,xz}\left(\mathbf{k}\right) & =-2t_{1}\cos k_{x}-2t_{2}\cos k_{y}-4t_{3}\cos k_{x}\cos k_{y}\nonumber \\
\epsilon_{yz,yz}\left(\mathbf{k}\right) & =-2t_{2}\cos k_{x}-2t_{1}\cos k_{y}-4t_{3}\cos k_{x}\cos k_{y}\nonumber \\
\epsilon_{xy,xy}\left(\mathbf{k}\right) & =-2t_{5}\left(\cos k_{x}+\cos k_{y}\right)-4t_{6}\cos k_{x}\cos k_{y}+\Delta_{\mathrm{CF}}\label{aux_3_orbitals_1}
\end{align}
as well as:

\begin{align}
\epsilon_{xz,yz}\left(\mathbf{k}\right) & =-4t_{4}\sin k_{x}\sin k_{y}\nonumber \\
\epsilon_{xz,xy}\left(\mathbf{k}\right) & =-2it_{7}\sin k_{x}-4it_{8}\sin k_{x}\cos k_{y}\nonumber \\
\epsilon_{yz,xy}\left(\mathbf{k}\right) & =-2it_{7}\sin k_{y}-4it_{8}\sin k_{y}\cos k_{x}\label{aux_3_orbitals_2}
\end{align}
where $\Delta_{\mathrm{CF}}$ is the crystal field splitting. Note
that $\epsilon_{\mu\nu}\left(\mathbf{k}\right)=\epsilon_{\nu\mu}^{*}\left(\mathbf{k}\right)$.
Although more complex than the 2-orbital model, the 3-orbital model
is still much simpler than the 5-orbital one, and may be conveniently
expressed in terms of the eight $3\times3$ Gell-Mann matrices.

\begin{figure}
\begin{centering}
\includegraphics[width=0.7\columnwidth]{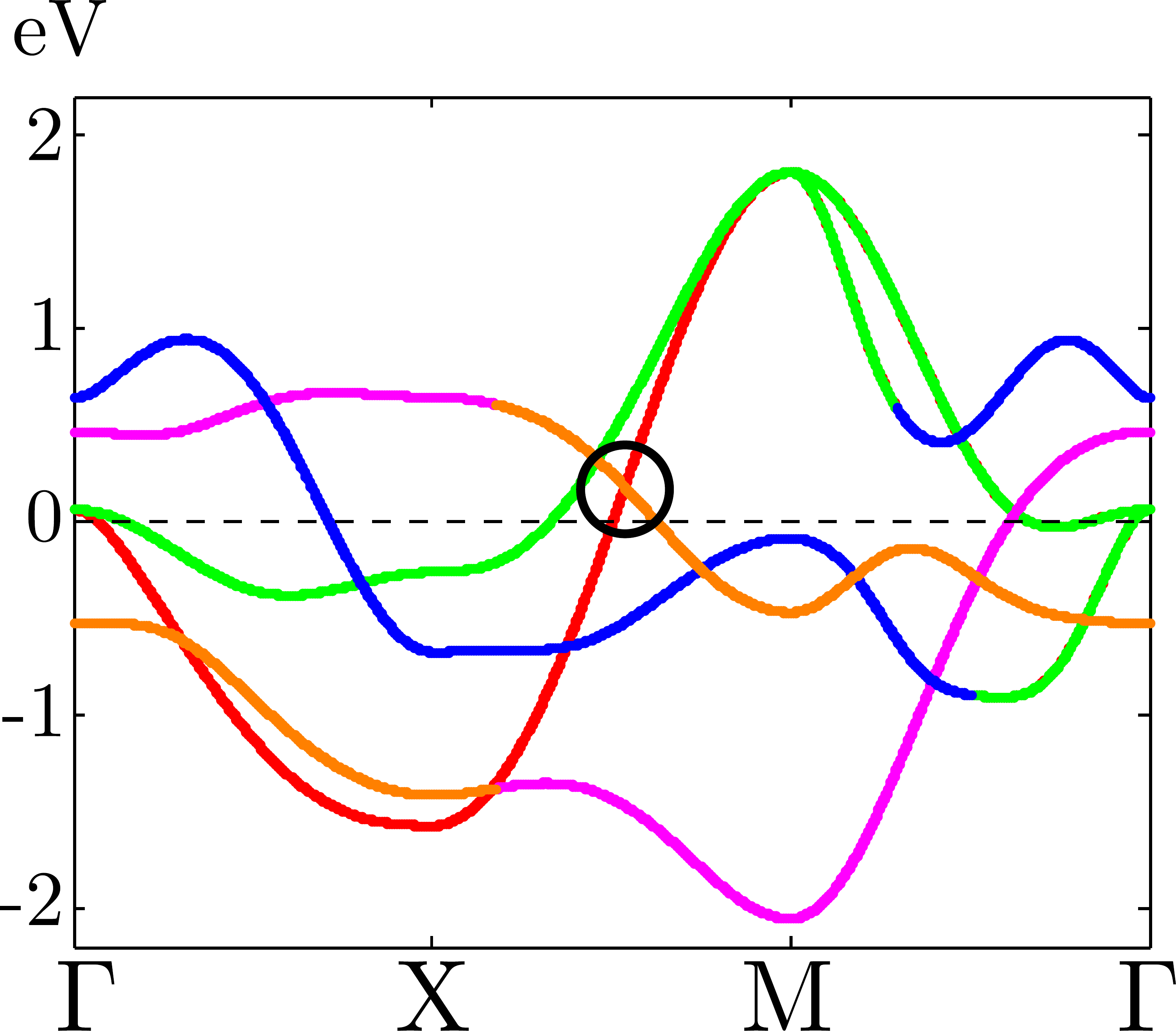} 
\par\end{centering}

\protect\protect\protect\protect\protect\protect\protect\caption{The five-orbital model of Fig. \ref{fig_5orbital_band} with the hybridization
between the $t_{2g}$ orbitals ($d_{xz},d_{yz},d_{xy}$) and the $e_{g}$
orbitals ($d_{x^{2}-y^{2}},d_{z^{2}}$) turned off. The orbital color
code is the same as Fig. \ref{fig_5orbital_band}. The absence of
hybridization leads to a spurious crossing of one of the $t_{2g}$
bands at the Fermi level (highlighted area). \label{fig_no_hybridization}}
\end{figure}

The main issue with restricting the orbitals to the $t_{2g}$ subspace
(i.e. $d_{xz}$, $d_{yz}$, and $d_{xy}$) is the presence of an additional,
spurious Fermi surface pocket due to the lack of hybridization with
the $e_{g}$ orbitals ($d_{x^{2}-y^{2}}$ and $d_{z^{2}}$) \cite{Lee_Wen08,Vafek13}.
To illustrate this point, we consider again the 5-orbital model of
Fig. \ref{fig_5orbital_band} but turn off the hybridization between
the $t_{2g}$ and $e_{g}$ orbitals. The result, shown in Fig. \ref{fig_no_hybridization},
reveals an additional hole-like pocket near the $M$ point due to
the fact that one of the (hybridized) $e_{g}$ bands and the $d_{xz}/d_{yz}$-dominated
band cross the Fermi level. A comparison with Fig. \ref{fig_5orbital_band}
shows that it is the hybridization between this $e_{g}$ band and
the $d_{xz}/d_{yz}$ band that prevents both bands from crossing the
Fermi level. This clearly indicates that all five orbitals are necessary
to obtain the correct geometry of the Fermi pockets, despite the fact
that the low-energy states in the correct geometry are composed only
from $t_{2g}$ orbitals.

\begin{figure}
\begin{centering}
\includegraphics[width=0.7\columnwidth]{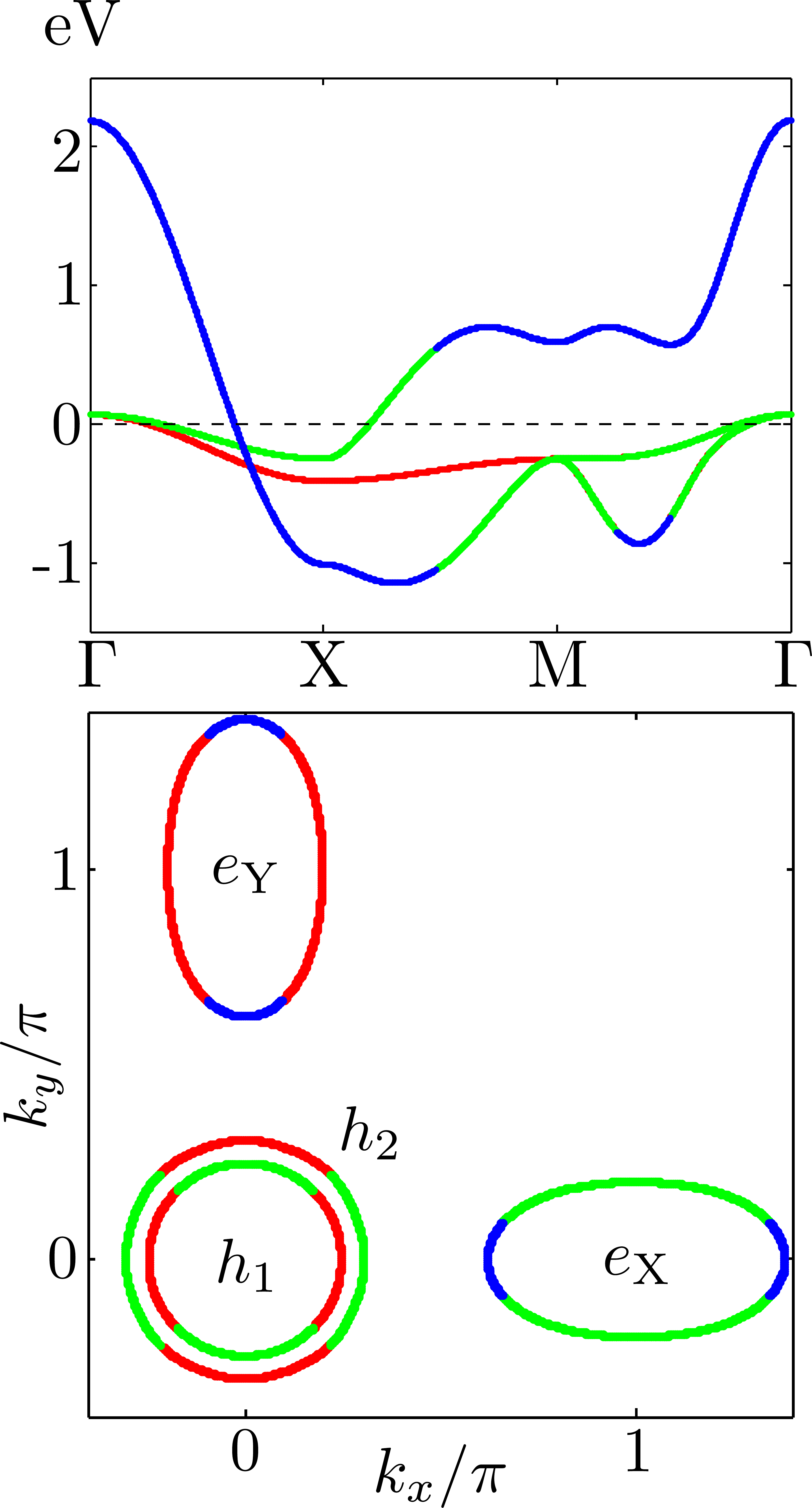} 
\par\end{centering}

\protect\protect\protect\protect\protect\protect\protect\caption{Three-orbital model of Ref. \cite{Daghofer10}: band dispersion (upper
panel) and Fermi surface (lower panel). In the latter, the Fermi surface
is colored according to the orbital that contributes the largest spectral
weight (red for $d_{xz}$, blue for $d_{xy}$, and green for $d_{yz}$).
\label{fig_3orbital}}
\end{figure}

This generic difficulty with the 3-orbital model can be overcome by
changing the tight-binding parameters in Eq. \ref{matrix_5orb} to
alter the ordering of the bands at the $M$ point. In particular,
one can move the $d_{xz}/d_{yz}$ bands below the Fermi level at the
$M$ point, while keeping the $d_{xy}$ band at $M$ above the Fermi
level \cite{Daghofer10}. As a result, the spurious Fermi pocket is
removed, as shown in Fig. \ref{fig_3orbital}. While this alternative
is appealing, it cannot capture the $d_{xy}$ hole pocket at the $M$
point without reintroducing the spurious $d_{xz}/d_{yz}$ pocket around
$M$.

\subsection{Order parameters}

\label{new_ac}

The order parameters whose condensation leads to density-waves, superconductivity,
and orbital order, are bilinear combinations of fermions in the particle-hole
and particle-particle channels, either with zero transferred momentum
(or total momentum, in the case of superconductivity), or with a finite
momentum. In general, each order parameter is a $5\times5$ matrix
in the orbital space \cite{Graser09,Daghofer10,Gastiasoro15,Christensen15}.
The CDW and SDW order parameters are 
\begin{eqnarray}
 &  & {\Delta}_{\mathrm{CDW,j}}^{\mu\nu}(\mathbf{k})=d_{\mu,\mathbf{k}\alpha}^{\dagger}{\delta}_{\alpha\beta}d_{\nu,\mathbf{k}+\mathbf{Q}_{j}\beta}+h.c.\nonumber \\
 &  & {\Delta}_{\mathrm{iCDW,j}}^{\mu\nu}(\mathbf{k})=id_{\mu,\mathbf{k}\alpha}^{\dagger}\delta_{\alpha\beta}d_{\nu,\mathbf{k}+\mathbf{Q}_{j}\beta}+h.c.\nonumber \\
 &  & \boldsymbol{\Delta}_{\mathrm{SDW},j}^{\mu\nu}(\mathbf{k})=d_{\mu,\mathbf{k}\alpha}^{\dagger}\boldsymbol{\sigma}_{\alpha\beta}d_{\nu,\mathbf{k}+\mathbf{Q}_{j}\beta}+h.c.\nonumber \\
 &  & \boldsymbol{\Delta}_{\mathrm{iSDW,j}}^{\mu\nu}(\mathbf{k})=id_{\mu,\mathbf{k}\alpha}^{\dagger}\boldsymbol{\sigma}_{\alpha\beta}d_{\nu,\mathbf{k}+\mathbf{Q}_{j}\beta}+h.c.\label{OP_orbital}
\end{eqnarray}
where $\mu,\nu$ label the orbitals and $j=X,Y$, with $\mathbf{Q}_{X}=(\pi,0)$
and $\mathbf{Q}_{Y}=(0,\pi)$. SC order parameters for spin-singlet
pairing with zero center-of-mass pair momentum are defined for a given
${\mathbf{k}}$ according to 
\begin{equation}
{\Delta}_{\mathrm{SC}}^{\mu\nu}(\mathbf{k})=d_{\mu,\mathbf{k}\alpha}^{\dagger}\left(i\boldsymbol{\sigma}_{\alpha\beta}^{y}\right)d_{\nu,\mathbf{-k}\beta}^{\dagger}f_{\mathrm{SC}}(\mathbf{k})+h.c.\label{SC_orbital}
\end{equation}
where $f_{\mathrm{SC}}(\mathbf{k})$ is an even function of $\mathbf{k}$
that has the full lattice symmetry, but can change sign between, e.g.,
$\mathbf{k}=0$ and $\mathbf{k}=(0,\pi)/(\pi,0)$. Out of these order
parameters one can construct the combinations that transform as $A_{1g}$,
$B_{1g}$, $B_{2g}$, and $A_{2g}$ irreducible representations of
the $D_{4h}$ group. For instance, $\Delta_{\mathrm{SC}}^{xz,xz}+\Delta_{\mathrm{SC}}^{yz,yz}$
belongs to the $A_{1g}$ representation, while $\Delta_{\mathrm{SC}}^{xz,xz}-\Delta_{\mathrm{SC}}^{yz,yz}$
belongs to the $B_{1g}$ representation. These two are often called
$s$-wave and $d$-wave, by analogy with isotropic systems. Alternatively,
one can classify linear combinations of the order parameters in the
orbital basis as orbitally in-phase and orbitally anti-phase \cite{Kotliar_antiphase,Chubukov_Borisenko}

The eigenfunctions from each representation can be further classified
into sub-classes depending on how $f_{\mathrm{SC}}(\mathbf{k})$ evolves
between the high-symmetry points $(0,0)$, $(0,\pi)/(\pi,0)$, and
$(\pi,\pi)$. These symmetry points coincide with the center of hole
and electron pockets, but their presence is not explicitly emphasized
in the analysis in the orbital basis. The two most known sub-classes,
called ``plus-plus'' and ``plus-minus'' \cite{Mazin08,Kuroki1,Chubukov09,Kontani_spp},
correspond to $f_{\mathrm{SC}}(0)=f_{\mathrm{SC}}(0,\pi)=f_{\mathrm{SC}}(\pi,0)=f_{\mathrm{SC}}(\pi,\pi)$
(plus-plus) and $f_{\mathrm{SC}}(0)=f_{\mathrm{SC}}(\pi,\pi)=-f_{\mathrm{SC}}(0,\pi)=-f_{\mathrm{SC}}(\pi,0)$
(plus-minus). In the $A_{1g}$ ($B_{1g}$) channels, these subclasses
are called $s^{++}$ ( $d^{++}$) and $s^{+-}$ ($d^{+-}$), respectively.

Orbital order is an instability in the charge channel. It gives rise
to a CDW if the order parameter has a finite momentum, in which case
the corresponding order parameter is a particular combination of the
terms from Eq. \ref{OP_orbital}. Orbital order with zero momentum
emerges as a Pomeranchuk instability, and the corresponding order
parameter is given by 
\begin{equation}
\boldsymbol{\Delta}_{\mathrm{POM}}^{\mu\nu}(\mathbf{k})=d_{\mu,\mathbf{k}\alpha}^{\dagger}{\delta}_{\alpha\beta}d_{\nu,\mathbf{k}\beta}f_{\mathrm{POM}}(\mathbf{k})
\end{equation}

Similarly to superconductivity, one can form linear combinations of
${\Delta}_{\mathrm{POM}}^{\mu\nu}(k)$ that transform as the $A_{1g}$,
$B_{1g}$, $B_{2g}$, and $A_{2g}$ irreducible representations of
the $D_{4h}$ space group. In particular, $\Delta^{xz,xz}+\Delta^{yz,yz}$
belongs to the $A_{1g}$ representation, $\Delta^{xz,xz}-\Delta^{yz,yz}$
belongs to the $B_{1g}$ representation, $\Delta^{xz,yz}+\Delta^{yz,xz}$
belongs to the $B_{2g}$ representation, and $\Delta^{xz,yz}-\Delta^{yz,xz}$
belongs to the $A_{2g}$ representation. In the literature, ferro-orbital
order \cite{Ku09,Kruger09,Devereaux10,Phillips10,WCLee_Philips,Kontani16}
is usually associated with the $B_{1g}$ order parameter $\Delta^{xz,xz}-\Delta^{yz,yz}$.
Again, each representation can be further classified into sub-classes,
depending on the symmetry properties of $f_{\mathrm{POM}}(\mathbf{k})$.
The notations ``plus-plus'' and ``plus-minus'' apply to the cases
$f_{\mathrm{POM}}(0)=f_{\mathrm{POM}}(0,\pi)=f_{\mathrm{POM}}(\pi,0)$
and $f_{\mathrm{POM}}(0)=-f_{\mathrm{POM}}(0,\pi)=-f_{\mathrm{POM}}(\pi,0)$,
respectively. In real space, plus-plus $B_{1g}$ (i.e. $d$-wave)
order is on-site ferro-orbital order, while plus-minus order is a
bond order. An $s$-wave charge order with zero momentum ($s^{++}$
or $s^{+-}$) does not break any symmetry and therefore does not represent
a true order parameter since the mean values of $\Delta^{xz,xz}+\Delta^{yz,yz}$
are non-zero at any temperature. Fluctuations in the $s^{++}$ Pomeranchuk
channel are frozen due to the constraint of a constant occupation
number (Luttinger's theorem). Fluctuations in the $s^{+-}$ Pomeranchuk
channel, however, are not frozen, and the corresponding susceptibility
can sharply increase around a certain temperature, mimicking the development
of a true order parameter. The $d$-wave Pomeranchuk order parameter,
on the other hand, can develop spontaneously, and its condensation
breaks the tetragonal symmetry of the system (i.e. the $x$ and $y$
spatial directions become inequivalent), but preserves the translational
symmetry.

\subsection{Interaction effects}

As we discussed above, the main goal of the studies of the effects
of interactions in the orbital basis is to understand the ordered
states which we just introduced, namely magnetism, superconductivity,
and orbital order, without focusing \emph{a priori }on the low-energy
states near the Fermi pockets. Another goal of these studies is to
find how strong the effects leading to electron localization are,
and how these effects differentiate between distinct orbitals.

Nearly all studies of the interaction effects in FeSC within the orbital
basis depart from the onsite interaction Hamiltonian from Eq. (\ref{H_int_orb}),
with inter-orbital and intra-orbital terms, and use mean-field (RPA)
self-consistent analysis. For magnetism, such an analysis revealed
magnetic instabilities towards a SDW order with momenta $\mathbf{Q}_{X}$
or $\mathbf{Q}_{Y}$, as well as a subleading instability towards
a Neel order with $\mathbf{Q}_{M}=\left(\pi,\pi\right)$ \cite{Graser09,Graser10,Moreo_2orbital,Bascones10,Brydon_Daghofer,Dagotto_Hartree_Fock,Gastiasoro15}.
The selection of magnetic order -- i.e. whether both $\mathbf{Q}_{X}$
and $\mathbf{Q}_{Y}$ are condensed in a double-\textbf{Q} tetragonal
state, or a single-\textbf{Q }stripe phase is stabilized -- has only
been considered more recently, for instance via unrestricted Hartree-Fock
calculations \cite{Gastiasoro15}. Although for a wide range of parameters
the magnetic ground state is stripe-like (single-\textbf{Q}), and
therefore breaks tetragonal symmetry, hole-doped systems have been
shown to display double-\textbf{Q }tetragonal magnetic states, consistent
with what is observed experimentally \cite{Avci14,Bohmer15,Allred16}.
In another set of studies of SDW order within the orbital basis, robust
nodes in the SDW gap have been found, which give rise to ``Dirac-like''
band dispersions in the magnetically ordered state \cite{DHLee_Dirac,Tohyama10,Knolle11}.

RPA calculations have also been employed to study the onset of on-site
ferro-orbital order characterized by unequal occupations of the $d_{xz}$
and $d_{yz}$ orbitals \cite{Yamase13,Nevidomskyy,Kontani16}. A spontaneous
ferro-orbital order is found within RPA, but only if $2U'-J>U$, i.e.,
when inter-orbital $U'$ is substantially strong. For smaller $U'$
ferro-orbital order does not develop. We return to this issue in Sec.
\ref{sec:Hybrid-band-orbital-models}, where we question the validity
of RPA for such an analysis.

The main issue with orbital-basis models is that they do not distinguish
high-energy and low-energy states, which makes it difficult to implement
methods beyond RPA within this approach. The proposed modification
of RPA relies on the assumption that magnetism comes from electronic
states at higher energies and can be reasonably well captured within
RPA in the orbital basis, while superconductivity and nematic order
originate from interactions between low-energy fermions, mediated
by already developed magnetic fluctuations. Along these lines, several
groups used RPA in the orbital basis to obtain the magnetic susceptibility,
and then focused on the low-energy sector to study magnetically-mediated
superconductivity within BCS theory (Refs. \cite{Kuroki1,Graser09,Ikeda10,Graser10,Mazin_review,Scalapino_review})
or magnetically-mediated nematicity \cite{Fanfarillo15,Christensen16}.
We will come back to these RPA studies in Section \ref{sec:Hybrid-band-orbital-models}.

A different approach to superconductivity is based on models mixing
localized spins interacting with itinerant electrons \cite{JPHu08,Si08,DHLee_Davis}.
One idea promoted by some of these studies is that the SC gap is present
everywhere in the BZ and its momentum-dependence closely follows one
of the $C_{4}$ symmetric lattice functions, e.g., $\cos k_{x}+\cos k_{y}$
~\cite{JPHu_Ding}. This is very far from BCS theory, in which the
gap is confined to the Fermi surface, because only there the pairing
interaction can be logarithmically enhanced. We believe that the presence
of a robust SC gap everywhere on the Fermi surface is highly unlikely
in the first place because the interactions in FeSC are not overly
strong, otherwise these systems would not display a metallic behavior.
Another possibility studied in orbital-basis models \cite{JPHu_antiphase,Kotliar_antiphase,Coleman16}
is an exotic pairing involving the combination of orbital and SC degrees
of freedom.

\section{Band-basis models \label{sec:Band-basis-models}}

We now discuss an alternative approach, which starts directly from
the band-basis representation and treats the band states as the fundamental
low-energy states instead of expressing them as linear combinations
of orbital states $d_{\mu,\mathbf{k}\sigma}^{\dagger}$. In the band
representation, the non-interacting Hamiltonian is diagonal in band
indices and describes excitations near hole and electron pockets:
\begin{equation}
\mathcal{H}_{0}=\sum_{m=1}^{5}\varepsilon_{m}\left(\mathbf{k}\right)c_{m,\mathbf{k}\sigma}^{\dagger}c_{m,\mathbf{k}\sigma}\label{H0_band_aux}
\end{equation}

The band dispersions are parametrized as simple tight-binding or parabolic
dispersions, according to the symmetries imposed by the positions
of the centers of the various Fermi pockets, with no reference to
their orbital content. The interacting Hamiltonian contains all possible
interactions between these low-energy electronic states. These interactions
were argued to contain angle-dependent terms, but in band-basis models
these angle dependencies are imposed by the underlying $C_{4}$ symmetry
and the locations of the Fermi pockets, rather than the orbital content
of the excitations \cite{Maiti10}. For example, all pairing interactions
contain $\cos{4n\theta}$ dependencies, because these angular dependencies
are consistent with $C_{4}$ symmetry. The pairing interactions involving
states near the electron pockets, however, also contain $\cos{(4n+2)\theta}$
terms, because the center of the electron pockets are not along the
diagonal directions in the 1-Fe BZ.

We emphasize that these band-basis models cannot be described as the
low-energy versions of the orbital-basis models from Section (\ref{sec:Orbital-basis-models}),
expressed in a different basis. In particular, these band models cannot
describe orbital order simply because they do not distinguish between
different orbitals. We will discuss the proper low-energy models later,
in Section \ref{sec:Hybrid-band-orbital-models}.

Band models were quite successful in the description of SDW and SC
orders and the interplay between them \cite{Chubukov09,Tesanovic09,Brydon09,RMF10,Vorontsov10,Eremin11}.
This success implies that, while the orbital composition of the low-energy
states does play some role for magnetism and superconductivity, it
does not provide the crucial ingredient for these two orders, as opposed
to orbital order.

Band-based models are constructed to capture the low-energy states
near the Fermi surfaces and their application to FeSCs is based on
the assumption that not only superconductivity but also SDW magnetism
are low-energy phenomena. Namely, SDW magnetism is viewed as the result
of near-nesting between hole-like and electron-like bands. In this
respect, the reasonings for band-basis models and for orbital-basis
models are different.

Because only low energies are involved in band-basis models, one can
go beyond RPA and, e.g., analyze the interplay not only between long-range
SDW and SC orders but also between SC and SDW fluctuations. Another
advantage of band-basis models is that they can be straightforwardly
extended to analyze composite Ising-nematic order \cite{RMF12,Brydon11,Una15},
which is related to the order parameter manifold of stripe SDW magnetism
rather than with the orbital composition of the excitations.

This discussion raises the question of whether it may be more appropriate
to use the band basis to describe magnetism and superconductivity
in FeSC \cite{Knolle11}. In this section we briefly review the results
of the models conceived entirely in the band basis. As in the previous
section, we first discuss the non-interacting Hamiltonian, then introduce
the order parameters, and then include interactions to discuss the
instabilities of these models at the mean-field level and beyond it.

\subsection{Non-interacting Hamiltonian \label{sub:Two-band-model}}

As discussed above, a generic FeSC contains two small hole pockets
at the $\Gamma$ point, two electron pockets at $X$ and $Y$ points
with similar sizes, and may also contain another hole pocket at the
$M$ point. The tetragonal symmetry requires that the hole pockets
must be $C_{4}$ symmetric (i.e. invariant under a $90^{\circ}$ rotation),
since they are centered at either the center or the corner of the
BZ, whereas the electron pockets only need to be $C_{2}$ symmetric
(i.e. invariant under a $180^{\circ}$ rotation), since they are centered
at the sides of the BZ. Note that the two electron pockets are related
to each other by a $90^{\circ}$ rotation. Also, because the pockets
are assumed to be small, their band dispersions can be expanded in
powers of the relative momentum with respect to the center of the
pockets, in which case one can assume parabolic dispersions.

\begin{figure}
\begin{centering}
\includegraphics[width=0.75\columnwidth]{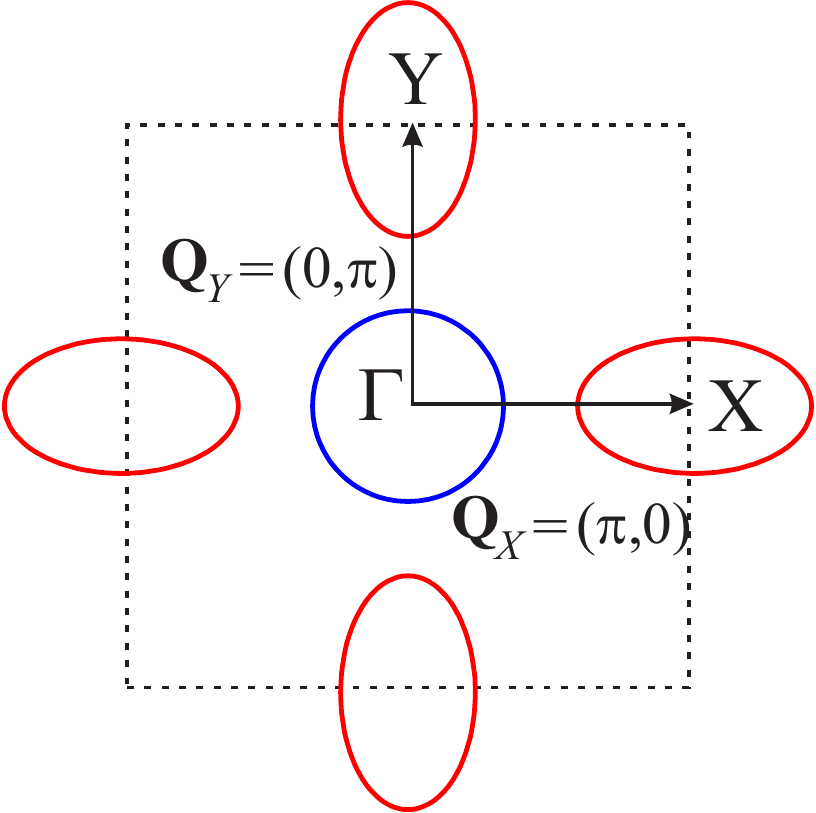} 
\par\end{centering}

\protect\protect\protect\protect\protect\protect\protect\caption{The effective three-band model: in the 1-Fe BZ, a circular hole pocket
(blue) is centered at $\Gamma$, whereas elliptical electron pockets
(red) are centered at $X$ and $Y$. Figure adapted from Ref. \cite{RMF12}.
\label{fig_3bands}}
\end{figure}

Under these conditions, one can write an effective 5-band model for
electronic states residing near the hole and electron pockets. Here,
we focus on a simplified model containing three bands \cite{Eremin11,RMF12}
-- one central hole pocket and two elliptical electron pockets centered
at the $X$ and $Y$ points, as shown in Fig. \ref{fig_3bands}. The
motivation to neglect the $M$ hole pocket is because it is not generically
present in all compounds \cite{Ding_review}. The restriction to a
single hole-pocket at the $\Gamma$ point is less justified, but the
argument is that in general one hole pocket has better nesting with
the electron pockets than the other \cite{Eremin11}.

The non-interacting Hamiltonian of the 3-band model is written as:
\begin{align}
\mathcal{H}_{0} & =\sum_{\mathbf{k}}\varepsilon_{h}\left(\mathbf{k}\right)c_{h,\mathbf{k}\sigma}^{\dagger}c_{h,\mathbf{k}\sigma}\nonumber \\
 & +\sum_{\mathbf{k},i=X,Y}\varepsilon_{e_{i}}\left(\mathbf{k}+\mathbf{Q}_{i}\right)c_{e_{i},\mathbf{k}+\mathbf{Q}_{i}\sigma}^{\dagger}c_{e_{i},\mathbf{k}+\mathbf{Q}_{i}\sigma}\label{H0_3band}
\end{align}
with parabolic band dispersions:

\begin{align}
\varepsilon_{h}\left(\mathbf{k}\right) & =\varepsilon_{h,0}-\frac{k^{2}}{2m}-\mu\nonumber \\
\varepsilon_{e_{X}}\left(\mathbf{k}+\mathbf{Q}_{X}\right) & =-\varepsilon_{e,0}+\frac{k_{x}^{2}}{2m_{x}}+\frac{k_{y}^{2}}{2m_{y}}-\mu\nonumber \\
\varepsilon_{e_{Y}}\left(\mathbf{k}+\mathbf{Q}_{Y}\right) & =-\varepsilon_{e,0}+\frac{k_{x}^{2}}{2m_{y}}+\frac{k_{y}^{2}}{2m_{x}}-\mu\label{3band_dispersions}
\end{align}

Similarly, one can consider effective tight-binding dispersions for
each band, as done in Refs. \cite{Brydon11,Knolle_Schmalian11}. The
main advantage of the parabolic dispersions is their simplicity and
convenience for analytical calculations. Hereafter, we simplify the
notation by leaving it implicit that the momenta of the electron-like
states are measured relative to the respective $\mathbf{Q}_{i}$.

One of the goals of the band-basis models is to relate SDW magnetism
to nesting properties of the band structure, as manifested in Fig.
\ref{fig_3bands}. Perfect nesting requires $\varepsilon_{h}\left(\mathbf{k}\right)=-\varepsilon_{e_{i}}\left(\mathbf{k}+\mathbf{Q}_{i}\right)$,
which is clearly not the case for the real materials, as the hole
and electron pockets do not have identical shapes. Instead, the FeSC
usually display pairs of points satisfying the condition $\varepsilon_{h}\left(\mathbf{k}_{\mathrm{hs}}\right)=\varepsilon_{e_{i}}\left(\mathbf{k}_{\mathrm{hs}}+\mathbf{Q}_{i}\right)=0$
-- the so-called hot spots. Yet, the hypothetical limit of perfect
nesting is very useful to gain insight into the generic properties
of the model, as we will show latter. In this regard, it is useful
to consider an alternative parametrization of the band dispersions
in terms of the angle $\theta$ measured with respect to the $k_{x}$
axis \cite{Vorontsov10}:

\begin{align}
\varepsilon_{h}\left(\mathbf{k}\right) & =-\varepsilon_{\mathbf{k}}\nonumber \\
\varepsilon_{e_{X}}\left(\mathbf{k}+\mathbf{Q}_{X}\right) & =\varepsilon_{\mathbf{k}}-\left(\delta_{\mu}+\delta_{m}\cos2\theta\right)\nonumber \\
\varepsilon_{e_{Y}}\left(\mathbf{k}+\mathbf{Q}_{Y}\right) & =\varepsilon_{\mathbf{k}}-\left(\delta_{\mu}-\delta_{m}\cos2\theta\right)\label{3band_dispersions_angle}
\end{align}

The parameter $\delta_{\mu}$ is proportional to the sum of the chemical
potential and the offset between the top of the hole pocket and the
bottom of the electron pocket, whereas the parameter $\delta_{m}$
is proportional to the ellipticity of the electron pockets. The condition
for perfect nesting is $\delta_{\mu}=\delta_{m}=0$. This parametrization
is convenient because the expansion near perfect nesting can be performed
in powers of these two parameters.

\subsection{Order parameters}

Before discussing the effects of interactions we introduce different
order parameters in the band basis. To avoid lengthy formulas, we
focus on the 3-band model. The existence of the nesting vectors $\mathbf{Q}_{X}$
and $\mathbf{Q}_{Y}$ allows one to introduce several density-wave
order parameters involving fermions from the electron and hole pockets
\cite{Tesanovic09,Chubukov09,Brydon09,Kang_Tesanovic}. We define:
\begin{align}
\Delta_{\mathrm{CDW},j}(\mathbf{k}) & \propto c_{h,\mathbf{k}\alpha}^{\dagger}\delta_{\alpha\beta}c_{e_{j},\mathbf{k}+\mathbf{Q}_{j}\beta}+h.c.\nonumber \\
\Delta{}_{\mathrm{iCDW},j}(\mathbf{k}) & \propto ic_{h,\mathbf{k}\alpha}^{\dagger}\delta_{\alpha\beta}c_{e_{j},\mathbf{k}+\mathbf{Q}_{j}\beta}+h.c.\nonumber \\
\boldsymbol{\Delta}{}_{\mathrm{SDW},j}(\mathbf{k}) & \propto c_{h,\mathbf{k}\alpha}^{\dagger}\boldsymbol{\sigma}_{\alpha\beta}c_{e_{j},\mathbf{k}+\mathbf{Q}_{j}\beta}+h.c.\nonumber \\
\boldsymbol{\Delta}{}_{\mathrm{iSDW},j}(\mathbf{k}) & \propto ic_{h,\mathbf{k}\alpha}^{\dagger}\boldsymbol{\sigma}_{\alpha\beta}c_{e_{j},\mathbf{k}+\mathbf{Q}_{j}\beta}+h.c.\label{band_instabilities}
\end{align}
where $j=X,\, Y$ and $c_{h},\, c_{e_{j}}$ label fermionic operators
near hole and electron pockets. The order parameters $\Delta_{\mathrm{CDW},j}$
and $\boldsymbol{\Delta}{}_{\mathrm{SDW},j}$ describe charge and
spin density-waves (CDW and SDW, respectively) with transferred momenta
$\mathbf{Q}_{j}$, whereas $\Delta{}_{\mathrm{iCDW},j}$ and $\boldsymbol{\Delta}{}_{\mathrm{iSDW},j}$
describe charge-current (iCDW) and spin-current (iSDW) density-waves.
It is also useful to introduce the order parameters with momentum
$\mathbf{Q}_{X}+\mathbf{Q}_{Y}=\left(\pi,\pi\right)$, which involve
fermions from the two electron pockets, e.g., the Neel order parameter:

\begin{equation}
\boldsymbol{\Delta}{}_{\mathrm{Neel}}(\mathbf{k})\propto c_{e_{X},\mathbf{k}+\mathbf{Q}_{X}\alpha}^{\dagger}\boldsymbol{\sigma}_{\alpha\beta}c_{e_{Y},\mathbf{k}+\mathbf{Q}_{Y}\beta}+h.c.
\end{equation}

However, because the two electron pockets are not nested, the instabilities
at momentum $\mathbf{Q}_{X}+\mathbf{Q}_{Y}$ are subleading to the
ones at momenta $\mathbf{Q}_{X}$ and $\mathbf{Q}_{Y}$, at least
at weak coupling.

We also introduce the SC order parameters. In principle, they have
angular dependencies already in the band basis due to the locations
and symmetries of the Fermi surfaces. In some cases, this dependence
can even lead to accidental nodes, particularly on electron pockets
\cite{Maiti10,Kemper10,Vorontsov10,Chubukov_Eremin_nodes,Vorontsov_Vekhter,Khodas_Chubukov2}.
We will not dwell into this issue here and focus instead on the angle-independent
parts of SC order parameters. It is useful to define the order parameters
for each pocket:

\begin{align}
\Delta_{h}(\mathbf{k}) & \propto c_{h,-\mathbf{k}\downarrow}c_{h,\mathbf{k}\uparrow}+h.c.\nonumber \\
\Delta_{e_{X}}(\mathbf{k}) & \propto c_{e_{X},-\mathbf{k}-\mathbf{Q}_{X}\downarrow}c_{e_{X},\mathbf{k}+\mathbf{Q}_{X}\uparrow}+h.c.\nonumber \\
\Delta_{e_{Y}}(\mathbf{k}) & \propto c_{e_{Y},-\mathbf{k}-\mathbf{Q}_{Y}\downarrow}c_{e_{Y},\mathbf{k}+\mathbf{Q}_{Y}\uparrow}+h.c.\label{3band_gaps}
\end{align}

Each SC order parameter $\Delta$ (often called the gap function)
has an amplitude and a phase. We define all gaps such that in the
ordered state they have the same global phase and will not consider
phase fluctuations. Because the system has tetragonal symmetry, the
three gap functions can be recast in terms of three different combinations,
two of which transform as $A_{1g}$ ($s$-wave) representation, and
one as $B_{1g}$ ($d$-wave) representation \cite{RMF_Millis}: 
\begin{align}
\Delta_{s^{++}} & =\sin\Psi\,\Delta_{h}+\frac{\cos\Psi}{\sqrt{2}}\left(\Delta_{e_{X}}+\Delta_{e_{Y}}\right)\nonumber \\
\Delta_{s^{+-}} & =\cos\Psi\,\Delta_{h}-\frac{\sin\Psi}{\sqrt{2}}\left(\Delta_{e_{X}}+\Delta_{e_{Y}}\right)\nonumber \\
\Delta_{d} & =\frac{1}{\sqrt{2}}\left(\Delta_{e_{X}}-\Delta_{e_{Y}}\right)\label{3band_gaps_irrep}
\end{align}

The mixing angle $\Psi$ depends on the strength of the pairing interactions
$V_{1}$ between the $h$ and the $e_{X/Y}$ pockets and $V_{2}$
between the $e_{X}$ and $e_{Y}$ pockets according to:

\begin{equation}
\tan\Psi=\frac{\sqrt{8V_{1}^{2}+V_{2}^{2}}-V_{2}}{2\sqrt{2}V_{1}}
\end{equation}

The interpretation of these three SC order parameters is straightforward:
in the $s^{++}$-wave state the gap functions on different pockets
all have the same sign; in the $s^{+-}$-state the gaps on the hole
and on the electron pockets have different signs; and in the $d$-wave
state the gaps on the two electron pockets have opposite signs. Note
that the absence of the $d$-wave component on the hole pockets is
just the result of our neglect of the angular dependencies. In reality,
a $d$-wave gap on the hole pocket behaves as $\cos2\theta$.

One can also define Pomeranchuk order parameters at zero momentum
transfer $\mathbf{Q}=0$. In the charge channel we have 
\begin{eqnarray}
 &  & \Delta_{\mathrm{POM}}^{h}(\mathbf{k})\propto c_{h,\mathbf{k}\alpha}^{\dagger}\delta_{\alpha\beta}c_{h,\mathbf{k}\beta}+h.c.\nonumber \\
 &  & \Delta_{\mathrm{POM}}^{X}(\mathbf{k})\propto c_{e_{X},\mathbf{k}+\mathbf{Q}_{X}\alpha}^{\dagger}\delta_{\alpha\beta}c_{e_{X},\mathbf{k}+\mathbf{Q}_{X}\beta}+h.c.\nonumber \\
 &  & \Delta_{\mathrm{POM}}^{Y}(\mathbf{k})\propto c_{e_{Y},\mathbf{k}+\mathbf{Q}_{Y}\alpha}^{\dagger}\delta_{\alpha\beta}c_{e_{Y},\mathbf{k}+\mathbf{Q}_{Y}\beta}+h.c.\label{new_5}
\end{eqnarray}
The development of a non-zero $\sum_{\mathbf{k}}\left\langle \Delta_{\mathrm{POM}}^{X}(\mathbf{k})-\Delta_{\mathrm{POM}}^{Y}(\mathbf{k})\right\rangle $
breaks $C_{4}$ lattice rotational symmetry down to $C_{2}$ and gives
rise to nematic order \cite{RMF12}, which in the band-only model
is not identified with any orbital order, but still gives rise to
a $d$-wave distortion of the electron Fermi surfaces. Another possibility
is the appearance of $s^{+-}$ Pomeranchuk order with $\sum_{\mathbf{k}}\left\langle \Delta_{\mathrm{POM}}^{X}(\mathbf{k})+\Delta_{\mathrm{POM}}^{Y}(\mathbf{k})\right\rangle $
and $\sum_{\mathbf{k}}\left\langle \Delta_{\mathrm{POM}}^{h}(\mathbf{k})\right\rangle $
with opposite signs. This leads to either shrinking or expansion of
the sizes of both hole and electron pockets, such that the total number
of charge carriers is preserved \cite{Ortenzi09,RMF_Chubukov_Khodas}.
Like we said, an order parameter of this kind does not break any symmetry
and is generally non-zero at any temperature, but it can be strongly
enhanced around a particular temperature.

\subsection{Interaction effects}

The interactions in the band-basis model are not necessarily given
by on-site terms only. Instead, they include all possible interactions
involving pairs of fermions from the same or from different bands.
As a result, the number of interaction terms increases with the number
of bands. For the three-band model introduced above, there are eight
distinct interaction terms \cite{Maiti10}: 
\begin{align}
\mathcal{H}_{\mathrm{int}} & =U_{1}\sum_{i}c_{h\alpha}^{\dagger}c_{e_{i}\beta}^{\dagger}c_{e_{i}\beta}c_{h\alpha}+U_{2}\sum_{i}c_{h\alpha}^{\dagger}c_{e_{i}\beta}^{\dagger}c_{h\beta}c_{e_{i}\alpha}\nonumber \\
 & +\frac{U_{3}}{2}\sum_{i}\left(c_{h\alpha}^{\dagger}c_{h\beta}^{\dagger}c_{e_{i}\beta}c_{e_{i}\alpha}+\mathrm{h.c.}\right)\nonumber \\
 & +\frac{U_{4}}{2}\sum_{i}c_{e_{i}\alpha}^{\dagger}c_{e_{i}\beta}^{\dagger}c_{e_{i}\beta}c_{e_{i}\alpha}+\frac{U_{5}}{2}\sum c_{h\alpha}^{\dagger}c_{h\beta}^{\dagger}c_{h\beta}c_{h\alpha}\nonumber \\
 & +U_{6}\sum c_{e_{X}\alpha}^{\dagger}c_{e_{Y}\beta}^{\dagger}c_{e_{Y}\beta}c_{e_{X}\alpha}+U_{7}\sum c_{e_{X}\alpha}^{\dagger}c_{e_{Y}\beta}^{\dagger}c_{e_{X}\beta}c_{e_{Y}\alpha}\nonumber \\
 & +\frac{U_{8}}{2}\sum\left(c_{e_{X}\alpha}^{\dagger}c_{e_{X}\beta}^{\dagger}c_{e_{Y}\beta}c_{e_{Y}\alpha}+\mathrm{h.c.}\right)\label{3band_Hint}
\end{align}

These terms correspond to density-density interactions ($U_{1}$,
$U_{4}$, $U_{5}$, $U_{6}$), spin-exchange interactions ($U_{2}$,
$U_{7}$), and pair-hopping interactions ($U_{3}$, $U_{8}$), all
of which have purely electronic origin. These interactions should
be viewed as input parameters rather than the combinations of Hubbard
and Hund interactions from Eq. (\ref{H_int_orb}). The reasoning is
that Hubbard and Hund interaction terms become angle-dependent once
we transform from orbital to band basis, due to the matrix elements
of Eq. (\ref{matrix_5orb}), while $U_{i}$ in Eq. (\ref{3band_Hint})
are taken to be angle-independent. We will come back to this point
in Section \ref{sec:Hybrid-band-orbital-models}.

\subsubsection{Renormalization Group (RG) analysis: the basics }

\label{sec:rg}

As we discussed above, the advantage of using the band basis is that
one can focus on the low-energy sector and go beyond mean-field (RPA)
analysis. To do this, in this and next Sections we apply the RG technique.
The RG machinery (either numerical functional RG or analytical parquet
RG) allows one to analyze how different interaction channels compete
with each other as one progressively integrates out fermions with
higher energies, starting from the upper energy cutoff $\Lambda$
of the low-energy sector (loosely defined as the scale at which corrections
to the parabolic dispersion near the $X$, $Y$, and $\Gamma$ points
become substantial) and moving down in energy \cite{Thomale09,Chubukov09,DHLee_fRG,DHLee_review,Thomale_review}.
The couplings in different interaction channels all evolve in this
process. The flow of the couplings is described by a set of differential
equations 
\begin{equation}
\frac{dU_{i}(L)}{dL}=a_{ijk}U_{j}(L)U_{k}(L)
\end{equation}
where $L\equiv\log\left(\frac{\Lambda}{E}\right)$ is a running RG
variable, which increases as the energy $E$ decreases away from the
cutoff $\Lambda$. The running interactions $U_{i}$ are all functions
of $L$. An instability develops at a critical RG scale $L_{c}=\log{\Lambda/E_{c}}$,
at which at least some of the couplings diverge. The critical temperature
for the instability is of the order of $E_{c}$.

One of the goals of the RG analysis is to verify whether the low-energy
behavior of a system is universal, i.e., that the running couplings
tend to the same values under RG for different initial interactions.
In the cases we discuss below, some couplings diverge upon approaching
the scale $L_{c}$, but their ratios tend to finite, fixed values
(the value of $L_{c}$ itself does depend on the bare values of the
interactions). In RG language, this is called a fixed trajectory.
There can be more than one fixed trajectory, in which case each has
a finite basin of attraction in the parameter space of initial interactions.
For each fixed trajectory, one can find with certainty what is the
leading and the subleading instability in the system.

To select what kind of order develops at $L=L_{c}$, one needs to
move the RG analysis to the next level and obtain the RG equations
for the flow of the vertices in different instability channels, $\Gamma_{j}$.
Each vertex is renormalized by a particular combination of the interactions
$U_{i}$. For the channels in which the vertex renormalizations are
logarithmical, like SC or SDW, vertex renormalizations are given by
the series of ladder diagrams with either particle-hole or particle-particle
bubble in every cross-section, and with interactions treated as the
running ones. To logarithmical accuracy the summation of these diagrams
is equivalent to solving the differential equations 
\begin{equation}
\frac{d\Gamma_{j}}{dL}=\Gamma_{j}u^{j}\label{aa_1}
\end{equation}
where $u^{j}$ is a dimensionless coupling in the channel $j$ (the
combination of $U_{i}N_{i}$, where $N_{i}$ is of the order of the
density of states $N_{F}$). The susceptibilities $\chi_{j}$ are
given by bubble diagrams with $\Gamma_{j}$ in the vertices and obey
\begin{equation}
\frac{d\chi_{j}}{dL}=\Gamma_{j}^{2}\label{aa_2}
\end{equation}
Solving Eq. (\ref{aa_1}) with the RG solution for $u_{j}$ as an
input, substituting the result into (\ref{aa_2}) and solving for
$\chi_{j}(L)$, one obtains $\chi_{j}(L)\propto1/(L_{c}-L)^{\alpha_{j}}$.
In general, the exponents $\alpha_{j}$ are all different. The channel
in which $\alpha_{j}$ has the largest value is the leading candidate
among the logarithmical channels to develop an order below the instability.

We will see, however, that the situation in at least some FeSC is
more involved because the susceptibility in initially non-logarithmical
channels, like Pomeranchuk channels, also flows with $L$ due to renormalizations
that involve the running couplings $U_{i}(L)$. We will show that
the corresponding susceptibilities scale with $L$ as $\chi_{j}(L)\propto1/(L_{c}-L)$.
If $\alpha_{j}$ in the leading logarithmical channel is smaller than
one, the susceptibilities in the Pomeranchuk channels diverge with
a higher exponent. In this situation, the system may actually develop
a Pomeranchuk order below the instability. We discuss this in more
detail in Sec. \ref{new_sec:ac}.

The advantage of the RG approach over mean-field approaches, such
as RPA, is that it allows one to analyze mutual feedbacks between
fluctuations in different channels (e.g., how superconducting fluctuations
modify SDW fluctuations, which in turn contribute to the pairing interaction).
The drawback of RG is that it is, by construction, a weak-coupling
analysis. Moreover, the selection of diagrams which are included into
RG analysis is justified only if vertex renormalizations in the particle-hole
channel (associated with density-wave instabilities) are logarithmic
at some transferred momentum, like in the Cooper channel (associated
with the superconducting instabilities). In FeSC, this condition is
generally satisfied because renormalizations in the particle-hole
channel with transfer momenta ${\mathbf{Q}}_{X}$ and/or ${\mathbf{Q}}_{Y}$
involve electronic states from hole-like and electron-like bands,
and a particle-hole bubble made of fermions from a hole and an electron
band depends logarithmically of the external frequency, much like
a Cooper bubble. But this only holds at energies larger than $|\delta_{m}|$
and $|\delta_{\mu}|$ in Eq. \ref{3band_dispersions_angle}, i.e.
RG can be rigorously justified down to the lowest energies only at
perfect nesting. Away from perfect nesting, the parquet RG flow of
the couplings towards one or another fixed trajectory holds between
the upper cutoff of the low-energy theory and, roughly, the largest
Fermi energy, $E_{F}$. At $E<E_{F}$ different channels no longer
``talk'' to each other. If the scale $L=L_{c}$ falls into this
range, the selection of the leading instability can be fully described
within parquet RG. If the RG scale $L$ reaches $L_{F}=\log{\Lambda/E_{F}}$
before the instability develops, parquet RG allows one to determine
the values of the running couplings at $L=L_{F}$. At smaller energies
(larger $L$) one can use, e.g., RPA with these couplings as inputs.
For a recent approach to extend RG equations to $L\rightarrow L_{F}$
see Ref. \cite{Murray14}.

\subsubsection{Two-band model}

To analyze how interactions select between different density-wave
and SC instabilities, we first consider a toy two-band model with
one hole and one of the two electron pockets, i.e. we consider SDW
and CDW orders with a single ordering vector. This model is blind
to $d$-wave superconductivity and $d-$wave Pomeranchuk order, yet
it offers interesting insights into the interplay between SDW, iCDW,
and $s^{+-}$ superconductivity. In terms of the interactions, this
toy model has five couplings $U_{1}-U_{5}$, while $U_{6}=U_{7}=U_{8}=0$.
\\

\paragraph{Perfect nesting, $E_{F}=0$\protect \protect \\
\protect \\
}

We first consider the limit of perfect nesting, $\delta_{\mu}=\delta_{m}=0$
in Eq. (\ref{3band_dispersions_angle}), when both hole and electron
bands just touch the Fermi level: $\epsilon_{e,h}=\pm k^{2}/(2m)$.
The masses will be absorbed into dimensional couplings. For free fermions,
the susceptibilities in the SDW and CDW channels with real and imaginary
order parameters and in $s^{++}$ and $s^{+-}$ SC channels are all
degenerate and scale as $\chi_{0}\propto\ln\left(\Lambda/T\right)$,
where $\Lambda$ is the bandwidth. Once the interactions from Eq.(\ref{3band_Hint})
are included, this degeneracy is lifted. Within RPA, different susceptibilities
become $\chi_{j}=\chi_{0}/\left(1-\Gamma_{j}\chi_{0}\right)$, where
\cite{Chubukov09}: 
\begin{align}
\Gamma_{\mathrm{SDW}}=U_{1}+U_{3}\quad;\; & \Gamma_{i\mathrm{SDW}}=U_{1}-U_{3}\nonumber \\
\Gamma_{\mathrm{CDW}}=U_{1}-U_{3}-2U_{2}\quad;\; & \Gamma_{i\mathrm{CDW}}=U_{1}+U_{3}-2U_{2}\nonumber \\
\Gamma_{s^{+-}}=-U_{4}+U_{3}\quad;\; & \Gamma_{s^{++}}=-U_{4}-U_{3}\label{2band_vertices}
\end{align}

When all $U_{i}$ are equal (Hubbard model in the band basis), the
leading instability within RPA is towards SDW magnetism. The interaction
in the SC $s^{++}$ channel is repulsive, and the one in the $s^{+-}$
channel vanishes ($\Gamma_{s^{+-}}=0$, $\Gamma_{s^{++}}<0$).

\begin{figure}
\begin{centering}
\includegraphics[width=0.8\columnwidth]{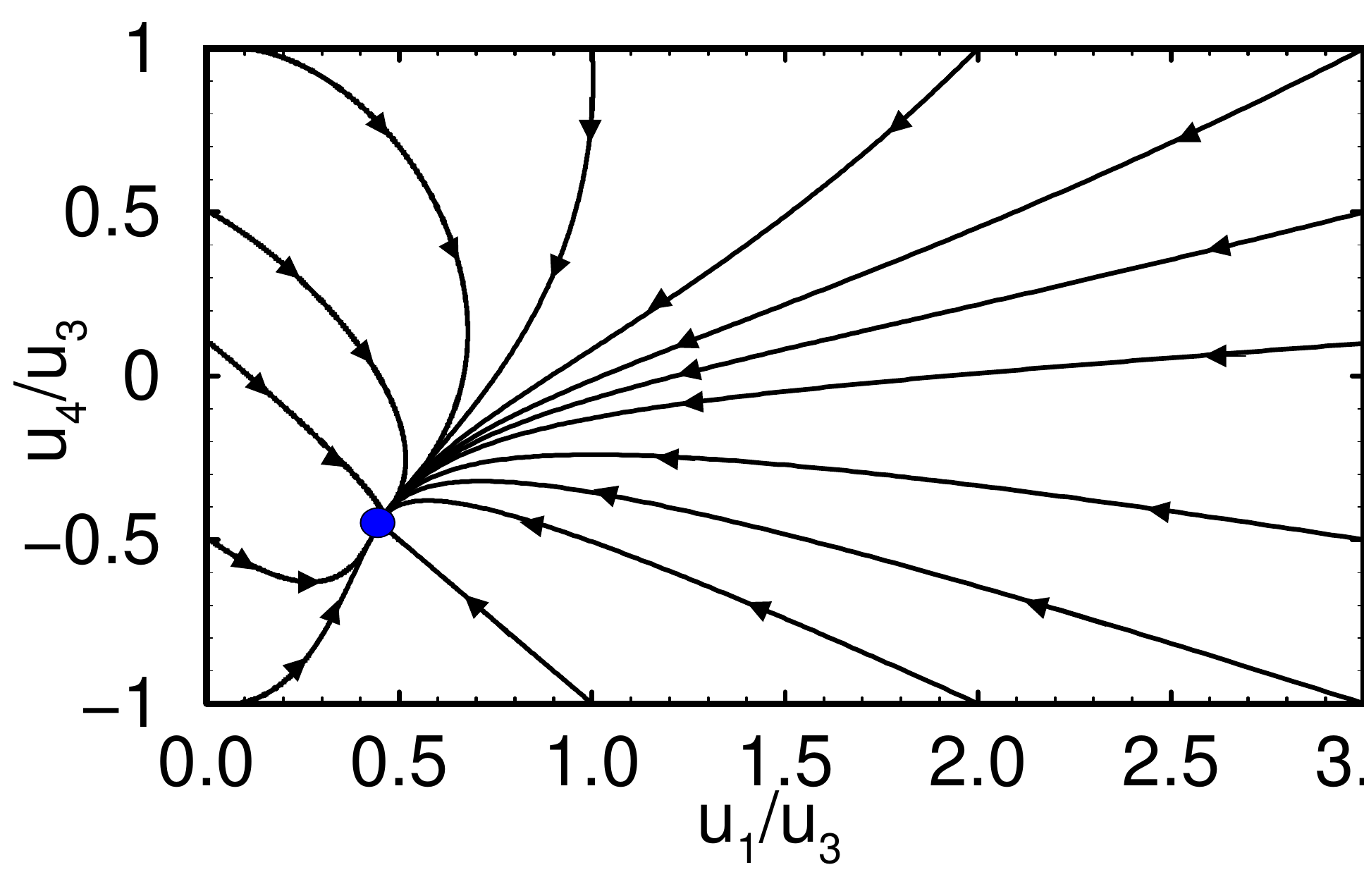} 
\par\end{centering}

\protect\protect\protect\protect\protect\protect\protect\caption{RG flows of the two-band model in the $\left(\frac{u_{1}}{u_{3}},\frac{u_{4}}{u_{3}}\right)$
plane. The fixed point is shown in blue. Figure from Ref. \cite{Chubukov09}.
\label{fig_RG_2band}}
\end{figure}

We now apply RG. We we do not give the details of this calculation,
and just list the results. The reader interested in details is referred
to the relevant literature \cite{Chubukov09}. There is one stable
fixed trajectory for positive (repulsive) interactions $U_{1}-U_{5}$.
All interactions diverge near $L=L_{c}$, but their ratios tend to
finite values. Specifically, the dimensionless $u_{j}=U_{j}N_{F}$,
where $N_{F}$ is the density of states, evolve near $L=L_{c}$ as
$u_{1}\propto1/(L_{c}-L)$, $u_{1}/u_{3}=-u_{4}/u_{3}=-u_{5}/u_{3}=1/\sqrt{5}$,
and $u_{2}/u_{3}=0$. Fig. \ref{fig_RG_2band} illustrates the RG
flow in the $\left(\frac{u_{1}}{u_{3}},\frac{u_{4}}{u_{3}}\right)$
plane, highlighting the stable fixed point $\left(\frac{1}{\sqrt{5}},-\frac{1}{\sqrt{5}}\right)$.

We now analyze use the running couplings $u_{i}$ as inputs and analyze
the flow of the vertices and susceptibilities in different channels.
Solving Eqs. (\ref{aa_1}) and (\ref{aa_2}) we obtain that the susceptibilities
in the SDW, iCDW, and $s^{+-}$ SC channel diverge as $1/(L_{0}-L)^{\alpha}$
with the same exponent $\alpha=(\sqrt{5}-2)/3=0.08$, while susceptibilities
in the iSDW, CDW, and $s^{++}$ SC channel do not diverge. The outcome
is that the system has an emergent enhanced $O(6)$ symmetry -- the
three order parameters form a $6$-dimensional super-vector $\mathbf{N}=\left(\boldsymbol{\Delta}{}_{\mathrm{SDW}},\,\Delta{}_{s^{+-}},\,\Delta{}_{i\mathrm{CDW}}\right)$
\cite{Chubukov09,Podolsky09}. This has important implications for
the competition between superconductivity and SDW, as we discuss below.\\

\paragraph{Away from perfect nesting, finite $E_{F}$.\protect \protect \protect \\
 \protect \\
}

\begin{figure}
\begin{centering}
\includegraphics[width=0.8\columnwidth]{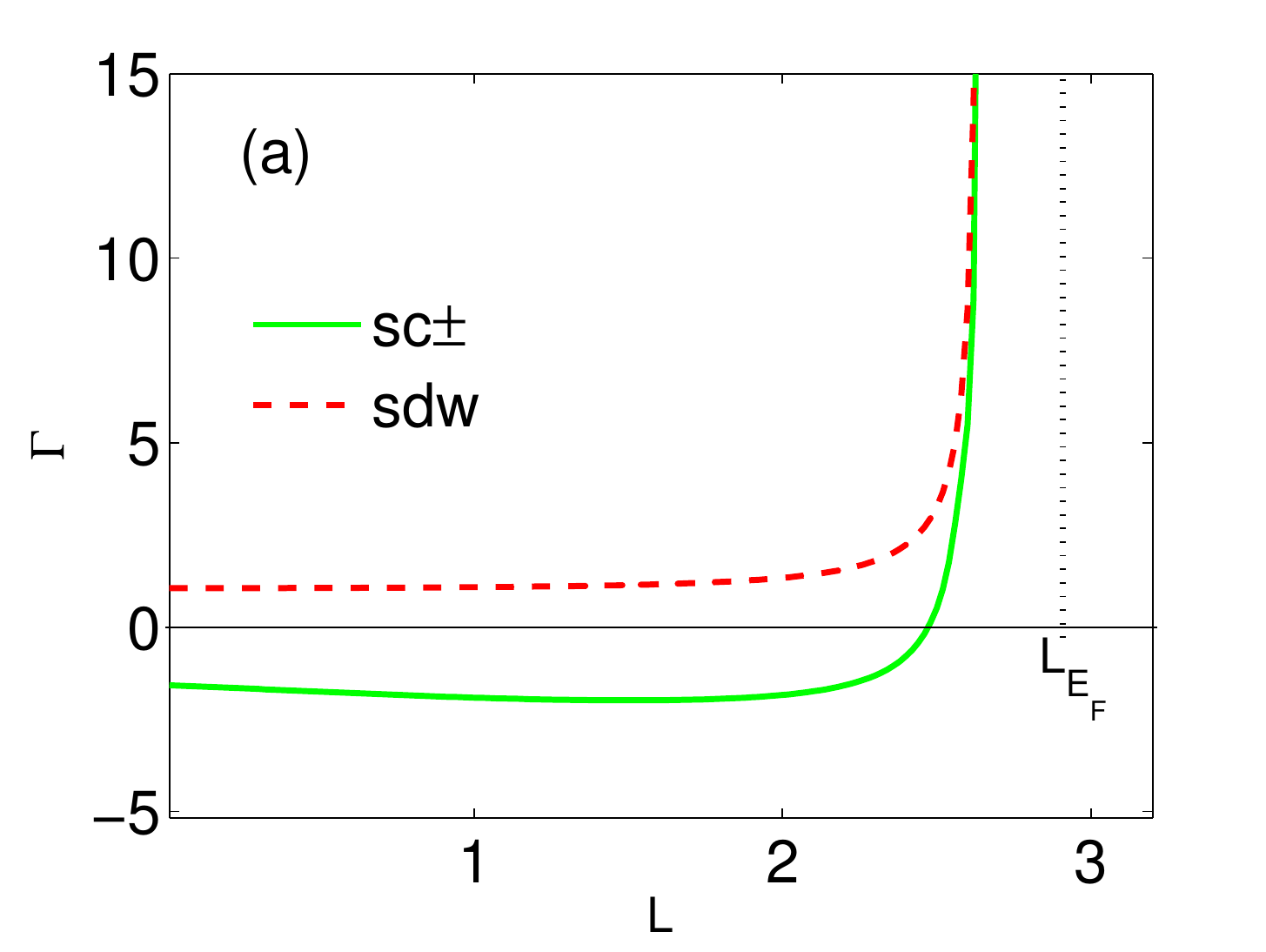} 
\par\end{centering}

\medskip{}

\begin{centering}
\includegraphics[width=0.8\columnwidth]{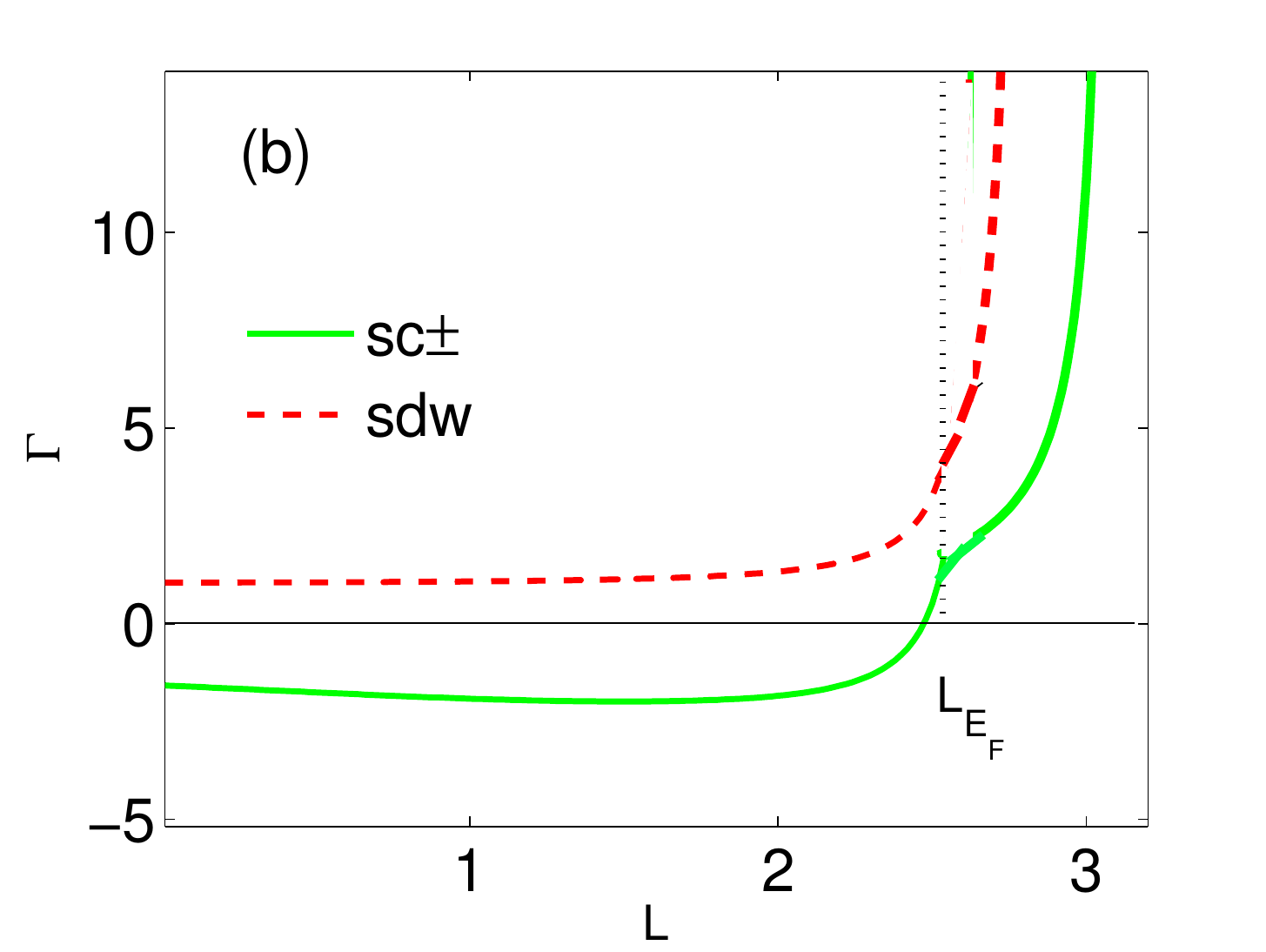} 
\par\end{centering}

\protect\protect\protect\protect\protect\protect\protect\caption{RG flow of the SDW (red/dashed curve) and $s^{+-}$ SC (green/solid
curve) vertices for the two-band model; (a) denotes the case $L_{F}>L_{c}$,
whereas (b) denotes the case $L_{F}<L_{c}$. Figure from Ref. \cite{Maiti10}.
\label{fig_RG_2bands_EF} }
\end{figure}

At non-zero $E_{F}$ (and hence non-zero $\delta_{\mu}$ and, in general,
also $\delta_{m}$), the parquet RG flow discussed above holds as
long as the running energy is larger than $E_{F}$, i.e. as long as
$L<L_{F}$. For $L>L_{F}$, the RG equations change as the six different
channels decouple. In most FeSC, the largest Fermi energy is about
$100$ meV, while $\Lambda$ is of order of eV. Thus, even though
$E_{F}\ll\Lambda$, it is likely that $L_{F}<L_{c}$, implying that
the instability is not really reached within parquet RG. At $L=L_{F}$,
the coupling in the SDW channel is the largest, and the ones in iCDW
and $s^{+-}$ channels are smaller (Refs. \cite{Chubukov09,Maiti10}).
At $L>L_{F}$, the only channel in which interactions continue to
grow logarithmically is the $s^{+-}$ SC channel, while the couplings
in SDW and iCDW channels eventually saturate. If the superconducting
$u^{SC}(L)$ is already attractive at $L=L_{F}$ and is close to $u^{SDW}$,
$s^{+-}$ superconductivity is the most likely outcome. If $u^{SDW}(L_{F})$
is large while the other $u^{j}$ are smaller, the system likely develops
SDW order, and if all $u^{j}$ are small at $L=L_{F}$ and the SC
interaction is repulsive, the system likely remains a metal down to
$T=0$ (see Fig. \ref{fig_RG_2bands_EF}). For some initial input
parameters, the system may also develop iCDW order \cite{Kang_Tesanovic,RMF_FeSe}.\\

\paragraph{Ginzburg-Landau free energy\protect \protect \protect \\
 \protect \\
}

Within the two-band model one can study the interplay between $s^{+-}$
superconductivity and SDW at $E<E_{F}$ in more detail by deriving
the Ginzburg-Landau free energy for the coupled $s^{+-}$ and SDW
order parameters $\Delta_{s^{+-}}$ and $\boldsymbol{\Delta}_{\mathrm{SDW}}$.
This is accomplished by performing a Hubbard-Stratonovich transformation
of the interaction terms in Eq. (\ref{3band_Hint}) in the SDW and
$s^{+-}$ channels. One can then integrate out the electronic degrees
of freedom and expand in powers of the two order parameters. This
yields \cite{RMF10_rapid,RMF10,Vorontsov10}:

\begin{align}
 & F\left(\Delta_{s^{+-}},\boldsymbol{\Delta}_{\mathrm{SDW}}\right)=\frac{a_{s}}{2}\,\Delta_{s^{+-}}^{2}+\frac{u_{s}}{4}\,\Delta_{s^{+-}}^{4}+\nonumber \\
 & \frac{a_{m}}{2}\,\Delta_{\mathrm{SDW}}^{2}+\frac{u_{m}}{4}\,\Delta_{\mathrm{SDW}}^{4}+\frac{\gamma}{2}\,\Delta_{s^{+-}}^{2}\Delta_{\mathrm{SDW}}^{2}\label{F_Delta_M}
\end{align}

All Ginzburg-Landau coefficients are given microscopically in terms
of the band dispersions (\ref{3band_dispersions}) and the effective
SDW and $s^{+-}$ SC interactions (see Refs. \cite{RMF10,Vorontsov10}
for details). Despite the fact that $\gamma>0$, indicating that the
two orders compete with each other, they coexist microscopically if
$\gamma<\sqrt{u_{s}u_{m}}$. Otherwise, if $\gamma>\sqrt{u_{s}u_{m}}$,
the SDW and $s^{+-}$ SC states phase-separate as the transition from
one phase to the other is first-order. Thus, it is convenient to define
the parameter $\bar{g}\equiv\gamma-\sqrt{u_{s}u_{m}}$.

\begin{figure}
\begin{centering}
\includegraphics[width=0.95\columnwidth]{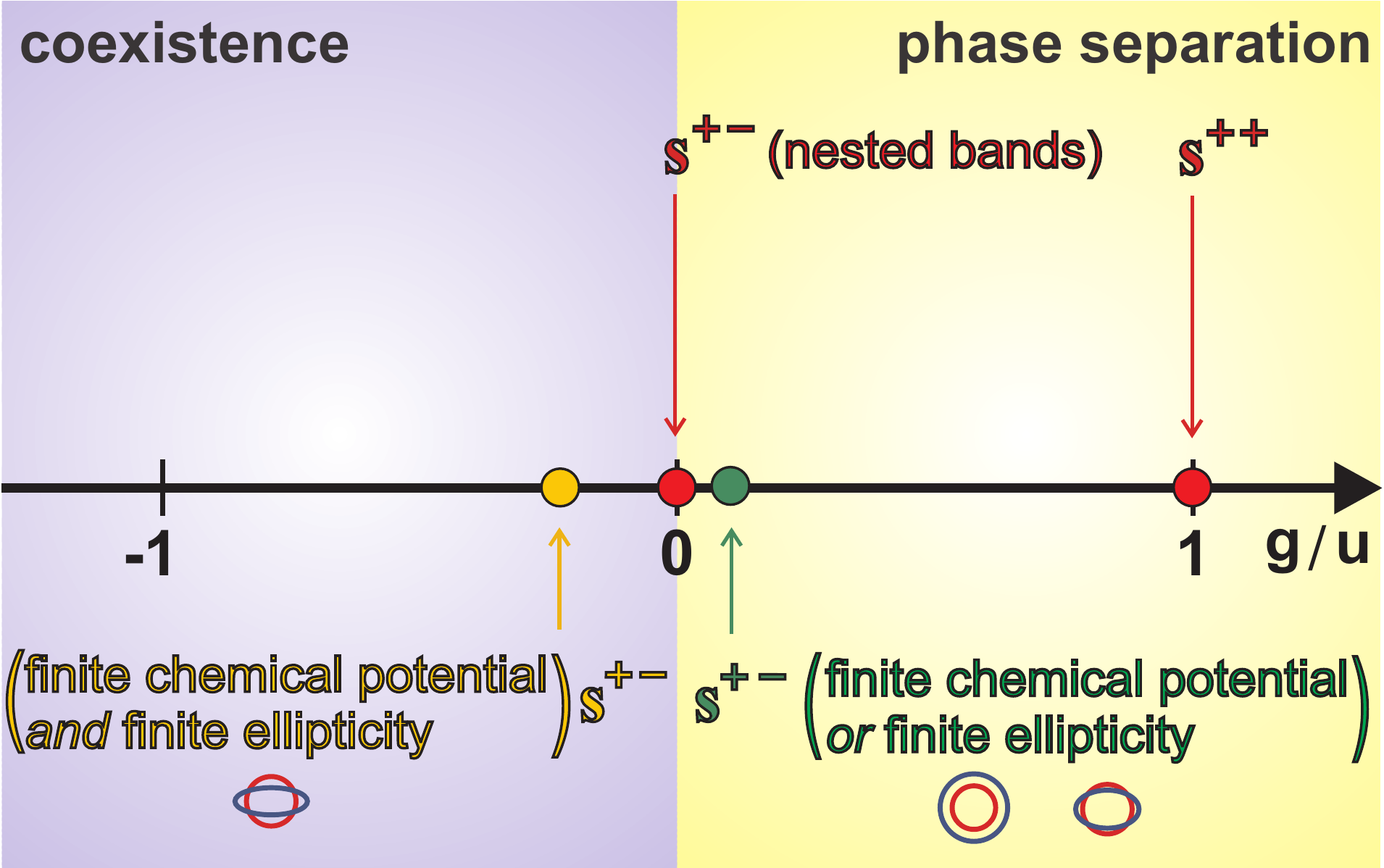} 
\par\end{centering}

\protect\protect\protect\protect\protect\protect\protect\caption{Schematic representation of the fate of the competing SC and SDW orders
in the two-band model: for perfect nesting, $s^{+-}$ is at the verge
of microscopic coexistence or macroscopic phase separation with SDW
($\bar{g}=0$), whereas $s^{++}$ is deep in the phase-separation
region ($\bar{g}>0$). Deviations from perfect nesting may take the
$s^{+-}$ state to either regime, whereas $s^{++}$ remains in the
phase-separation region. Figure adapted from Ref. \cite{RMF10_rapid}.
\label{fig_2band_coexistence}}
\end{figure}

The microscopic calculation reveals that, for perfect nesting, $\bar{g}=0,$
i.e. the system is at the edge between coexistence and phase separation
\cite{RMF10_rapid}. In this case, it is clear that the free energy
near the multi-critical point has an emergent $O\left(5\right)$ symmetry,
since only the combination $\left(\Delta_{s^{+-}}^{2}+\Delta_{\mathrm{SDW}}^{2}\right)$
appears in Eq. (\ref{F_Delta_M}). This is another manifestation of
the degeneracy between SDW and $s^{+-}$ SC at perfect nesting. Deviations
from perfect nesting may tip the balance to either $\bar{g}<0$ (promoting
microscopic coexistence) or $\bar{g}>0$ (promoting macroscopic phase
separation), as shown in Fig. \ref{fig_2band_coexistence}.

\subsubsection{Three-band model \label{sub:Three-band-model}}

Despite all the interesting insights offered by the 2-band model,
it has a major drawback: by considering the coupling between one hole-
and one electron-pocket only, it assumes that the selected magnetic
order has a single ordering vector and is therefore insensitive to
spontaneous tetragonal symmetry breaking, which is present in the
phase diagram of the FeSC (see Fig. \ref{fig:_phase_diagram}). The
tetragonal symmetry breaking can be captured within the 3-band model
as there are two possibilities for SDW order there, with ordering
vectors $\mathbf{Q}_{X}$ and $\mathbf{Q}_{Y}$. A spontaneous selection
of one of these orders breaks $C_{4}$ symmetry down to $C_{2}$.\\

\paragraph{RG analysis\protect \protect \protect \\
 \protect \\
}

\begin{figure}
\begin{centering}
\includegraphics[width=0.75\columnwidth]{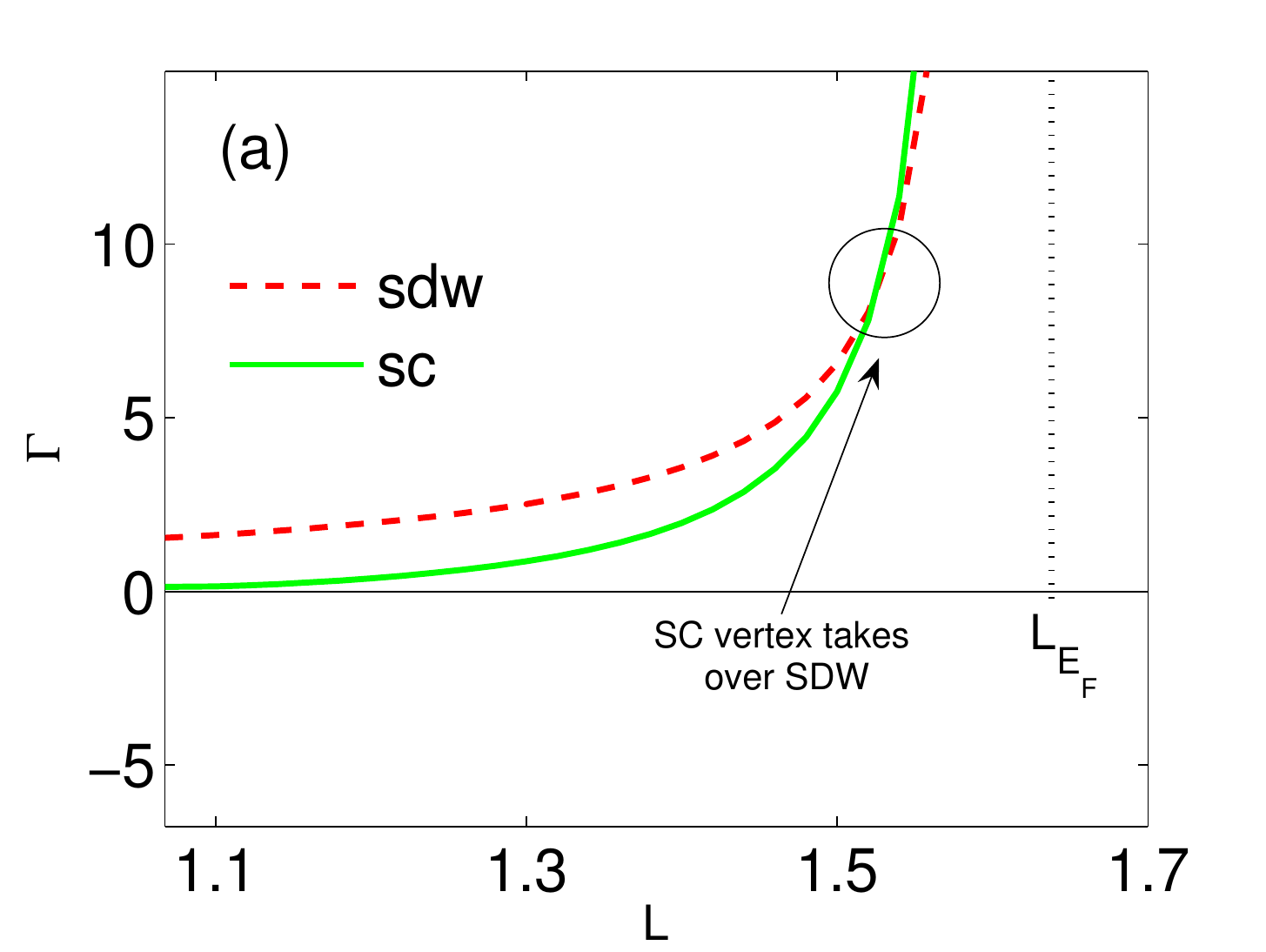} 
\par\end{centering}

\medskip{}

\begin{centering}
\includegraphics[width=0.75\columnwidth]{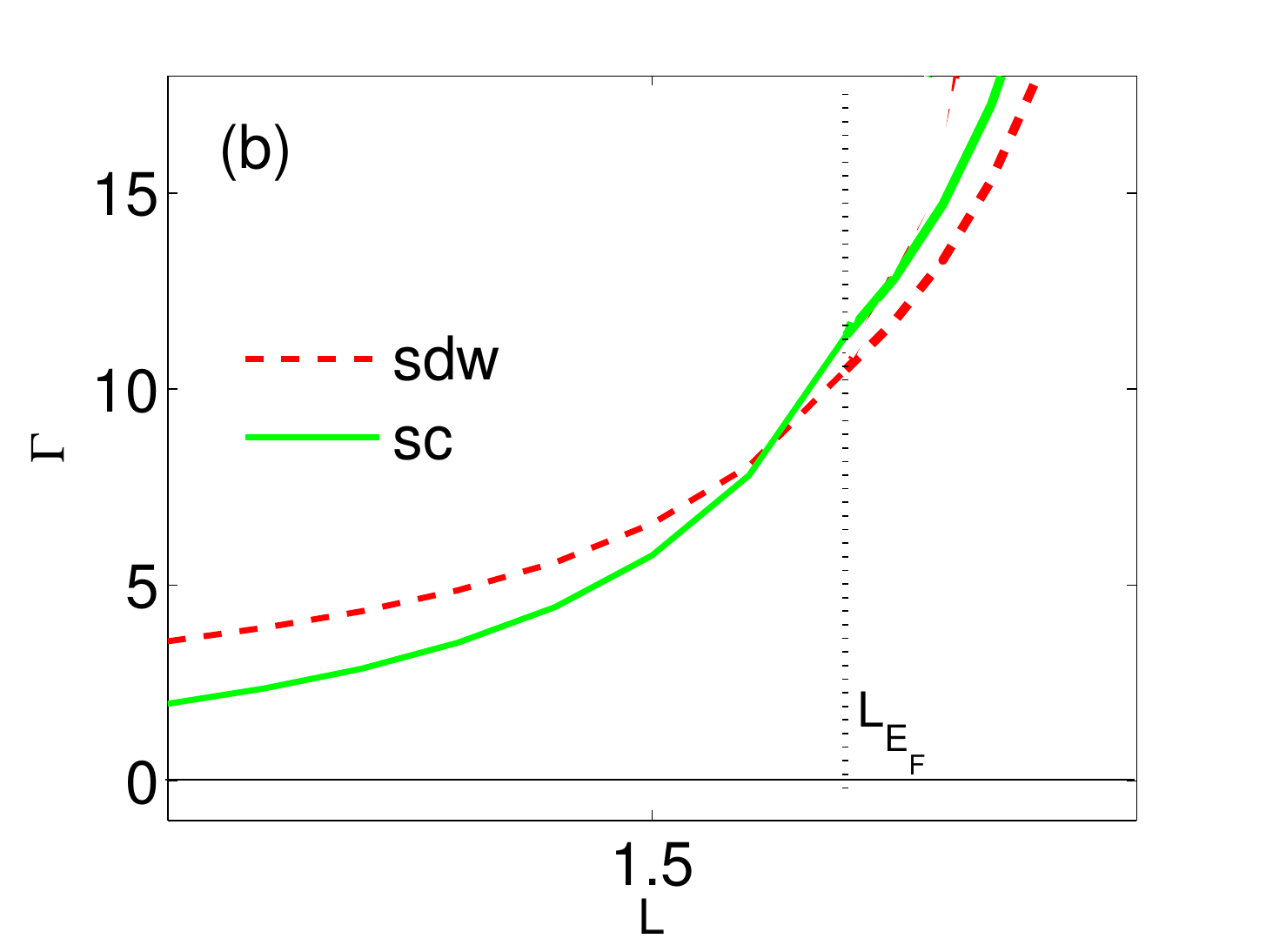} 
\par\end{centering}

\medskip{}

\begin{centering}
\includegraphics[width=0.75\columnwidth]{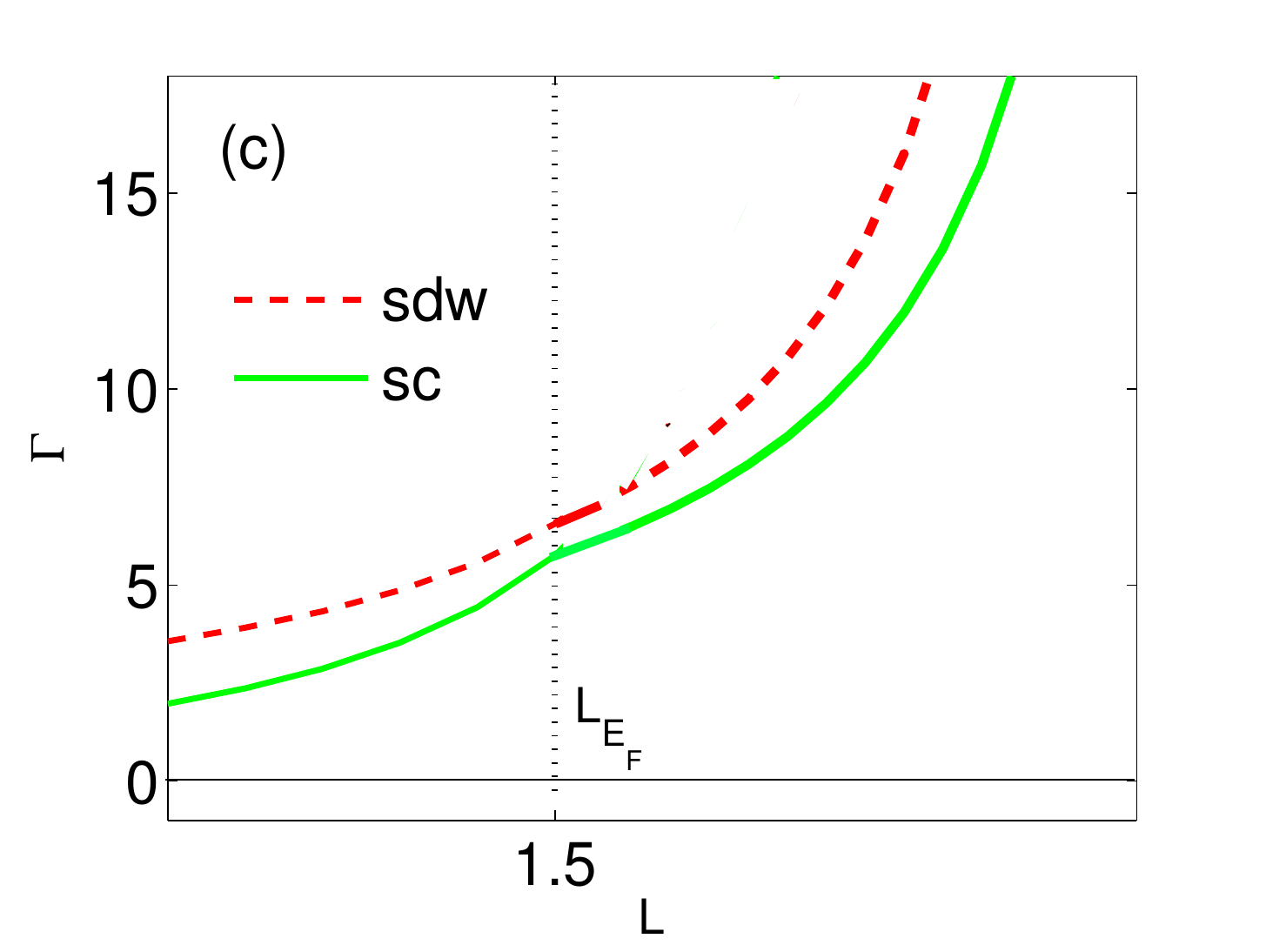} 
\par\end{centering}

\protect\protect\protect\protect\protect\protect\protect\caption{RG flows of the $s^{+-}$ SC (green/solid) and SDW (red/dashed) vertices
for the three-band model for different values of the ratio $L_{F}/L_{c}$.
Figure from Ref. \cite{Maiti10}. \label{fig_3band_RG}}
\end{figure}

We first briefly discuss how the RG results of the 2-band model are
modified in the 3-band case \cite{Maiti10}. Again, we just quote
the results and refer to Ref. \cite{Maiti10} for details. Similarly
to the 2-band case, one finds a divergence of the vertices $\Gamma_{\mathrm{SDW}}$
and $\Gamma_{s^{+-}}$ at a finite running coupling $L_{c}\equiv\log\left(\frac{\Lambda}{E_{c}}\right)$.
However, in contrast to the 2-band model, where the two instabilities
are degenerate, here $u^{s^{+-}}$ becomes larger than $u^{SDW}$
as the instability is approached, as shown in Fig. \ref{fig_3band_RG}.
This happens by purely geometrical reasons (two electron pockets instead
of one). As a result, $s^{+-}$ superconductivity wins over SDW even
in the case of perfect nesting. The situation however changes if the
Fermi energy is not small, i.e. if $L_{F}<L_{c}$. Then, the leading
instability is given by whichever vertex is larger at the scale $L_{F}$.
As shown in the same figure, for large enough Fermi energy values,
the leading instability becomes the SDW and not the SC one \cite{Maiti10}.

To complete the RG analysis, one should use the results for the flow
of the couplings and compute the susceptibilities in the SDW and SC
channels, and also in the channels with $\mathbf{Q}=0$ order. This
analysis shows that not only $s^{+-}$ superconductivity wins over
SDW, but also that the growth of the SDW susceptibility is halted
due to the negative feedback effect from increasing SC fluctuations.
It also shows that the susceptibilities in the $\mathbf{Q}=0$ channels
for the order parameters in Eq. (\ref{new_5}), which either break
$C_{4}$ symmetry down to $C_{2}$ (leading to a nematic order) or
simultaneously shrink/expand hole and electron pockets, grow with
larger exponents than the SC susceptibility. Thus, these $\mathbf{Q}=0$
orders may occur before superconductivity, if the system parameters
allow the RG flow run long enough. We will not focus on this physics
here but discuss it in detail in Sec. \ref{sec:Hybrid-band-orbital-models}
where we include the orbital composition of the Fermi pockets.\\

\paragraph{The selection of the magnetic order\protect \protect \protect \\
\protect \\
 }

An important question is which type of magnetic order is selected,
if the SDW instability occurs prior to superconductivity. Because
there are two possible SDW order parameters, $\boldsymbol{\Delta}_{\mathrm{SDW},X}$
and $\boldsymbol{\Delta}_{\mathrm{SDW},Y}$, with two different ordering
vectors, $\mathbf{Q}_{X}$ and $\mathbf{Q}_{Y}$, there are two possibilities
for the magnetically ordered state: either both order parameters condense
simultaneously, giving rise to a double-\textbf{Q }magnetic phase,
or only one of the order parameters condense, giving rise to a single-\textbf{Q
}stripe magnetic phase \cite{Lorenzana08,Eremin11,Giovannetti11,Brydon11,XWang15,Gastiasoro15,RMF_Berg}.
This issue has important implications for the onset of nematic order,
as we will discuss shortly. The selection of the magnetic order cannot
be determined coming from the magnetically-disordered phase because
the susceptibilities $\chi_{\mathbf{Q}_{X}}$ and $\chi_{\mathbf{Q}_{Y}}$
are identical in this regime. To select the type of SDW order one
needs to go into the ordered phase and analyze quartic couplings between
the two magnetic order parameters $\boldsymbol{\Delta}{}_{\mathrm{SDW},X}$
and $\boldsymbol{\Delta}{}_{\mathrm{SDW},Y}$.

To do this, one can derive the Ginzburg-Landau free energy for $\boldsymbol{\Delta}{}_{\mathrm{SDW},X}$
and $\boldsymbol{\Delta}{}_{\mathrm{SDW},Y}$ from the low-energy
fermionic model. The procedure is similar to the one described above
to study the competition between magnetism and superconductivity.
Here we only focus on the SDW component. The interacting terms in
Eq. (\ref{3band_Hint}) are decoupled in the SDW channel via appropriate
Hubbard-Stratonovich transformations. After integrating out the electronic
degrees of freedom in the partition function and expanding in powers
of the order parameters, we obtain the magnetic free energy (the SDW
subscript is omitted for simplicity) \cite{RMF12}:

\begin{align}
F\left[\boldsymbol{\Delta}_{X},\boldsymbol{\Delta}_{Y}\right] & =\frac{a}{2}\left(\Delta_{X}^{2}+\Delta_{Y}^{2}\right)+\frac{u}{4}\left(\Delta_{X}^{2}+\Delta_{Y}^{2}\right)^{2}\nonumber \\
 & -\frac{g}{4}\left(\Delta_{X}^{2}-\Delta_{Y}^{2}\right)^{2}+w\left(\boldsymbol{\Delta}_{X}\cdot\boldsymbol{\Delta}_{Y}\right)^{2}\label{3band_Fmag}
\end{align}

Before discussing the values of the Ginzburg-Landau coefficients obtained
from the microscopic model, we discuss the possible ground states
of Eq. (\ref{3band_Fmag}). The first two terms depend only on the
combination $\left(\Delta_{X}^{2}+\Delta_{Y}^{2}\right)$, and therefore
do not distinguish between single-\textbf{Q }or double-\textbf{Q}
phases. The last two terms do: $g>0$ favors a state in which either
$\Delta_{X}$ or $\Delta_{Y}$ vanish (single-\textbf{Q}), whereas
$g<0$ favors a state in which $\Delta_{X}=\Delta_{Y}$ (double-\textbf{Q}).
Within the double-\textbf{Q }subspace, $w>0$ favors the configuration
in which $\boldsymbol{\Delta}_{X}\perp\boldsymbol{\Delta}_{Y}$, whereas
$w<0$ favors the configuration in which $\boldsymbol{\Delta}_{X}\parallel\boldsymbol{\Delta}_{Y}$.

\begin{figure}
\begin{centering}
\includegraphics[width=0.9\columnwidth]{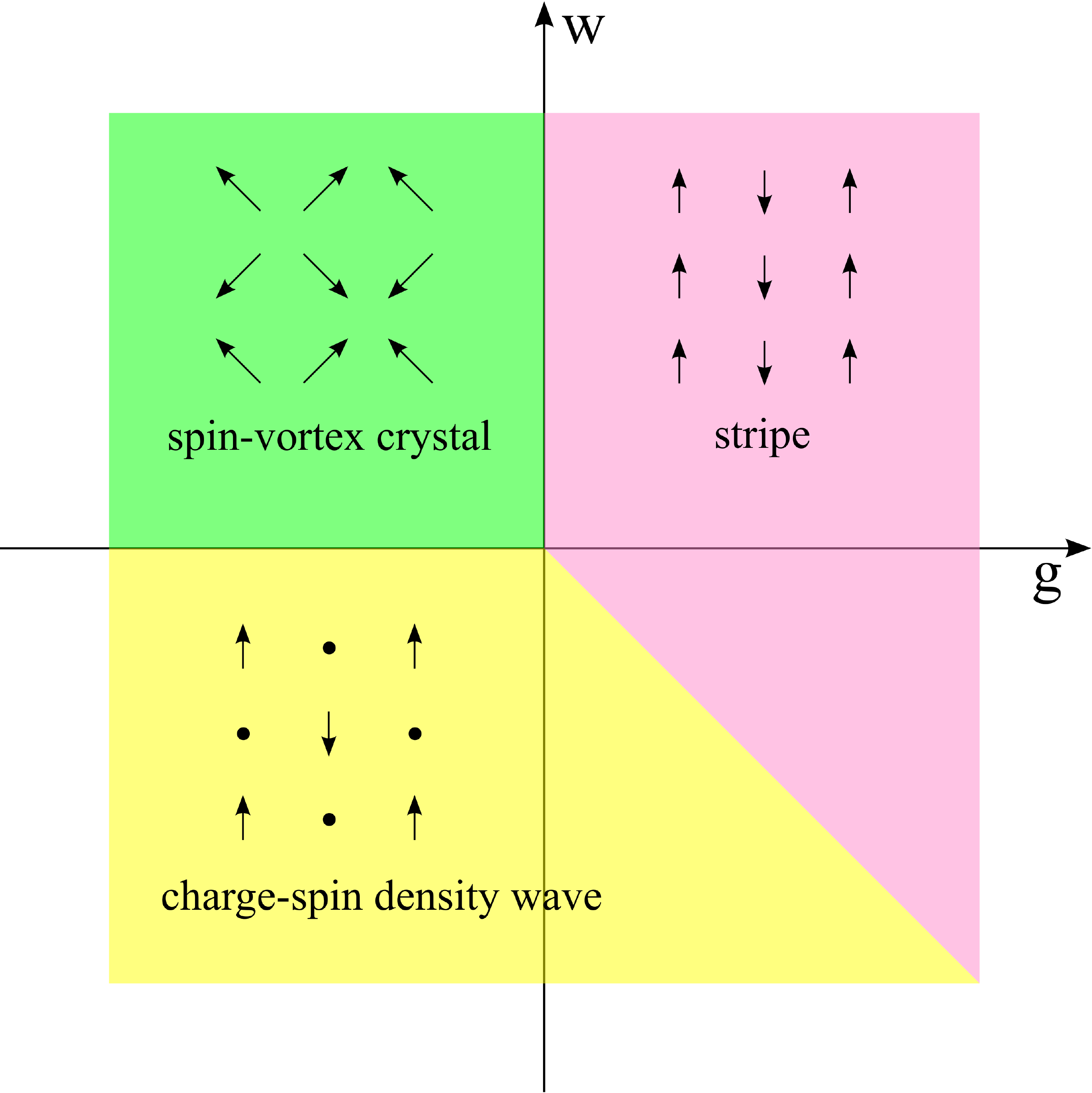} 
\par\end{centering}

\protect\protect\protect\protect\protect\protect\protect\caption{Phase diagram of the SDW free energy (\ref{3band_Fmag}) displaying
the double-\textbf{Q} spin-vortex crystal and charge-spin density-wave
phases, as well as the stripe single-\textbf{Q} phase. Figure adapted
from Ref. \cite{Wang_RMF}. \label{fig_3band_SDW}}
\end{figure}

Fig. \ref{fig_3band_SDW} shows the complete phase diagram in the
$\left(g,w\right)$ plane, together with the depictions of different
ground states in real space \cite{Wang_RMF}. For $g>\max\left(0,-w\right)$,
the system develops a stripe-type magnetic state in which either $\Delta_{X}\neq0$
or $\Delta_{Y}\neq0$. As it is apparent in the figure, this states
breaks the $C_{4}$ tetragonal symmetry of the system down to orthorhombic
$C_{2}$. For $g<-w$ and $w<0$, the ground state is the so-called
charge-spin density-wave (CSDW) \cite{RMF_Berg}, characterized by
$\Delta_{X}=\Delta_{Y}$ and $\boldsymbol{\Delta}_{X}\parallel\boldsymbol{\Delta}_{Y}$.
This is a double-\textbf{Q }state that preserves the tetragonal symmetry
of the system and displays a non-uniform magnetization in the Fe sites.
Finally, for $g<0$ and $w>0$, the magnetic configuration is a non-collinear
one, called a spin-vortex crystal (SVC) \cite{RMF_Berg}, in which
$\Delta_{X}=\Delta_{Y}$ and $\boldsymbol{\Delta}_{X}\perp\boldsymbol{\Delta}_{Y}$.
This is another double-\textbf{Q }state that preserves the tetragonal
symmetry.

The Ginzburg-Landau coefficients $u$, $g$, and $w$ are expressed
via fermionic propagators and depend on the band dispersions of the
underlying fermionic model. The terms $u$ and $g$ are given by \cite{RMF12}:

\begin{align}
u & =\frac{1}{2}\sum_{n}\int\frac{d^{2}\mathbf{k}}{\left(2\pi\right)^{2}}\, G_{h,\mathbf{k}}^{2}\left(G_{e_{X},\mathbf{k}}+G_{e_{Y},\mathbf{k}}\right)^{2}\nonumber \\
g & =-\frac{1}{2}\sum_{n}\int\frac{d^{2}\mathbf{k}}{\left(2\pi\right)^{2}}\, G_{h,\mathbf{k}}^{2}\left(G_{e_{X},\mathbf{k}}-G_{e_{Y},\mathbf{k}}\right)^{2}\label{GL_coefficients}
\end{align}
where $G_{j,\mathbf{k}}^{-1}=i\omega_{n}-\varepsilon_{j,\mathbf{k}}$
is the free-fermion Green's function of band $j$. The coefficient
$w$ vanishes due to phase space and momentum conservation constraints.

For perfect nesting, $g=0$, since $G_{e_{X},\mathbf{k}}=G_{e_{Y},\mathbf{k}}$.
In this case, the magnetic ground state manifold has a larger $O(6)$
symmetry, because only the combination $\left(\Delta_{X}^{2}+\Delta_{Y}^{2}\right)$
appears in the free energy. Small deviations from perfect nesting
yield $g>0$, which implies the existence of a single-\textbf{Q }stripe
magnetic state, as observed experimentally. Stronger deviations from
perfect nesting, however, may change the sign of $g$ and promote
a double-\textbf{Q }phase, as shown in Ref. \cite{XWang15}. Experimentally,
tetragonal double-\textbf{Q }phases have been recently observed in
hole-doped FeSC \cite{Avci14,Bohmer15,Allred16}. Note that, within
this model, $w=0$ and the two types of double-\textbf{Q }phases are
degenerate. To lift this degeneracy, one needs to include effects
beyond those arising from the electronic structure. In particular,
residual interactions in Eq. (\ref{3band_Hint}) that do not contribute
to the SDW instability favor $w>0$ (and therefore a spin-vortex crystal)
\cite{Eremin11,XWang15} whereas coupling to disorder favors $w<0$
(and hence a charge-spin density-wave) \cite{Hoyer16}. Recent Mossbauer
experiments have shown that, at least in some of the compounds where
the double-\textbf{Q }phase has been reported, it is of the charge-spin
density-wave type \cite{Allred16}.\\

\paragraph{Ising-nematic order\protect \protect \protect \\
 }

The 3-band model also offers a suitable platform to study the onset
of nematic order \cite{JCDavis10,Yi11,Chu12,Kasahara12,Gallais13,Rosenthal14,PCDai14,Degiorgi15,Bohmer_review,Gallais_review},
by which we mean the order which breaks $C_{4}$ symmetry down to
$C_{2}$ but does not break spin-rotational symmetry. In the case
where the magnetic ground state is the single-\textbf{Q }stripe one
$(g>0)$, the system actually has a doubly-degenerate ground state
corresponding to either $\Delta_{Y}=0$ ($\mathbf{Q}_{X}$ order)
or $\Delta_{X}=0$ ($\mathbf{Q}_{Y}$ order). In the former, the stripes
are parallel to the $y$ axis, whereas in the latter they are parallel
to the $x$ axis. Therefore, these two ground states are not related
by an overall rotation of the spins, but rather by a $90^{\circ}$
rotation. As a result, the ground state manifold in the $g>0$ case
is $O(3)\times Z_{2}$, with $O(3)$ referring to the spin-rotational
symmetry and $Z_{2}$ to the tetragonal symmetry of the system. In
a mean-field approach both symmetries are broken simultaneously, but
fluctuations in general suppress the continuous $O(3)$ symmetry-breaking
transition to lower temperatures than the discrete $Z_{2}$ symmetry-breaking
transition, particularly in anisotropic layered systems. As a result,
there appears an intermediate phase where the $Z_{2}$ symmetry is
broken (i.e. the system is orthorhombic) but the $O(3)$ symmetry
is preserved (i.e. the system is paramagnetic). In other words, the
stripe SDW phase melts in two stages, giving rise to an intermediate
phase with $O(3)$ symmetry restored but $Z_{2}$ broken \cite{Chandra90,Kivelson08,Sachdev08,Mazin09,RMF10_nematic,Dagotto13,RMF12,RMF14}.
In analogy with liquid crystals, the stripe SDW phase can be viewed
as a smectic phase and the intermediate $Z_{2}$ phase as a nematic
phase.

The spin-driven nematic order parameter $\varphi$ (also often called
Ising-nematic order parameter to underline that it breaks $Z_{2}$
symmetry) is a composite operator made out of products of two SDW
order parameters.

\begin{equation}
\varphi\propto\Delta_{X}^{2}-\Delta_{Y}^{2}\label{nematic}
\end{equation}

When the mean value of $\varphi$ is non-zero, the tetragonal symmetry
is broken, because magnetic fluctuations around $\mathbf{Q}_{X}$
become larger or smaller than magnetic fluctuations around $\mathbf{Q}_{Y}$,
as shown in Fig. \ref{fig_3band_nematic}.

The properties of the nematic phase can be obtained directly from
the microscopically-derived free energy (\ref{3band_Fmag}) by either
computing the susceptibility for the $\varphi$ field within RPA,
or using bosonic RG or large-$N$. We refer the interested reader
to the relevant literature \cite{Kivelson08,Sachdev08,Xu_Qi09,RMF12,RMF14}.

The calculation of the static nematic susceptibility within RPA yields
\cite{RMF10_nematic}:

\begin{figure}
\begin{centering}
\includegraphics[width=1\columnwidth]{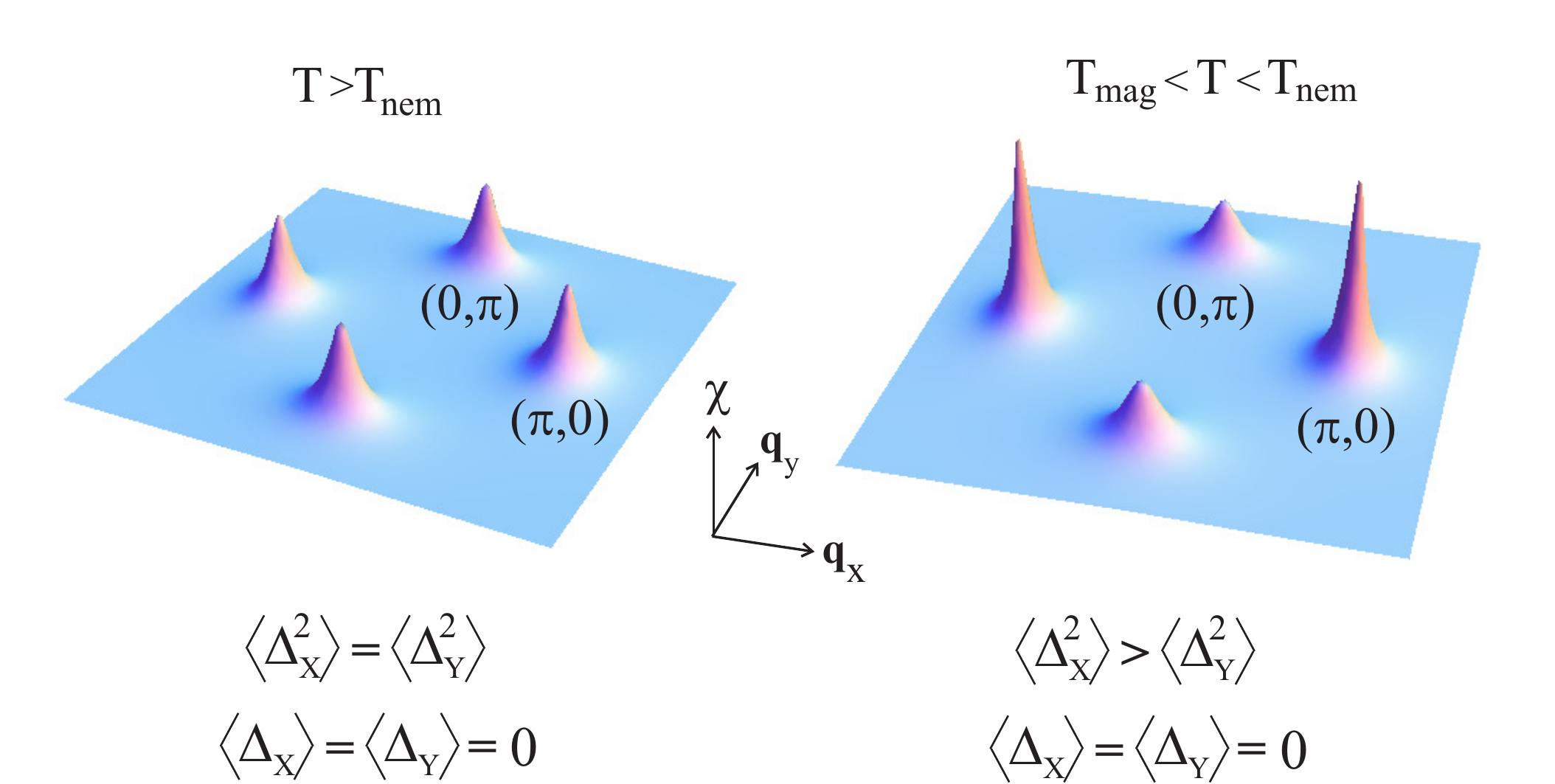} 
\par\end{centering}

\protect\protect\protect\protect\protect\protect\protect\caption{Schematic representation of the nematic phase promoted by the partial
melting of the stripe SDW state. Below the nematic transition temperature
$T_{\mathrm{nem}}$ but above the magnetic transition temperature
$T_{\mathrm{mag}}$, the inelastic magnetic peaks become different
around $\mathbf{Q}_{X}=\left(\pi,0\right)$ and $\mathbf{Q}_{Y}=\left(0,\pi\right)$.
Figure adapted from Ref. \cite{RMF12}. \label{fig_3band_nematic}}
\end{figure}

\begin{equation}
\chi_{\mathrm{nem}}=\frac{\int_{k}\chi_{\mathrm{SDW}}^{2}\left(k\right)}{1-g\int_{k}\chi_{\mathrm{SDW}}^{2}\left(k\right)}\label{chi_nem}
\end{equation}
where we introduced the notation $k=\left(\omega_{n},\mathbf{k}\right)$.
To understand the meaning of this expression, consider that the system
is approaching an SDW transition from high temperatures. At the SDW
transition, the quantity $\int_{k}\chi_{\mathrm{SDW}}^{2}\left(k\right)$
must diverge. However, before it diverges, it will reach the value
$1/g$, no matter how small $g$ is, as long as $g>0$. Thus, $\chi_{\mathrm{nem}}\rightarrow\infty$
before $\chi_{\mathrm{SDW}}\rightarrow\infty$. The close relationship
between these two transitions is evident: if the magnetic transition
temperature is suppressed, then $\int_{k}\chi_{\mathrm{SDW}}^{2}\left(k\right)$
will only reach the value $1/g$ at a lower temperature. Because of
this, the nematic transition line follows the magnetic transition
line, in agreement with the experimental phase diagrams of most FeSC
(except for FeSe, see next section).

Note that in more sophisticated RG or large-$N$ approaches, the nematic
transition is not always a second order transition. It can be first-order,
with a jump in $\varphi$ from zero to a finite value. Furthermore,
in some cases the jump in $\varphi$ triggers the magnetic order,
giving rise to a simultaneous first-order magnetic-nematic transition
(see Ref. \cite{RMF12} for details).

As for the interplay between superconductivity and SDW, the 3-band
model reveals a new ingredient absent in the 2-band model. Similarly
to the 2-band model (see Eq. (\ref{F_Delta_M})), we can derive from
the microscopic model the Ginzburg-Landau free energy for the magnetic
($\Delta_{X}$, $\Delta_{Y}$) and SC ($\Delta_{s^{+-}}$, $\Delta_{s^{++}}$,
$\Delta_{d}$) order parameters. The competition between SDW and superconductivity
is still present, as the biquadratic couplings $\Delta_{s^{+-}}^{2}\Delta_{X/Y}^{2}$
have positive coefficients. But besides this, a new coupling appears
in the free energy \cite{RMF_Millis,Kang14}:

\begin{equation}
\tilde{F}=\lambda\left(\Delta_{s^{+-}}^{*}\Delta_{d}+\Delta_{s^{+-}}\Delta_{d}^{*}\right)\left(\Delta_{X}^{2}-\Delta_{Y}^{2}\right)\label{trilinear}
\end{equation}

This term can be interpreted as a trilinear coupling between the $s^{+-}$
and $d$-wave SC order parameters and the nematic order parameter.
The consequences of this term are interesting: an obvious one is that
long-range nematic order leads to an admixture of the $s^{+-}$ and
$d$-wave gaps. This is not unexpected, since in the orthorhombic
phase these two gaps no longer belong to different irreducible representations.
What is more interesting is that $T_{c}$ can actually increase in
the presence of nematic order, because the pairing frustration between
$s^{+-}$ or $d$-wave is lifted by long-range nematic order \cite{RMF_Millis,DHLee_pairing}.
This is particularly relevant when $s^{+-}$ and $d$-wave channels
are nearly degenerate~\cite{Graser09,Thomale_splusid}. Analogously,
if the system condenses in a single-\textbf{Q }stripe phase, the suppression
of $T_{c}$ due to the competition between SDW and superconductivity
may be alleviated by this effect. This is to be contrasted with the
case of a double-\textbf{Q }phase, in which $\Delta_{X}^{2}=\Delta_{Y}^{2}$,
and the term (\ref{trilinear}) does not contribute to an energy gain
\cite{Kang15}.

\begin{figure}
\begin{centering}
\includegraphics[width=0.8\columnwidth]{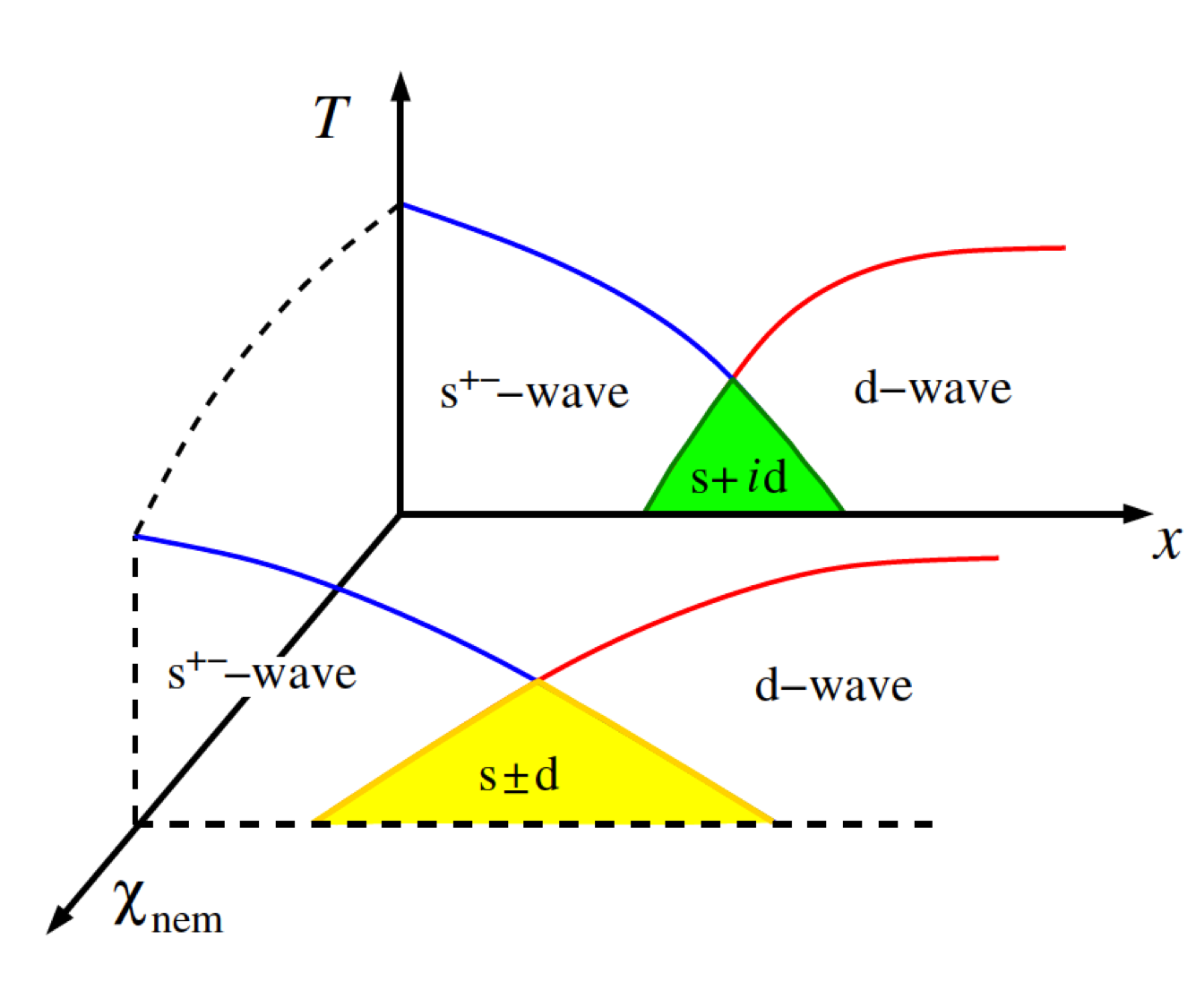} 
\par\end{centering}

\protect\protect\protect\protect\protect\protect\protect\caption{Interplay between $d$-wave and $s^{+-}$-wave superconductivity as
function of the intensity of nematic fluctuations ($\chi_{\mathrm{nem}}$):
for weak nematic fluctuations, the coexistence state $s\pm id$ breaks
time-reversal symmetry, whereas for moderate fluctuations, the coexistence
state $s\pm d$ breaks tetragonal symmetry. \label{fig_nematic_SC}}
\end{figure}

Even in the tetragonal phase, where there is no long-range nematic
order, the trilinear coupling (\ref{trilinear}) can become important
-- as long as the $s^{+-}$ and $d$-wave SC states have comparable
energies and nematic fluctuations are strong. After integrating out
nematic fluctuations, we find that nematic fluctuations promote an
effective attraction between the $s^{+-}$ and $d$-wave channels.
As a result, an exotic nematic-SC state $s\pm d$ that spontaneously
breaks tetragonal symmetry can be stabilized \cite{RMF_Millis,Livanas15}
instead of the $s\pm id$ state that would appear in the absence of
nematic fluctuations \cite{WCLee_splusid}, see Fig. \ref{fig_nematic_SC}.

To summarize, the analysis of the 2-band and 3-band models reveal
that, despite their simplicity, they offer deep insights into the
rich physics of the FeSC, particularly the interplay between SDW and
SC orders, the selection of SDW order by fluctuations, and vestigial
Ising-nematic order. The main drawback of the band models is the neglect
of orbital degrees of freedom, e.g. band models cannot describe the
phenomena associated with spontaneous orbital order. They also cannot
detect specific orbital-induced features in the SDW and SC phases,
such as nodal SDW and orbital anti-phase pairing state.

\section{Orbital-projected band models \label{sec:Hybrid-band-orbital-models}}

We now discuss recent works that aim to capture the low-energy physics
of FeSC by focusing on band excitations near the Fermi surface, while
fully keeping the orbital content of these excitations \cite{Vafek13,Christensen15,RMF_Chubukov_Khodas}.
The inputs for this approach are the Fermi surface geometry (the location
of the Fermi surfaces near $\Gamma$, $X$, $Y$, and $M$ points
in the 1-Fe BZ) and the fact that the excitations near the Fermi surfaces
are composed predominantly of three orbitals -- $d_{xz},d_{yz}$,
and $d_{xy}$. The electronic states near each pocket are treated
as separate excitations, as in the band-basis approach of the previous
section. However, the interactions between the low-energy electronic
states are not treated phenomenologically. Instead, they are obtained
directly from the underlying orbital model and contain information
about the orbital composition of the low-energy states via the orbital-band
matrix elements from Eq (\ref{band_orbital}).

As discussed above, with the $d_{xz},d_{yz}$, and $d_{xy}$ orbitals,
one can successfully describe the low-energy sector of the electronic
dispersion, but one cannot describe the dispersions over the whole
BZ. Accordingly, the restriction to states near the Fermi pockets
is justified if the excitations with momenta far from the high-symmetry
points of the BZ have high enough energy and do not contribute to
the low-energy physics. This generally requires all Fermi pockets
to be small and all excitations at, say, half the distance between
different pockets, to be high in energy. The first condition is satisfied
in most FeSC, particularly the ones with only two hole and two electron
pockets. The second condition needs to be verified for each specific
material as some bands remain rather flat between the $\Gamma$-centered
hole pockets and the $X$-and $Y$-centered electron pockets.

In the discussion below we assume that the conditions for the separation
into low-energy states near the pockets and high-energy states between
the pockets are met and analyze how the orbital content of the excitations
affects the hierarchy of instabilities towards SC, density-wave, and
orbital orders.

\subsection{Non-interacting Hamiltonian}

There are different ways to construct low-energy excitations near
the Fermi pockets. One way is to exploit the properties of the $P4/nmm$
space group of a single FeAs layer and construct the minimal model
using the Luttinger's method of invariants \cite{Vafek13}. The free
parameters of the non-interacting part of the model can then be extracted
from the fit to first-principle calculations. Another way is to start
directly from the five-orbital model of Eq. (\ref{matrix_5orb}) 
\begin{equation}
\mathcal{H}_{0}=\sum_{\mu\nu}\left[\epsilon_{\mu\nu}-{\bar{\mu}}\delta_{\mu\nu}\right]\left(\mathbf{k}\right)d_{\mathbf{k}\mu\sigma}^{\dagger}d_{\mathbf{k}\nu\sigma}\label{H0_repeat}
\end{equation}
where $\epsilon_{\mu\nu}\left(\mathbf{k}\right)$ is the $5\times5$
dispersion matrix, and restrict $\epsilon_{\mu\nu}\left(\mathbf{k}\right)$
to the subspace of $d_{xz}$, $d_{yz}$, and $d_{xy}$ orbitals that
dominate the low-energy states near the desired high-symmetry point
\cite{Christensen15,RMF_Chubukov_Khodas}. Expanding near each high-symmetry
point and diagonalizing the quadratic Hamiltonian, one obtains the
dispersion of low-energy excitations in the band basis. In this case,
the band dispersion parameters are given in terms of the original
tight-binding parameters of the orbital model. The drawback of this
procedure is that the actual low-energy dispersions, extracted from
ARPES experiments, generally differ from the ones obtained from the
truncated tight-binding model due to interaction-driven renormalizations
involving high-energy states \cite{Ortenzi09,Fanfarillo16}. According
to ARPES, such renormalizations change the hopping parameters by orbital-dependent
numerical factors, which range between one and three in most of FeSC,
but can be as high as seven \cite{Borisenko16}. In other words, to
obtain the actual low-energy dispersion from the underlying $5\times5$
orbital model, one has to integrate out high-energy states (including
the ones from the other orbitals) rather than just neglect them. It
is therefore more convenient to fit the expansion parameters to the
experiments rather than to first-principle calculations.

We start by considering the region near the $\Gamma$ point. As shown
previously in Fig. \ref{fig_5orbital_spectrum}, the spectral weight
of the low-energy states arises mainly from the $d_{xz}$ and $d_{yz}$
orbitals. In the absence of spin-orbit coupling, these two orbitals
are degenerate at the $\Gamma$ point, i.e., $\epsilon_{xz,xz}(k=0)=\epsilon_{yz,yz}(k=0)$
and $\epsilon_{xz,yz}(k=0)=0$. The degeneracy is exact and stems
from the fact that in group theoretical language $d_{xz}$ and $d_{yz}$
states form the two-dimensional $E_{g}$ irreducible representation
of the $D_{4h}$ group. Introducing the spinor 
\begin{equation}
\psi_{\Gamma,\mathbf{k}}=\left(\begin{array}{c}
\phantom{-}d_{yz,\mathbf{k}\sigma}\\
-d_{xz,\mathbf{k}\sigma},
\end{array}\right)\label{Gamma_spinor}
\end{equation}
one can write the kinetic energy part of the Hamiltonian as 
\begin{equation}
\mathcal{H}_{0,\Gamma}=\sum_{\mathbf{k}}\psi_{\Gamma,\mathbf{k}}^{\dagger}h_{\Gamma}\left(\mathbf{k}\right)\psi_{\Gamma,\mathbf{k}}^{\phantom{\dagger}}\label{H0_Gamma}
\end{equation}

To obtain the elements of the $2\times2$ matrix $h_{\Gamma}\left(\mathbf{k}\right)$
one can either expand the $2\times2$ matrix $\epsilon_{\mu\nu}\left(\mathbf{k}\right)$
for small $\mathbf{k}$ (with $\mu,\nu=d_{xz},d_{yz}$) or write down
all the trigonometric invariants that satisfy the symmetry property
that one orbital transforms into the other under a rotation by $\pi/2$.
In both cases, we obtain \cite{Vafek13,Christensen15} 
\begin{align}
 & h_{\Gamma}(\mbf{k})=\nonumber \\
 & \begin{pmatrix}\epsilon_{\Gamma}+\frac{k^{2}}{2m_{\Gamma}}+bk^{2}\cos2\theta & ck^{2}\sin2\theta\\
ck^{2}\sin2\theta & \epsilon_{\Gamma}+\frac{k^{2}}{2m_{\Gamma}}-bk^{2}\cos2\theta
\end{pmatrix}\otimes\sigma^{0}\label{aux_H0_Gamma}
\end{align}
where the Pauli matrix $\sigma^{0}$ refers to the spin space and
the angle $\theta$ is measured with respect to the $k_{x}$ axis.
The parameters $\epsilon_{\Gamma}$, $m_{\Gamma}$, $b$, and $c$
could be related to the tight-binding parameters of $\epsilon_{\mu\nu}(\mathbf{k})$.
However, due to the reasons discussed above, they should better be
understood as input parameters that can be obtained from fits to ARPES
data. The Hamiltonian is diagonalized by transforming to hole-band
operators $c_{h_{1},\mathbf{k}\sigma}$ and $c_{h_{2},\mathbf{k}\sigma}$
via the rotation 
\begin{align}
c_{h_{1},\mathbf{k}\sigma} & =\cos\theta_{\mathbf{k}}d_{xz,\mathbf{k}\sigma}-\sin\theta_{\mathbf{k}}d_{yz,\mathbf{k}\sigma}\nonumber \\
c_{h_{2},\mathbf{k}\sigma} & =\cos\theta_{\mathbf{k}}d_{yz,\mathbf{k}\sigma}+\sin\theta_{\mathbf{k}}d_{xz,\mathbf{k}\sigma}\label{hole_op_aux}
\end{align}

The Hamiltonian in the band basis is: 
\begin{equation}
\mathcal{H}_{0,\Gamma}=\sum_{i=1,2}\sum_{\mathbf{k}\sigma i}\varepsilon_{h_{i}}\left(\mathbf{k}\right)c_{h_{i},\mathbf{k}\sigma}^{\dagger}c_{h_{i},\mathbf{k}\sigma}^{\phantom{\dagger}}\label{H0_Gamma_band}
\end{equation}
where 
\begin{equation}
\varepsilon_{h_{1,2}}\left(\mathbf{k}\right)=\epsilon_{\Gamma}+\frac{k^{2}}{2m_{\Gamma}}\mp k^{2}\sqrt{b^{2}\cos^{2}\theta+c^{2}\sin^{2}\theta}
\end{equation}

The angle $\theta_{k}$ is related to the polar angle $\theta$ by:
\begin{equation}
\tan2\theta_{\mathbf{k}}=\frac{c}{b}\,\tan2\theta
\end{equation}

The two angles satisfy a simple relationship when $c^{2}=b^{2}$.
Then $\theta_{\mathbf{k}}=-\mathrm{sign}\left(c\right)\,\theta$ if
$b<0$ and $\theta_{\mathbf{k}}=\mathrm{sign}\left(c\right)\,\theta+\frac{\pi}{2}$
if $b>0$. In either case, the condition $c^{2}=b^{2}$ implies that
the dispersions $\varepsilon_{h_{1,2}}\left(\mathbf{k}\right)$ are
isotropic, i.e. the two hole Fermi surfaces are circles of different
radii: 
\begin{equation}
\varepsilon_{h_{1,2}}\left(\mathbf{k}\right)=\epsilon_{\Gamma}+\frac{k^{2}}{2m_{1,2}}\label{epsilon_h}
\end{equation}
with: 
\begin{equation}
m_{1,2}=\frac{m_{\Gamma}}{1\mp2\left|c\right|m_{\Gamma}}\label{aux_epsilon_h}
\end{equation}

Consider now the momentum range near the $X$ pocket. The low-energy
orbital excitations in this region are composed out of $d_{yz}$ and
$d_{xy}$ orbitals (see Fig. \ref{fig_5orbital_spectrum} above).
By this reason, we can restrict the analysis to the $2\times2$ subspace
spanned by the $d_{yz}$ and $d_{xy}$ orbitals and express the kinetic
energy in the orbital space in terms of a spinor 
\begin{equation}
\psi_{X,\mathbf{k}}=\left(\begin{array}{c}
d_{yz,\mathbf{k}+\mathbf{Q}_{X}\sigma}\\
d_{xy,\mathbf{k}+\mathbf{Q}_{X}\sigma}
\end{array}\right)\label{spinor_X}
\end{equation}
as 
\begin{equation}
\mathcal{H}_{0,X}=\sum_{\mathbf{k}}\psi_{X,\mathbf{k}}^{\dagger}h_{X}\left(\mathbf{k}\right)\psi_{X,\mathbf{k}}^{\phantom{\dagger}}\label{H0_X}
\end{equation}

Note that ${\bf k}$ in $h_{X}\left(\mathbf{k}\right)$ is measured
relative to ${\bf Q}_{X}$. The elements of the matrix $h_{X}(\mathbf{k})$
obey certain symmetry conditions, which can also be obtained by expanding
the $2\times2$ matrix $\epsilon_{\mu\nu}\left(\mathbf{k}\right)$
for small $\mathbf{k}+\mathbf{Q}_{X}$ (with $\mu,\nu=d_{yz},d_{xy}$)
\cite{Vafek13,Christensen15}: 
\begin{align}
 & h_{X}(\mathbf{k})=\nonumber \\
 & \begin{pmatrix}\epsilon_{1}+\frac{k^{2}}{2m_{1}}-a_{1}k^{2}\cos2\theta & -2ivk\sin\theta\\
2ivk\sin\theta & \epsilon_{3}+\frac{k^{2}}{2m_{3}}-a_{3}k^{2}\cos2\theta
\end{pmatrix}\otimes\sigma^{0}\label{aux_H0_X}
\end{align}

Because the non-diagonal terms in $h_{X}(\mathbf{k})$ are imaginary,
the transformation to band operators involves complex factors 
\begin{align}
c_{e_{X1},\mathbf{k}+\mathbf{Q}_{X}\sigma} & =\cos\theta_{\mathbf{k}}d_{yz,\mathbf{k}+\mathbf{Q}_{X}\sigma}-i\sin\theta_{\mathbf{k}}d_{xy,\mathbf{k}+\mathbf{Q}_{X}\sigma}\nonumber \\
c_{e_{X2},\mathbf{k}+\mathbf{Q}_{X}\sigma} & =\cos\theta_{\mathbf{k}}d_{xy,\mathbf{k}+\mathbf{Q}_{X}\sigma}-i\sin\theta_{\mathbf{k}}d_{yz,\mathbf{k}+\mathbf{Q}_{X}\sigma}.\nonumber \\
\label{hole_op_aux_1}
\end{align}

The diagonal Hamiltonian in the band basis is 
\begin{equation}
\mathcal{H}_{0,X}=\sum_{\mathbf{k}\sigma i}\varepsilon_{e_{Xi}}\left(\mathbf{k}+\mathbf{Q}_{X}\right)c_{e_{Xi},\mathbf{k}+\mathbf{Q}_{X}\sigma}^{\dagger}c_{e_{Xi},\mathbf{k}+\mathbf{Q}_{X}\sigma}^{\phantom{\dagger}}\label{H0_X_band}
\end{equation}
where

\begin{align}
\varepsilon_{e_{X1,X2}}\left(\mathbf{k}+\mathbf{Q}_{X}\right) & =\frac{A_{1}+A_{3}}{2}\label{new_1}\\
 & \pm\sqrt{\left(\frac{A_{1}-A_{3}}{2}\right)^{2}+4k^{2}v^{2}\sin^{2}{\theta}}\nonumber 
\end{align}

Here, $A_{1}=\epsilon_{1}+k^{2}/(2m_{1})-a_{1}k^{2}\cos2\theta$ and
$A_{3}=\epsilon_{3}+k^{2}/(2m_{3})-a_{3}k^{2}\cos2\theta$ are the
diagonal elements of the matrix $h_{X}(\mathbf{k})$. The angle $\theta_{\mathbf{k}+\mathbf{Q}_{X}}$
is related to the polar angle $\theta$ by 
\begin{equation}
\tan{2\theta_{\mathbf{k}+\mathbf{Q}_{X}}}=\frac{4vk\sin2\theta}{A_{1}-A_{3}}
\end{equation}

Out of the two dispersions in (\ref{new_1}), only one crosses the
Fermi level. Let us first consider the angles $\theta=0,\pi$, for
which the hybridization between the $d_{xy}$ and $d_{yz}$ orbitals
vanishes. ARPES measurements show that at $k=0$ (i.e. at the $X$
point) both $A_{1}$ and $A_{3}$ are negative and the $d_{xy}$ orbital
has a lower energy, i.e. $A_{3}<A_{1}$ \cite{Coldea_FeSe1,Borisenko16_FeSe}.
However, for $k=k_{F}$, the band that crosses the Fermi level has
a pure $d_{xy}$ character. As a result, $A_{3}<A_{1}$ for $k=0$
and $A_{3}>A_{1}$ for $k=k_{F}$ and $\theta=0,\pi$. Consequently,
the band that crosses the Fermi level at these angles must be $\varepsilon_{e_{X1}}\left(\mathbf{k}\right)=\left(\frac{A_{1}+A_{3}}{2}\right)+\left|\frac{A_{1}-A_{3}}{2}\right|$,
which interpolates between pure $d_{yz}$ character at $k=0$ ($A_{3}<A_{1}$)
and pure $d_{xy}$ character at $k=k_{F}$ ($A_{3}>A_{1}$).

For any other value of $\theta$, the $d_{yz}$ and $d_{xy}$ orbital
dispersions become hybridized. By continuity, the dispersion which
crosses the Fermi level must be $\varepsilon_{e_{X1}}\left(\mathbf{k}\right)$.
Hereafter, we drop the subscript and denote this dispersion by $\varepsilon_{e_{X}}\left(\mathbf{k}\right)$
and the corresponding band operator by $c_{e_{X},{\mathbf{k}}+{\mathbf{Q}}_{X}\sigma}$.
The second dispersion $\varepsilon_{e_{X2}}\left(\mathbf{k}\right)$
does not cross the Fermi level and we assume that it does not belong
to the low-energy sector.

Similarly, for the electron pocket at $Y$ we consider the $2\times2$
subspace spanned by the $d_{xz}$ and $d_{xy}$ orbitals, define the
spinor: 
\begin{equation}
\psi_{Y,\mathbf{k}}=\left(\begin{array}{c}
d_{xz,\mathbf{k}+\mathbf{Q}_{Y}\sigma}\\
d_{xy,\mathbf{k}+\mathbf{Q}_{Y}\sigma}
\end{array}\right)\label{spinor_Y}
\end{equation}
and write the kinetic energy as 
\begin{equation}
\mathcal{H}_{0,Y}=\sum_{\mathbf{k}}\psi_{Y,\mathbf{k}}^{\dagger}h_{Y}\left(\mathbf{k}\right)\psi_{Y,\mathbf{k}}^{\phantom{\dagger}}\label{H0_Y}
\end{equation}
with 
\begin{align}
 & h_{Y}(\mathbf{k})=\nonumber \\
 & \begin{pmatrix}\epsilon_{1}+\frac{k^{2}}{2m_{1}}+a_{1}k^{2}\cos2\theta & -2ivk\cos\theta\\
2ivk\cos\theta & \epsilon_{3}+\frac{k^{2}}{2m_{3}}+a_{3}k^{2}\cos2\theta
\end{pmatrix}\otimes\sigma^{0}\label{aux_H0_Y}
\end{align}

The dispersion that crosses the Fermi level is 
\begin{eqnarray}
 &  & \varepsilon_{e_{Y}}\left(\mathbf{k}+\mathbf{Q}_{Y}\right)=\varepsilon_{e_{Y1}}\left(\mathbf{k}+\mathbf{Q}_{Y}\right)=\nonumber \\
 &  & \frac{{\bar{A}}_{1}+{\bar{A}}_{3}}{2}+\sqrt{\left(\frac{{\bar{A}}_{1}-{\bar{A}}_{3}}{2}\right)^{2}+4k^{2}v^{2}\cos^{2}{\theta}}.
\end{eqnarray}
where $\bar{A}_{1}=\epsilon_{1}+k^{2}/(2m_{1})+a_{1}k^{2}\cos2\theta$
and $\bar{A}_{3}=\epsilon_{3}+k^{2}/(2m_{3})+a_{3}k^{2}\cos2\theta$
are diagonal components of $h_{Y}(\mathbf{k})$. We label the corresponding
band operator as $c_{e_{Y},{\mathbf{k}}+{\mathbf{Q}}_{Y}\sigma}$.

Combining Eqs. (\ref{H0_Gamma}), (\ref{H0_X}), and (\ref{H0_Y}),
we obtain the the free-fermion part of the low-energy Hamiltonian:
\begin{equation}
\mathcal{H}_{0}=\sum_{\mbf{k}}\Psi_{\mathbf{k}}^{\dagger}\left[\hat{H}_{0}(\mbf{k})-\mu\hat{1}\right]\Psi_{\mathbf{k}}\label{H0_final_qux}
\end{equation}
where $\Psi_{\mathbf{k}}$ is the enlarged spinor 
\begin{equation}
\Psi_{\mathbf{k}}=\left(\begin{array}{c}
\psi_{Y,\mathbf{k}}\\
\psi_{X,\mathbf{k}}\\
\psi_{\Gamma,\mathbf{k}}
\end{array}\right)\label{aux_spinor}
\end{equation}
and the Hamiltonian in the matrix form is 
\begin{equation}
\hat{H}_{0}(\mbf{k})=\begin{pmatrix}h_{Y}(\mbf{k}) & 0 & 0\\
0 & h_{X}(\mbf{k}) & 0\\
0 & 0 & h_{\Gamma}(\mbf{k})
\end{pmatrix}\label{aux_H0_final}
\end{equation}

\begin{figure}
\begin{centering}
\includegraphics[width=0.8\columnwidth]{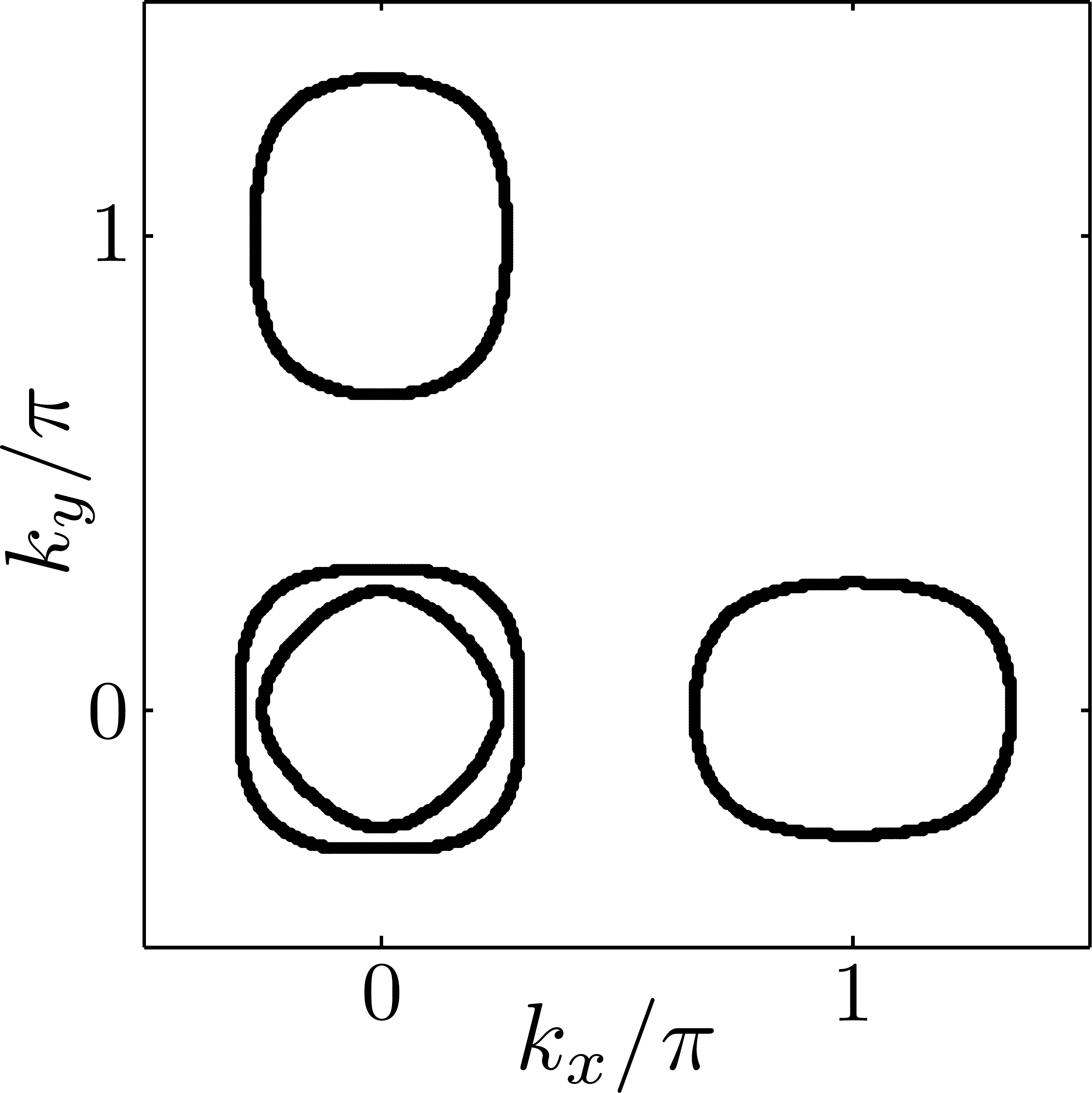} 
\par\end{centering}

\protect\protect\protect\protect\protect\protect\protect\caption{Fermi surface of the orbital-projected band model in the 1-Fe Brillouin
zone. \label{fig_hybrid_FS}}
\end{figure}

Fig. \ref{fig_hybrid_FS} shows the resulting Fermi surfaces from
this model. For this figure, the input parameters were obtained from
the fit to first-principle calculations \cite{Vafek13} (see Table
\ref{tab_hybrid} in Appendix A; higher-order off-diagonal terms have
been included to yield a better looking Fermi surface). As discussed
above, alternatively one can treat $\epsilon_{\Gamma}$, $\epsilon_{1}$,
$\epsilon_{3}$, $m_{\Gamma}$, $m_{1}$, $m_{3}$, $a_{1},a_{3},b,c$,
and $v$ as input parameters and obtain them from fits to ARPES data.
The advantage of this last procedure is that it deals with the actual
measured dispersion and hence includes all regular renormalizations
from high-energy fermions, which shrink and move the bands \cite{Ortenzi09,Medici_book,Valenti_LiFeAs}.
Note also that although the number of input parameters (11 total)
is not small, it is still much smaller than the number of input parameters
for the full-fledged five-orbital model from Sec. \ref{sub:Five-orbital-model}

We emphasize that the model presented here is not equivalent to the
3-orbital model which we considered in Sec. \ref{sec:Orbital-basis-models}.
To be more precise, the 3-orbital model considered here describes
the low-energy sector of the lattice model made out of $d_{xz}$,
$d_{yz}$, and $d_{xy}$ orbitals near points $\Gamma$, $X$, and
$Y$. We remind that the 3-orbital lattice model has additional Fermi
surfaces, not observed in the experiments. In the present analysis
we take as an input the fact that additional Fermi surfaces are eliminated
by the hybridization between the $t_{2g}$ and $e_{g}$ subsets, and
focus on the experimentally-observed Fermi surface geometry.

If the Fermi surface geometry is such that there exists an additional
hole pocket at the $M$ point, the analysis can be straightforwardly
extended to include it. Because this pocket is made out of the $d_{xy}$
orbital, we just introduce an additional operator $\psi_{M,\mathbf{k}}\equiv d_{xy,\mathbf{k}+\mathbf{Q}_{X}+\mathbf{Q}_{Y}\sigma}$
and write an additional kinetic energy term: 
\begin{equation}
\mathcal{H}_{0,M}=\sum_{\mathbf{k}}\psi_{M,\mathbf{k}}^{\dagger}h_{M}\left(\mathbf{k}\right)\psi_{M,\mathbf{k}}^{\phantom{\dagger}}\label{H0_M}
\end{equation}
with

\begin{equation}
h_{M}(\mathbf{k})=\epsilon_{M}+\frac{k^{2}}{2m_{M}}-b_{M}k^{4}\sin^{2}2\theta\label{H0_M_aux}
\end{equation}

With this extra term, the free-fermion Hamiltonian describes all five
Fermi pockets in terms of three distinct orbital states.

\subsection{Order parameters}

The order parameters can be defined either in the orbital or in the
band basis, similarly to how it was done in the previous two sections.
The difference with respect to the purely orbital models is that now
the momenta are confined to the vicinity of the $\Gamma$, $X$, $Y$,
and $M$ points. Consequently, some order parameters that seem different
when viewed in the full BZ become indistinguishable (see below). Conversely,
the difference with respect to the purely band models is that now
the order parameters do depend on the angles along the Fermi pockets
due to variation of the orbital content of the low-energy excitations.
The order parameters can be straightforwardly converted from one basis
to the other using the transformations from Eqs. (\ref{hole_op_aux})
and (\ref{hole_op_aux_1}). The full list of potential order parameters
is rather long, and for briefness we list below only the order parameters
composed of combinations of $d_{xz}$ and $d_{yz}$ orbitals.

\subsubsection{SDW and CDW orders}

There are four possible order parameters describing SDW order with
momenta $\mathbf{Q}_{X}$ and $\mathbf{Q}_{Y}$~\cite{RMF_Chubukov_Khodas}:
\begin{align}
\boldsymbol{\Delta}_{\mathrm{SDW},Y}(\mathbf{k}) & =d_{xz,{\mathbf{k}}+{\mathbf{Q}}_{Y}\alpha}^{\dag}\bm{\sigma}_{\alpha\beta}d_{xz,{\mathbf{k}}\beta}+h.c.\notag\\
\boldsymbol{\Delta}_{\mathrm{SDW},X}(\mathbf{k}) & =d_{yz,{\mathbf{k}}+{\mathbf{Q}}_{X}\alpha}^{\dag}\bm{\sigma}_{\alpha\beta}d_{yz,{\mathbf{k}}\beta}+h.c.\notag\\
\boldsymbol{\Delta}_{\mathrm{iSDW},Y}(\mathbf{k}) & =id_{xz{\mathbf{k}}+{\mathbf{Q}}_{Y}\sigma}^{\dag}\bm{\sigma}_{\alpha\beta}d_{xz{\mathbf{k}}\beta}+h.c.\notag\\
\boldsymbol{\Delta}_{\mathrm{iSDW},X}(\mathbf{k}) & =id_{yz{\mathbf{k}}+{\mathbf{Q}}_{X}\sigma}^{\dag}d_{yz{\mathbf{k}}\sigma}+h.c.\label{del_r}
\end{align}

The momentum ${\bf k}$ is assumed to be small, what implies that
the relevant electronic states are near the $\Gamma$ and the $X$
or $Y$ points. These order parameters are diagonal in the orbital
index, and correspond to real SDW or imaginary SDW (i.e. spin-current
density-wave). In addition, there are four possible orbital off-diagonal
SDW order parameters: 
\begin{align}
\bar{\boldsymbol{\Delta}}_{\mathrm{SDW},Y}(\mathbf{k}) & =d_{xz,{\mathbf{k}}+{\mathbf{Q}}_{Y}\alpha}^{\dag}\bm{\sigma}_{\alpha\beta}d_{yz,{\mathbf{k}}\beta}+h.c.\notag\\
\bar{\boldsymbol{\Delta}}_{\mathrm{SDW},X}(\mathbf{k}) & =d_{yz,{\mathbf{k}}+{\mathbf{Q}}_{X}\alpha}^{\dag}\bm{\sigma}_{\alpha\beta}d_{xz,{\mathbf{k}}\beta}+h.c.\notag\\
\bar{\boldsymbol{\Delta}}_{\mathrm{iSDW},Y}(\mathbf{k}) & =id_{xz,{\mathbf{k}}+{\mathbf{Q}}_{Y}\alpha}^{\dag}\bm{\sigma}_{\alpha\beta}d_{yz,{\mathbf{k}}\beta}+h.c.\notag\\
\bar{\boldsymbol{\Delta}}_{\mathrm{iSDW},X}(\mathbf{k}) & =id_{yz,{\mathbf{k}}+{\mathbf{Q}}_{X}\alpha}^{\dag}\bm{\sigma}_{\alpha\beta}d_{xz,{\mathbf{k}}\beta}+h.c.\label{bar_del}
\end{align}

In the band basis, these order parameters are bilinear combinations
of $c_{h,{\mathbf{k}}\sigma}$ and $c_{e_{X},{\mathbf{k}}+{\mathbf{Q}}_{X}\sigma}$
or $c_{h,{\mathbf{k}}\sigma}$ and $c_{e_{Y},{\mathbf{k}}+{\mathbf{Q}}_{Y}\sigma}$.
Below we consider the effects of interactions for the simplified model
with electron pockets consisting entirely of $d_{xz}$ and $d_{yz}$
orbitals. For this model, we obtain in the band basis: 
\begin{align}
\boldsymbol{\Delta}_{\mathrm{SDW},Y/X}(\mathbf{k}) & =c_{h_{1/2},{\mathbf{k}}\alpha}^{\dagger}\bm{\sigma}_{\alpha\beta}c_{e_{Y/X},{\mathbf{k}}+{\mathbf{Q}}_{Y/X}\beta}\cos{\theta}\nonumber \\
 & \pm c_{h_{2/1},{\mathbf{k}}\alpha}^{\dagger}\bm{\sigma}_{\alpha\beta}c_{e_{Y/X},{\mathbf{k}}+{\mathbf{Q}}_{Y/X}\beta}\sin{\theta}+h.c.\nonumber \\
\bar{\boldsymbol{\Delta}}_{\mathrm{SDW},Y/X}(\mathbf{k}) & =c_{h_{2/1},{\mathbf{k}}\alpha}^{\dagger}\bm{\sigma}_{\alpha\beta}c_{e_{Y/X},{\mathbf{k}}+{\mathbf{Q}}_{Y/X}\beta}\cos{\theta}\nonumber \\
 & \mp c_{h_{1/2},{\mathbf{k}}\alpha}^{\dagger}\bm{\sigma}_{\alpha\beta}c_{e_{Y/X},{\mathbf{k}}+{\mathbf{Q}}_{Y/X}\beta}\sin{\theta}+h.c.
\end{align}
where the upper sign is for $Y$ and the lower for $X$. The expressions
for the imaginary SDW order parameters are analogous. CDW order parameters
can be constructed by just replacing $\bm{\sigma}_{\alpha\beta}\rightarrow\delta_{\alpha\beta}$
in the expressions above.

\subsubsection{SC order}

We consider only spin-singlet pairing. There are four possible pairing
channels with non-zero order parameters: $A_{1g}$, $B_{1g}$, $B_{2g}$,
and $A_{2g}$. The order parameter in the $A_{2g}$ channel vanishes
under simultaneous interchange of orbital indices and spin projections.
The $A_{1g}$, $B_{1g}$, and $B_{2g}$ order parameters in the orbital
basis are \cite{RMF_Chubukov_Khodas}: 
\begin{eqnarray}
 &  & \Delta_{e}^{A_{1}}=d_{xz,{\mathbf{k}}+{\mathbf{Q}}_{Y}\uparrow}d_{xz,-{\mathbf{k}}-{\mathbf{Q}}_{Y}\downarrow}+d_{yz,{\mathbf{k}}+{\mathbf{Q}}_{X}\uparrow}d_{yz,-{\mathbf{k}}-{\mathbf{Q}}_{X}\downarrow}\nonumber \\
 &  & \Delta_{h}^{A_{1}}=d_{xz,{\mathbf{k}}\uparrow}d_{xz,-{\mathbf{k}}\downarrow}+d_{yz,{\mathbf{k}}\uparrow}d_{yz,-{\mathbf{k}}\downarrow}\nonumber \\
 &  & \Delta_{e}^{B_{1}}=d_{xz,{\mathbf{k}}+{\mathbf{Q}}_{Y}\uparrow}d_{xz,-{\mathbf{k}}-{\mathbf{Q}}_{Y}\downarrow}-d_{yz,{\mathbf{k}}+{\mathbf{Q}}_{X}\uparrow}d_{yz,-{\mathbf{k}}-{\mathbf{Q}}_{X}\downarrow}\nonumber \\
 &  & \Delta_{h}^{B_{1}}=d_{xz,{\mathbf{k}}\uparrow}d_{xz,-{\mathbf{k}}\downarrow}-d_{yz,{\mathbf{k}}\uparrow}d_{yz,-{\mathbf{k}}\downarrow}\nonumber \\
 &  & \Delta_{e}^{B_{2}}=0\nonumber \\
 &  & \Delta_{h}^{B_{2}}=d_{xz,{\mathbf{k}}\uparrow}d_{yz,-{\mathbf{k}}\downarrow}+d_{yz,{\mathbf{k}}\uparrow}d_{xz,-{\mathbf{k}}\downarrow}
\end{eqnarray}

In the band basis, these order parameters become:

\begin{align}
\Delta_{e}^{A_{1}} & =c_{e_{Y},{\mathbf{k}}+{\mathbf{Q}}_{Y}\uparrow}c_{e_{Y},-{\mathbf{k}}-{\mathbf{Q}}_{Y}\downarrow}+c_{e_{X},{\mathbf{k}}+{\mathbf{Q}}_{X}\uparrow}c_{e_{X},-{\mathbf{k}}-{\mathbf{Q}}_{X}\downarrow}\nonumber \\
\Delta_{h}^{A_{1}} & =c_{h_{1},{\mathbf{k}}\uparrow}c_{h_{1},-{\mathbf{k}}\downarrow}+c_{h_{2},{\mathbf{k}}\uparrow}c_{h_{2},-{\mathbf{k}}\downarrow}\nonumber \\
\Delta_{e}^{B_{1}} & =c_{e_{Y},{\mathbf{k}}+{\mathbf{Q}}_{Y}\uparrow}c_{e_{Y},-{\mathbf{k}}-{\mathbf{Q}}_{Y}\downarrow}-c_{e_{X},{\mathbf{k}}+{\mathbf{Q}}_{X}\uparrow}c_{e_{X},-{\mathbf{k}}-{\mathbf{Q}}_{X}\downarrow}\nonumber \\
\Delta_{h}^{B_{1}} & =\left(c_{h_{1},{\mathbf{k}}\uparrow}c_{h_{1},-{\mathbf{k}}\downarrow}-c_{h_{2},{\mathbf{k}}\uparrow}c_{h_{2},-{\mathbf{k}}\downarrow}\right)\cos2\theta\nonumber \\
 & +\left(c_{h_{1},{\mathbf{k}}\uparrow}c_{h_{2},-{\mathbf{k}}\downarrow}-c_{h_{2},{\mathbf{k}}\uparrow}c_{h_{1},-{\mathbf{k}}\downarrow}\right)\sin2\theta\nonumber \\
\Delta_{e}^{B_{2}} & =0\nonumber \\
\Delta_{h}^{B_{2}} & =\left(c_{h_{2},{\mathbf{k}}\uparrow}c_{h_{2},-{\mathbf{k}}\downarrow}-c_{h_{1},{\mathbf{k}}\uparrow}c_{h_{1},-{\mathbf{k}}\downarrow}\right)\sin2\theta\nonumber \\
 & +\left(c_{h_{1},{\mathbf{k}}\uparrow}c_{h_{2},-{\mathbf{k}}\downarrow}+c_{h_{2},{\mathbf{k}}\uparrow}c_{h_{1},-{\mathbf{k}}\downarrow}\right)\cos2\theta\label{new_4}
\end{align}

For non-circular hole pockets, there are additional order parameters
in each representation. They have the same structure as the ones above,
but contain additional powers of $C_{4}$-symmetric factors $\cos4\theta$
either on the hole or on the electron pockets. When the $d_{xy}$
orbital content on the two electron pockets is included, certain gaps
acquire additional contributions that depend on the angle along the
electron pockets as $\cos(4n+2)\theta$. These additional terms, when
large enough, give rise to the emergence of accidental nodes in an
$s-$wave gap \cite{Chubukov_nodes,Maiti10}.

\subsubsection{$\mathbf{Q}=0$ orbital order}

As discussed in Sec. \ref{sec:Orbital-basis-models}, the order parameters
with zero momentum transfer in the particle-hole charge channel are
\begin{align}
\Delta_{\mathrm{POM,}\mu\mu'}=d_{\mu\sigma}^{\dag}d_{\mu'\sigma}\label{rho}
\end{align}
where $\mu,\,\mu'=xz,\, yz$ and the summation over spin indices is
assumed. We label the corresponding combinations near the hole and
the electron pockets as $\Delta_{\mu\mu'}^{e}$ and $\Delta_{\mu\mu'}^{h}$.
The bilinear combinations, which are even under inversion, can be
classified by irreducible representations of the $D_{4h}$ group.
The most relevant ones for comparison with experiments are in the
$A_{1g}$ and $B_{1g}$ channels \cite{RMF_Chubukov_Khodas}:

\begin{align}
\Delta_{\mathrm{POM,}A_{1g}/B_{1g}}^{e} & =d_{xz,{\mathbf{k}}+{\mathbf{Q}}_{Y}\sigma}^{\dagger}d_{xz,{\mathbf{k}}+{\mathbf{Q}}_{Y}\sigma}\nonumber \\
 & \pm d_{yz,{\mathbf{k}}+{\mathbf{Q}}_{X}\sigma}^{\dagger}d_{yz,{\mathbf{k}}+{\mathbf{Q}}_{X}\sigma}\nonumber \\
\Delta_{\mathrm{POM,}A_{1g}/B_{1g}}^{h} & =d_{xz,{\mathbf{k}}\sigma}^{\dagger}d_{xz,{\mathbf{k}}\sigma}\pm d_{yz,{\mathbf{k}}\sigma}^{\dagger}d_{yz,{\mathbf{k}}\sigma}\label{rho_channels}
\end{align}

In the band basis, they become

\begin{align}
\Delta_{\mathrm{POM,}A_{1g}}^{e} & =c_{e_{Y},{\mathbf{k}}+{\mathbf{Q}}_{Y}\sigma}^{\dagger}c_{e_{Y},{\mathbf{k}}+{\mathbf{Q}}_{Y}\sigma}+c_{e_{X},{\mathbf{k}}+{\mathbf{Q}}_{X}\sigma}^{\dagger}c_{e_{X},{\mathbf{k}}+{\mathbf{Q}}_{X}\sigma}\nonumber \\
\Delta_{\mathrm{POM,}A_{1g}}^{h} & =c_{h_{1},{\mathbf{k}}\sigma}^{\dagger}c_{h_{1},{\mathbf{k}}\sigma}+c_{h_{2},{\mathbf{k}}\sigma}^{\dagger}c_{h_{2},{\mathbf{k}}\sigma}\nonumber \\
\Delta_{\mathrm{POM,}B_{1g}}^{e} & =c_{e_{Y},{\mathbf{k}}+{\mathbf{Q}}_{Y}\sigma}^{\dagger}c_{e_{Y},{\mathbf{k}}+{\mathbf{Q}}_{Y}\sigma}-c_{e_{X},{\mathbf{k}}+{\mathbf{Q}}_{X}\sigma}^{\dagger}c_{e_{X},{\mathbf{k}}+{\mathbf{Q}}_{X}\sigma}\nonumber \\
\Delta_{\mathrm{POM,}B_{1g}}^{h} & =\left(c_{h_{1},{\mathbf{k}}\sigma}^{\dagger}c_{h_{1},{\mathbf{k}}\sigma}-c_{h_{2},{\mathbf{k}}\sigma}^{\dagger}c_{h_{2},{\mathbf{k}}\sigma}\right)\cos{2\theta}\nonumber \\
 & +\left(c_{h_{1},{\mathbf{k}}\sigma}^{\dagger}c_{h_{2},{\mathbf{k}}\sigma}+c_{h_{2},{\mathbf{k}}\sigma}^{\dagger}c_{h_{1},{\mathbf{k}}\sigma}\right)\sin{2\theta}\label{rho_channels_1}
\end{align}

The order parameters in the band basis describe the distortions of
the Fermi surface and can be classified as Pomeranchuk order parameters
in either the $s$-wave ($A_{1g}$) or $d$-wave ($B_{1g}$) channels.
In general, there is no requirement that $\Delta_{\mathrm{POM,}A_{1g}/B_{1g}}^{e}$
and $\Delta_{\mathrm{POM,}A_{1g}/B_{1g}}^{h}$ are the same. To make
this point explicit, we introduce symmetric and antisymmetric combinations
of $\Delta_{j}^{e}$ and $\Delta_{j}^{h}$ in different irreducible
channels. In analogy to the SC case, we label these combinations ``plus-plus''
and ``plus-minus'':

\begin{align}
\Delta_{s,\mathrm{POM}}^{++} & =\Delta_{\mathrm{POM,}A_{1g}}^{e}+\Delta_{\mathrm{POM,}A_{1g}}^{h}\nonumber \\
\Delta_{s,\mathrm{POM}}^{+-} & =\Delta_{\mathrm{POM,}A_{1g}}^{e}-\Delta_{\mathrm{POM,}A_{1g}}^{h}\nonumber \\
\Delta_{d,\mathrm{POM}}^{++} & =\Delta_{\mathrm{POM,}B_{1g}}^{e}+\Delta_{\mathrm{POM,}B_{1g}}^{h}\nonumber \\
\Delta_{d,\mathrm{POM}}^{+-} & =\Delta_{\mathrm{POM,}B_{1g}}^{e}-\Delta_{\mathrm{POM,}B_{1g}}^{h}\label{rho_channels_2}
\end{align}

\begin{figure}
\begin{centering}
\includegraphics[width=0.9\columnwidth]{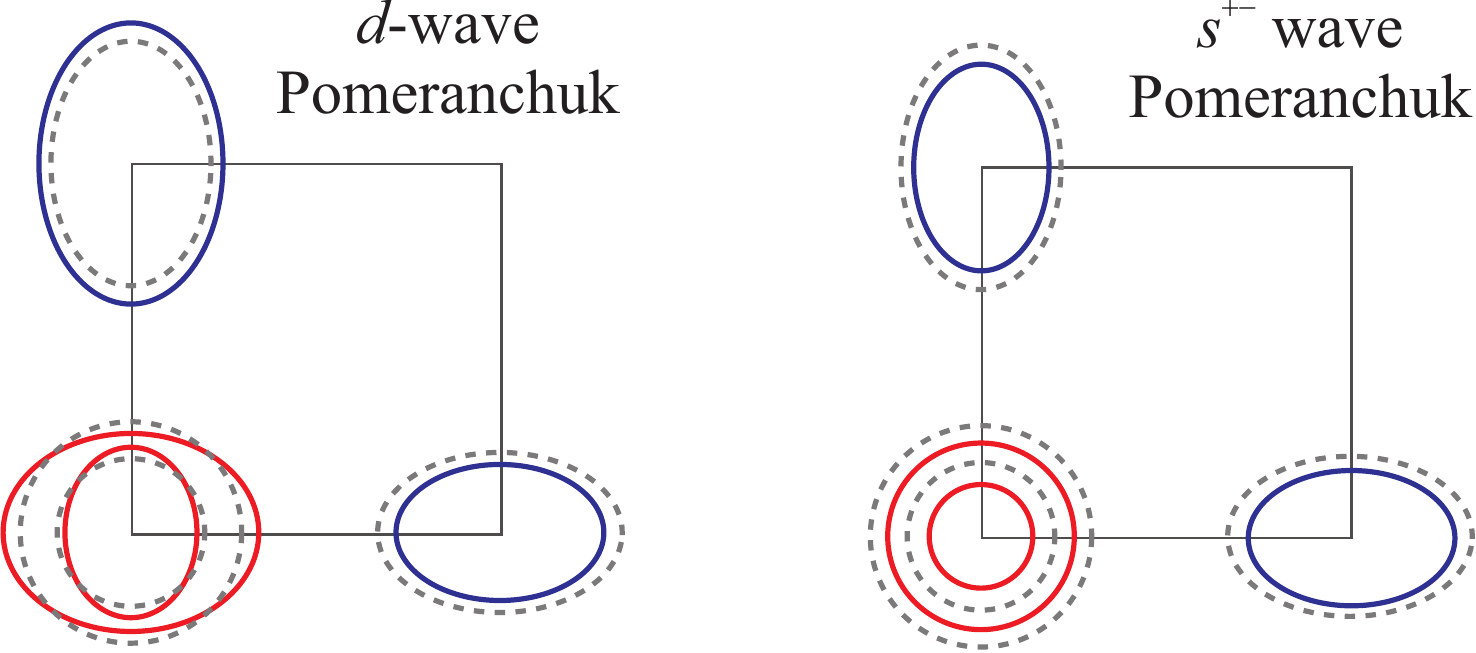} 
\par\end{centering}

\protect\protect\protect\protect\protect\protect\caption{Pomeranchuk instabilities of the orbital-projected band model. Below
the $d$-wave Pomeranchuk transition temperature, the originally circular
hole pockets (dashed lines) are distorted into ellipses of opposite
ellipticities (solid red lines), whereas the two electron pockets
become inequivalent (solid blue lines). A non-zero $s^{+-}$-wave
Pomeranchuk order parameter shrinks or expands all Fermi pockets equally,
keeping the occupation number constant. \label{fig_Pomeranchuk}}
\end{figure}

The average value $\left\langle \Delta_{s,\mathrm{POM}}^{++}\right\rangle $
is never zero and just reflects the fact that the chemical potential
varies with the interaction. A non-zero average $\left\langle \Delta_{s,\mathrm{POM}}^{+-}\right\rangle $
accounts for an interaction-driven simultaneous shrinking (or expansion)
of hole and electron pockets that does not affect charge conservation
(see Fig. \ref{fig_Pomeranchuk}). Because this order parameter does
not break any symmetry of the system, it is generally non-zero at
any temperature \cite{Ortenzi09,Fanfarillo16}, as we discussed in
Sec. \ref{new_ac}. Yet, the susceptibility towards an $s^{+-}$ Pomeranchuk
instability may have a strong temperature dependence. This seems to
be the case for FeSe and, possibly, other materials \cite{Dhaka13,Brouet13,Borisenko16}.

On the other hand, the onset of either $\left\langle \Delta_{d,\mathrm{POM}}^{++}\right\rangle $
or $\left\langle \Delta_{d,\mathrm{POM}}^{+-}\right\rangle $ does
break tetragonal symmetry, distorting the hole Fermi pockets into
ellipses and changing the relative sizes of the two electron pockets
(see Fig. \ref{fig_Pomeranchuk}). As a result, these two order parameters
can only appear below a particular temperature $T_{\mathrm{nem}}$.
Because the two are not orthogonal to each other, both are generally
non-zero below $T_{\mathrm{nem}}$. An equivalent way to state this
is to define the $d$-wave Pomeranchuk order parameter in the orbital
basis as 
\begin{equation}
\Delta_{d,\mathrm{POM}}(\mathbf{k})=\left(d_{xz,{\mathbf{k}}\sigma}^{\dagger}d_{xz,{\mathbf{k}}\sigma}-d_{yz,{\mathbf{k}}\sigma}^{\dagger}d_{yz,{\mathbf{k}}\sigma}^{\dagger}\right)f(\mathbf{k})
\end{equation}
By construction, $f({\mathbf{Q}}_{X})=f({\mathbf{Q}}_{Y})=f({\mathbf{Q}})$.
The non-equivalence of the $B_{1g}$ order parameters on the hole
and on the electron pockets implies that, in general, $f(0)\neq f({\mathbf{Q}})$.
Although the momentum range is confined to the vicinities of the $\Gamma$
and $X/Y$ points, one can still argue that in real space such an
order parameter has on-site and bond components (between nearest neighbors,
and, in general, also further neighbors). If $f(0)\approx f({\mathbf{Q}})$,
i.e., $\Delta_{d,\mathrm{POM}}^{++}\gg\Delta_{d,\mathrm{POM}}^{+-}$,
the on-site component is the largest, whereas if $\Delta_{d,\mathrm{POM}}^{++}\ll\Delta_{d,\mathrm{POM}}^{+-}$,
the bond component is the dominant one.

Other forms of $d$-wave orbital order have been proposed \cite{JPHu_FeSe,TaoLi15},
but in the low-energy sector they are indistinguishable from the ones
we introduced here -- of course, as long as these order parameters
do not mix the $d$-wave and $s$-wave symmetries, which remain strictly
orthogonal within the model we discuss in this section due to the
tetragonal symmetry of the system. To illustrate this, consider the
$d$-wave orbital order with zero transferred momentum proposed in
Ref. \cite{JPHu_FeSe}: 
\begin{equation}
\bar{\Delta}_{d,\mathrm{POM}}=\left(d_{xz,{\mathbf{k}}\alpha}^{\dagger}d_{xz,{\mathbf{k}}\alpha}+d_{yz,{\mathbf{k}}\alpha}^{\dagger}d_{yz,{\mathbf{k}}\alpha}^{\dagger}\right)\left(\cos{k_{x}}-\cos{k_{y}}\right)
\end{equation}

One can readily verify that, if one restricts it to the low-energy
sector, such an order parameter is the same as those in Eqs. (\ref{rho_channels})
- (\ref{rho_channels_2}), and it corresponds to $\Delta_{B_{1g}}^{h}\ll\Delta_{B_{1g}}^{e}$,
i.e. $\Delta_{d,\mathrm{POM}}^{++}\approx\Delta_{d,\mathrm{POM}}^{+-}$.

\subsection{Interaction effects}

We now turn to the analysis of the role of interactions. As before,
our goal is to understand what kind of instability (if any) develops
in the system upon lowering the temperature, and whether different
orders can coexist at the lowest temperature. We briefly review three
approaches. Two fall into the ``spin-fluctuation scenario''. The
first is based on RPA and is not, strictly speaking, a low-energy
approach. The second is a semi-phenomenological approach based on
the low-energy spin-fluctuation model. The third approach is a low-energy
one, based on RG. We consider these three approaches separately.

\subsubsection{RPA approach}

This approach follows a similar analysis previously done for cuprate
superconductors \cite{Scalapino_review}. Its main goal is to understand
the origin of SC pairing and the interplay between different pairing
channels. The idea is to start with the full orbital model (no low-energy
expansion) with on-site Hubbard and Hund interactions, split the interaction
into the spin and charge channels, and use RPA to compute the effective
spin-mediated pairing interaction between the electrons \cite{Graser09,Kuroki1,Kuroki2,Maier_gapanisotropy,Ikeda10,Kemper10,Hirschfeld_Borisenko}.
This procedure is uncontrolled but is generally justified on physics
grounds because magnetism and superconductivity are close to each
other on the phase diagram. The effective, magnetically-mediated pairing
interaction can then be decomposed into different pairing channels
and analyzed separately within each channel. This last analysis is
done within the low-energy subset, by taking the pairing interaction
as static but assuming that the upper cutoff for the pairing is much
smaller than the bandwidth.

\begin{figure}
\begin{centering}
\includegraphics[width=0.9\columnwidth]{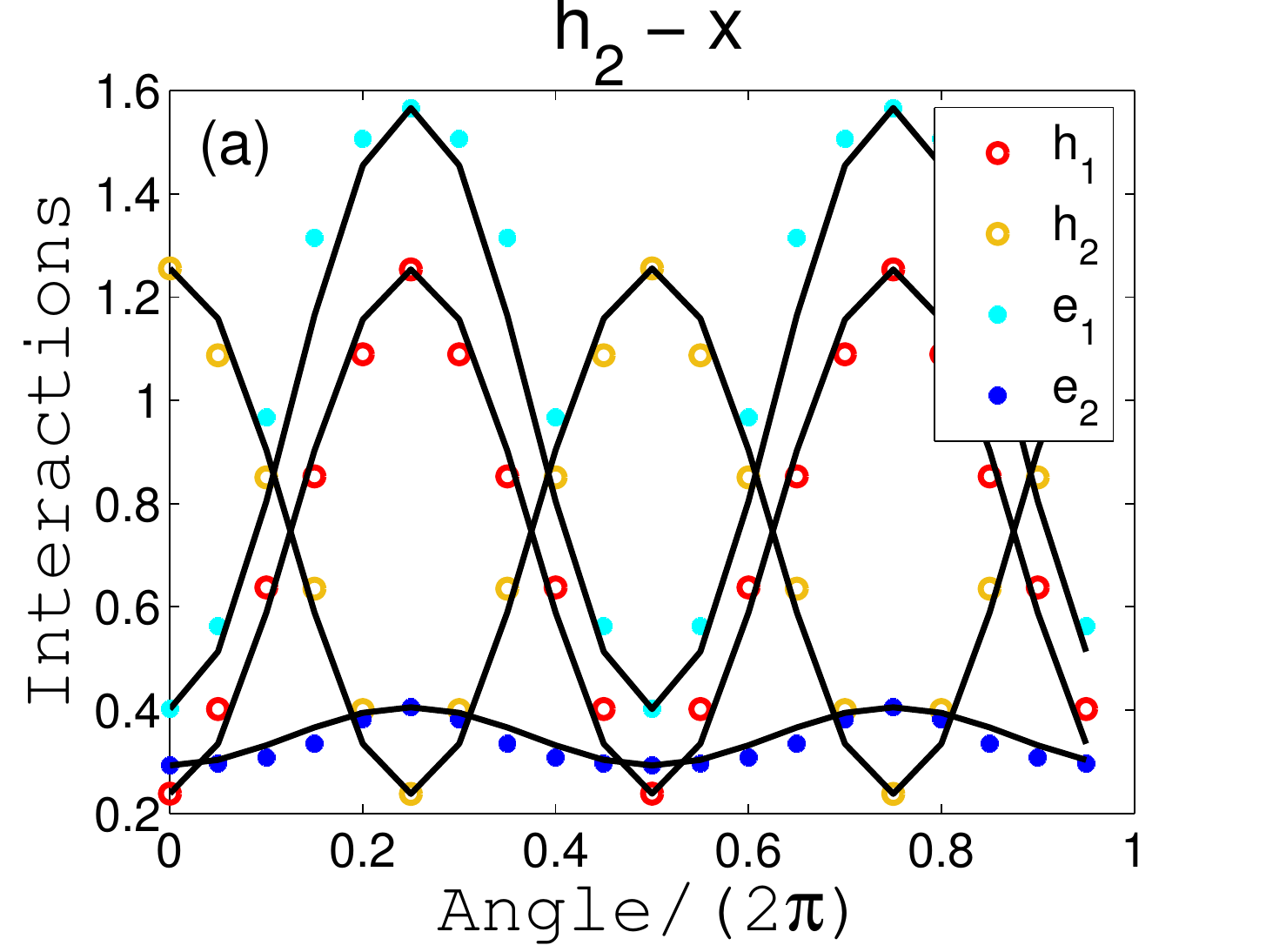} 
\par\end{centering}

\protect\protect\protect\protect\caption{The pairing interactions $\Gamma_{h_{2}\alpha}\left(0,\theta\right)$
involving the Fermi pocket $h_{2}$ at the Fermi momentum corresponding
to $\theta_{h_{2}}=0$ and the Fermi pocket $\alpha=h_{1},h_{2},e_{X},e_{Y}$
at the Fermi momentum corresponding to the polar angle $\theta$.
Solid lines are the leading angular harmonic approximations in Eq.
(\ref{eq_LAHA}) whereas the symbols are the RPA results. Figure from
Ref. \cite{Maiti_LAHA}. \label{fig_LAHA}}
\end{figure}

The RPA approach has been reviewed before \cite{Mazin_review,Scalapino_review}
and here we will just provide a brief description of the solution
of the pairing problem with spin-mediated interaction. The pairing
problem can be analyzed either numerically, by solving a large-size
matrix equation for the eigenvalues in each pairing channel using
the actual tight-binding band structure, or analytically. The analytical
approach is based on the assumption that the pairing interaction $\Gamma_{lm}(k,-k;p-p)\equiv\Gamma_{lm}(k,p)$,
where $l,m$ label Fermi pockets, can be approximated by the lowest-order
harmonics in the angular expansion, i.e., by the products of the terms
that we listed in Eq. (\ref{new_4}) (one for ${\bf k}$, another
for ${\bf p}$), and the terms with $\cos2\theta$ dependence along
the electron pockets \cite{Maiti_LAHA}. We show in Fig. \ref{fig_LAHA}
the comparison between the harmonic expansion and the actual interaction,
showing that the two are close. Accordingly, we approximate $\Gamma_{lm}(k,p)$
as 
\begin{widetext}
\begin{align}
\Gamma_{h_{i}h_{j}}(\theta_{h},\theta'_{h}) & =U_{h_{i}h_{j}}+\tilde{U}_{h_{i}h_{j}}\cos2\theta_{h}\cos2\theta'_{h}\nonumber \\
\Gamma_{h_{i}e_{j}}(\theta,\theta_{e}) & =U_{h_{i}e}(1\pm2\alpha_{h_{i}e}\cos2\theta_{e})+\tilde{U}_{h_{i}e}(\pm1+2{\tilde{\alpha}}_{h_{1}e}\cos2\theta_{e})\cos2\theta_{h}\nonumber \\
\Gamma_{e_{i}e_{i}}(\theta_{e},\theta'_{e}) & =U_{ee}\left[1\pm2\alpha_{ee}(\cos2\theta_{e}+\cos2\theta'_{e})+4\beta_{ee}\cos2\theta_{e}\cos2\theta'_{e}\right]+{\tilde{U}}_{ee}\left[1\pm2{\tilde{\alpha}}_{ee}(\cos2\theta_{e}+\cos2\theta'_{e})+4{\tilde{\beta}}_{ee}\cos2\theta_{e}\cos2\theta'_{e}\right]\nonumber \\
\Gamma_{e_{1}e_{2}}(\theta_{e},\theta'_{e}) & =U_{ee}\left[1+2\alpha_{ee}(\cos2\theta_{e}-\cos2\theta'_{e})-4\beta_{ee}\cos2\theta_{e}\cos2\theta'_{e}\right]+{\tilde{U}}_{ee}\left[-1-2{\tilde{\alpha}}_{ee}(\cos2\theta_{e}-\cos2\theta'_{e})+4{\tilde{\beta}}_{ee}\cos2\theta_{e}\cos2\theta'_{e}\right]\label{eq_LAHA}
\end{align}

\end{widetext}

Here the upper sign is for the electron pocket at $Y$ (pocket $e_{Y}$)
and the lower sign is for the electron pocket at $X$ (pocket $e_{X}$).
The indices $i,j=1,2$ for the hole pockets and $Y,X$ for the electron
pockets. The angles $\theta_{h}$ and $\theta_{e}$ are along hole
and electron pockets, respectively.

The coefficients are obtained by matching this $\Gamma_{lm}(k,p)$
to the full RPA expression for spin-mediated pairing interaction.
Once the prefactors are known, the pairing problem can be easily analyzed
analytically within BCS theory. One cannot obtain the SC transition
temperature $T_{c}$ in this way, because the pairing interaction
is taken as static, but one can compare eigenvalues in different channels.
The first instability will be in the channel with the largest eigenvalue,
at least at weak coupling.

Using this RPA-based spin-fluctuation approach, one can compare the
eigenvalues in the $s^{++}$, $s^{+-}$, $d_{x^{2}-y^{2}}$ and $d_{xy}$
channels and also analyze the angular dependence of the gap function
(the eigenfunction) corresponding to the largest eigenvalue. One also
can analyze how many channels are attractive. In general, one finds
that the leading SC instability is towards an $s^{+-}$ state, but
that the $d$-wave state is very close in energy \cite{Graser09,Maiti_LAHA,Hackl14}.
The same approach can be adapted to study pairing mediated by orbital
fluctuations, if somehow the interaction in the CDW channel becomes
attractive. The CDW-mediated interaction generally leads to superconducting
instability in the $s^{++}$ channel \cite{Kontani_spp,Kontani_Borisenko}.

The RPA approach clearly has advantages but also has its limitations.
By construction, it analyzes the development of superconductivity
prior to SDW magnetism, i.e. it does not address the issue of coexistence
of magnetism and superconductivity (although the RPA approach can
be modified to include this). It also neglects the feedback effect
from SC fluctuations on the magnetic propagator. Finally, the approach
has been designed to study only pairing and cannot be straightforwardly
modified to study orbital order.

\subsubsection{Spin-fermion model}

An alternative reasoning is to abandon RPA and treat the static part
of the magnetically-mediated interaction as an input for the low-energy
model, with parameters taken from the experiment. The dynamical part
of the magnetically-mediated interaction comes from fermions with
low energies, and can be explicitly computed within the low-energy
sector. One then use the full dynamical interaction to obtain $T_{c}$.

Such an approach has been applied to the cuprates and, more generally,
to systems with interaction mediated by near-critical soft fluctuations
(for a review, see Ref. \cite{Abanov_review}). In cases where the
bosonic dynamics is dominated by Landau damping, bosons can be viewed
as slow compared to fermions. In this situation, one can use Eliashberg
theory to compute $T_{c}$ and also the fermionic self-energy. Whether
the same holds for FeSC needs further analysis because the bosonic
dynamics is more complex than Landau damping due to the fact that
both hole-like and electron-like excitations are present.

\subsubsection{RG analysis}

A third approach is to treat magnetism, superconductivity, and orbital
order on equal footing and use the RG technique described in Section
\ref{sub:Two-band-model} to study the hierarchy of instabilities
caused by interactions.

In contrast to the purely band-basis model discussed in that section,
the orbital-projected band model contains information about the orbital
content of the low-energy states. As a result, besides CDW, SDW, and
SC, one can also study within RG the onset of orbital orders of different
types \cite{RMF_Chubukov_Khodas}.

One unavoidable complication is that the number of symmetry-allowed
couplings between the low-energy states of the orbital-projected model
is much larger than that in the purely band model. There, the maximum
number of couplings was $8$. Here the number of couplings for a generic
model with two hole and two electron pockets made out of $d_{xz},d_{yz}$,
and $d_{xy}$ orbitals is 30 \cite{Vafek13}. Once the $d_{xy}$ pocket
at the $M$ point is added, the number grows to 40. These are also
much larger numbers than the number of parameters $U,U',J,J'$ in
the onsite interaction Hamiltonian, Eq. (\ref{H_int_orb}). As we
discussed in Sec. \ref{sec:rg}, the additional couplings can be viewed
as interactions between Fe atoms on different sites.

The existence of 30 (or even 40) distinct couplings, which all flow
under RG, complicates the analysis but also raises questions about
the validity of RPA (or mean-field) approaches, which neglect the
fact that the actual number of distinct couplings is much larger than
those four in Eq. (\ref{H_int_orb}) -- or, equivalently, that interactions
between orbitals at nearest and further neighbors must be included
into the theory, even if they are not present at the bare level. We
argue below that non-onsite interactions are fundamental to describe
the low-energy physics of FeSC.

To illustrate how the RG approach works and how its results differ
from RPA, we consider below the simplified model in which the partial
$d_{xy}$ content of the $X$ and $Y$ electron pockets is neglected,
i.e. we identify the electron pockets centered at $X$ as purely $d_{yz}$
and the one centered at $Y$ as purely $d_{xz}$. To avoid repeating
the RG analysis of the band models, we focus on the novel aspect of
the RG analysis of the orbital-projected model, namely on the possibility
that spontaneous orbital order may be a competitor to SDW and SC.
Specifically, we discuss whether one can obtain attraction in the
orbital channel despite starting with purely repulsive interactions,
and, if this is the case, whether orbital order can become the leading
instability of the system. For details of this calculation we refer
to Ref. \cite{RMF_Chubukov_Khodas}.

\subsubsection{RG for the 4-pocket model without $d_{xy}$ orbital contribution
\label{new_sec:ac}}

Without the $d_{xy}$ contribution, the electron-pocket operators
$c_{e_{i},\mathbf{k}+\mathbf{Q}_{i}\sigma}$ are pure orbital operators:

\begin{align}
c_{e_{X},\mathbf{k}+\mathbf{Q}_{X}\sigma} & =d_{yz,\mathbf{k}+\mathbf{Q}_{X}\sigma}\nonumber \\
c_{e_{Y},\mathbf{k}+\mathbf{Q}_{Y}\sigma} & =d_{xz,\mathbf{k}+\mathbf{Q}_{Y}\sigma}\label{electron_op_aux}
\end{align}
and the kinetic energy near the $X$ and $Y$ points is given by:

\begin{align}
\mathcal{H}_{0,X} & =\sum_{\mathbf{k}\sigma}\varepsilon_{e_{X}}\left(\mathbf{k}+\mathbf{Q}_{X}\right)c_{e_{X},\mathbf{k}+\mathbf{Q}_{X}\sigma}^{\dagger}c_{e_{X},\mathbf{k}+\mathbf{Q}_{X}\sigma}^{\phantom{\dagger}}\nonumber \\
\mathcal{H}_{0,Y} & =\sum_{\mathbf{k}\sigma}\varepsilon_{e_{Y}}\left(\mathbf{k}+\mathbf{Q}_{Y}\right)c_{e_{Y},\mathbf{k}+\mathbf{Q}_{Y}\sigma}^{\dagger}c_{e_{Y},\mathbf{k}+\mathbf{Q}_{Y}\sigma}^{\phantom{\dagger}}\label{H0_X_Y}
\end{align}
with effective band dispersions:

\begin{align}
\varepsilon_{e_{X}}\left(\mathbf{k}+\mathbf{Q}_{X}\right) & =-\varepsilon_{e,0}+\frac{k_{x}^{2}}{2m_{x}}+\frac{k_{y}^{2}}{2m_{y}}\nonumber \\
\varepsilon_{e_{Y}}\left(\mathbf{k}+\mathbf{Q}_{X}\right) & =-\varepsilon_{e,0}+\frac{k_{x}^{2}}{2m_{y}}+\frac{k_{y}^{2}}{2m_{x}}\label{aux_H0_X_Y}
\end{align}

The kinetic energy operator $\mathcal{H}_{0,\Gamma}$, presented in
Eq. (\ref{H0_Gamma_band}), remains unchanged because it does not
have contributions from the $d_{xy}$ orbital. To write down $\mathcal{H}_{\mathrm{int}}$,
we assemble all distinct interactions between low-energy fermions
in the orbital basis. One can verify that there are 14 distinct electronic
interactions involving the low-energy $d_{xz}/d_{yz}$ orbital states
near $\Gamma$, $X$, and $Y$. We present all 14 in the formula below,
where for simplicity of notation the momentum index is omitted and
$\tilde{d}$ operators are shorthand notations for $\tilde{d}_{yz,\sigma}\equiv d_{yz,\mathbf{k}+\mathbf{Q}_{X}\sigma}$
and $\tilde{d}_{xz,\sigma}\equiv d_{xz,\mathbf{k}+\mathbf{Q}_{Y}\sigma}$: 
\begin{widetext}
\begin{align}
\mathcal{H}_{\mathrm{int}}= & U_{1}\sum\left[\tilde{d}_{xz,\sigma}^{\dag}\tilde{d}_{xz,\sigma}d_{xz,\sigma'}^{\dag}d_{xz,\sigma'}+\tilde{d}_{yz,\sigma}^{\dag}\tilde{d}_{yz,\sigma}d_{yz,\sigma'}^{\dag}d_{yz,\sigma'}\right]+\bar{U}_{1}\sum\left[\tilde{d}_{yz,\sigma}^{\dag}\tilde{d}_{yz,\sigma}d_{xz,\sigma'}^{\dag}d_{xz,\sigma'}+\tilde{d}_{xz,\sigma}^{\dag}\tilde{d}_{xz,\sigma}d_{yz,\sigma'}^{\dag}d_{yz,\sigma'}\right]\nonumber \\
+ & U_{2}\sum\left[\tilde{d}_{xz,\sigma}^{\dag}d_{xz,\sigma}d_{xz,\sigma'}^{\dag}\tilde{d}_{xz,\sigma'}+\tilde{d}_{yz,\sigma}^{\dag}d_{yz,\sigma}d_{yz,\sigma'}^{\dag}\tilde{d}_{yz,\sigma'}\right]+\bar{U}_{2}\sum\left[\tilde{d}_{xz,\sigma}^{\dag}d_{yz,\sigma}d_{yz,\sigma'}^{\dag}\tilde{d}_{xz,\sigma'}+\tilde{d}_{yz,\sigma}^{\dag}d_{xz,\sigma}d_{xz,\sigma'}^{\dag}\tilde{d}_{yz,\sigma'}\right]\nonumber \\
+ & \frac{U_{3}}{2}\sum\left[\tilde{d}_{xz,\sigma}^{\dag}d_{xz,\sigma}\tilde{d}_{xz,\sigma'}^{\dag}d_{xz,\sigma'}+\tilde{d}_{yz,\sigma}^{\dag}d_{yz,\sigma}\tilde{d}_{yz,\sigma'}^{\dag}d_{yz,\sigma'}\right]+\frac{\bar{U}_{3}}{2}\sum\left[\tilde{d}_{xz,\sigma}^{\dag}d_{yz,\sigma}\tilde{d}_{xz,\sigma'}^{\dag}d_{yz,\sigma'}+\tilde{d}_{yz,\sigma}^{\dag}d_{xz,\sigma}\tilde{d}_{yz,\sigma'}^{\dag}d_{xz,\sigma'}\right]+h.c.\nonumber \\
+ & \frac{U_{4}}{2}\sum\left[d_{xz,\sigma}^{\dag}d_{xz,\sigma}d_{xz,\sigma'}^{\dag}d_{xz,\sigma'}+d_{yz,\sigma}^{\dag}d_{yz,\sigma}d_{yz,\sigma'}^{\dag}d_{yz,\sigma'}\right]+\frac{\bar{U}_{4}}{2}\sum\left[d_{xz,\sigma}^{\dag}d_{yz,\sigma}d_{xz,\sigma'}^{\dag}d_{yz,\sigma'}+d_{yz,\sigma}^{\dag}d_{xz,\sigma}d_{yz,\sigma'}^{\dag}d_{xz,\sigma'}\right]\nonumber \\
+ & \tilde{U}_{4}\sum d_{xz,\sigma}^{\dag}d_{xz,\sigma}d_{yz,\sigma'}^{\dag}d_{yz,\sigma'}+\tilde{\tilde{U}}_{4}\sum d_{xz,\sigma}^{\dag}d_{yz,\sigma}d_{yz,\sigma'}^{\dag}d_{xz,\sigma'}\nonumber \\
+ & \frac{U_{5}}{2}\sum\left[\tilde{d}_{xz,\sigma}^{\dag}\tilde{d}_{xz,\sigma}\tilde{d}_{xz,\sigma'}^{\dag}\tilde{d}_{xz,\sigma'}+\tilde{d}_{yz,\sigma}^{\dag}\tilde{d}_{yz,\sigma}\tilde{d}_{yz,\sigma'}^{\dag}\tilde{d}_{yz,\sigma'}\right]+\frac{\bar{U}_{5}}{2}\sum\left[\tilde{d}_{xz,\sigma}^{\dag}\tilde{d}_{yz,\sigma}\tilde{d}_{xz,\sigma'}^{\dag}\tilde{d}_{yz,\sigma'}+\tilde{d}_{yz,\sigma}^{\dag}\tilde{d}_{xz,\sigma}\tilde{d}_{yz,\sigma'}^{\dag}\tilde{d}_{xz,\sigma'}\right]\nonumber \\
+ & \tilde{U}_{5}\sum\tilde{d}_{xz,\sigma}^{\dag}\tilde{d}_{xz,\sigma}\tilde{d}_{yz,\sigma'}^{\dag}\tilde{d}_{yz,\sigma'}+\tilde{\tilde{U}}_{5}\sum\tilde{d}_{xz,\sigma}^{\dag}\tilde{d}_{yz,\sigma}\tilde{d}_{yz,\sigma'}^{\dag}\tilde{d}_{xz,\sigma'}\label{Hint_hybrid}
\end{align}

\end{widetext}

If one departs from the model of Eq. (\ref{H_int_orb}) with only
onsite interactions, the initial (bare) values of all 14 couplings
are expressed in terms of $U$, $U'$, $J$, and $J'$: 
\begin{align}
U_{1} & =U_{2}=U_{3}=U_{4}=U_{5}=U,\notag\label{Hubbard_relation}\\
\bar{U}_{1} & =\tilde{U}_{4}=\tilde{U}_{5}=U',\notag\\
\bar{U}_{2} & =\tilde{\tilde{U}}_{4}=\tilde{\tilde{U}}_{5}=J,\notag\\
\bar{U}_{3} & =\bar{U}_{4}=\bar{U}_{5}=J'
\end{align}
However, as we said, different couplings evolve differently under
RG. This can be interpreted as if the system generates interactions
between $d_{xz}$ and $d_{yz}$ orbitals at neighboring sites.

Because the non-interacting Hamiltonian $\mathcal{H}_{0}$ is diagonal
in the band basis, it is useful to change the interacting part $\mathcal{H}_{\mathrm{int}}$
to the band basis as well. From Eqs. (\ref{hole_op_aux}) and (\ref{electron_op_aux}),
it is clear that the effect of this change of basis is to dress the
interactions with form factors that depend on the position at the
Fermi pockets, i.e. to induce angle-dependent interactions enforced
by the orbital contents of the Fermi pockets. Note in passing that
the total number of different terms in the band basis is 152. They
are clustered into 14 combinations and each combination flows as a
whole under RG.

Before we discuss the results of the RG analysis, we briefly review
the results of a mean-field approach. Within mean-field, different
channels do not talk to each other and the susceptibility in each
channel behaves as 
\begin{equation}
\chi_{j}(T)=\frac{\chi_{j,0}(T)}{1-\Gamma_{j}\chi_{j,0}(T)}
\end{equation}
where $j$ labels different channels: SDW, CDW, SC, Pomeranchuk, etc
(positive $\Gamma_{j}$ implies attraction). For the orbital-projected
model with onsite interactions only, the couplings in the SDW, $s^{+-}$
SC and $d$-wave Pomeranchuk channels are 
\begin{equation}
\Gamma_{\mathrm{SDW}}=2U,~\Gamma_{\mathrm{SC}}=0,~\Gamma_{\mathrm{POM}}=2U'-U-J
\end{equation}

We see that coupling in the $s^{+-}$ SC channel vanishes, while the
one in SDW channel is attractive and strong. The coupling in the Pomeranchuk
channel is attractive if $2U'>U+J$ (or $U>5J$ if we further impose
spin-rotational invariance, $U'=U-2J$). Given that the susceptibility
in the SDW channel is logarithmically enhanced and the one in the
Pomeranchuk channel is just the density of states, it is obvious that
SDW is the leading instability within mean-field.

We now turn to the RG analysis, which was formally explained above
in Section \ref{sec:rg}. It turns out that for positive (repulsive)
$U_{i}$ in Eq. (\ref{Hint_hybrid}), there exists one stable fixed
trajectory. Along this trajectory, the interactions $\tilde{\tilde{U}}_{4,5}$
and $\tilde{U}_{4,5}$ flow to zero, whereas $\bar{U}_{i}$ and $U_{i}$
($i=1,...,5$) keep increasing and diverge at the same scale $L_{c}=\log\left(\frac{\Lambda}{E_{c}}\right)$.
The ratios of the couplings approach \textit{\emph{universal}} numbers
on a fixed trajectory, no matter what these ratios are at the bare
level. In particular, in our case we obtain $\bar{U}_{i}=U_{i}$ ($i=1,...,5$)
and universal values for the ratios $U_{i}/U_{1}$. All couplings
flow as 
\begin{equation}
U_{i}\left(L\right),{\bar{U}}_{i}\left(L\right)\sim\frac{1}{L_{c}-L}\label{U1}
\end{equation}

We emphasize again that this result implies that commonly neglected
non-onsite interactions become sizable and relevant.

The running couplings near the fixed trajectory are then used as inputs
to compute the fully renormalized vertices in different channels and
the corresponding susceptibilities $\chi_{j}$. These calculations
show that the susceptibilities in the $s^{+-}$ SC channel, the SDW
channel, and the $d$-wave Pomeranchuk channel behave as\textbf{ }\cite{RMF_Chubukov_Khodas}:
\begin{equation}
\chi_{j}\sim\frac{1}{\left(L_{c}-L\right)^{\alpha_{j}}}\label{suscept}
\end{equation}

\begin{figure}
\begin{centering}
\includegraphics[width=0.9\columnwidth]{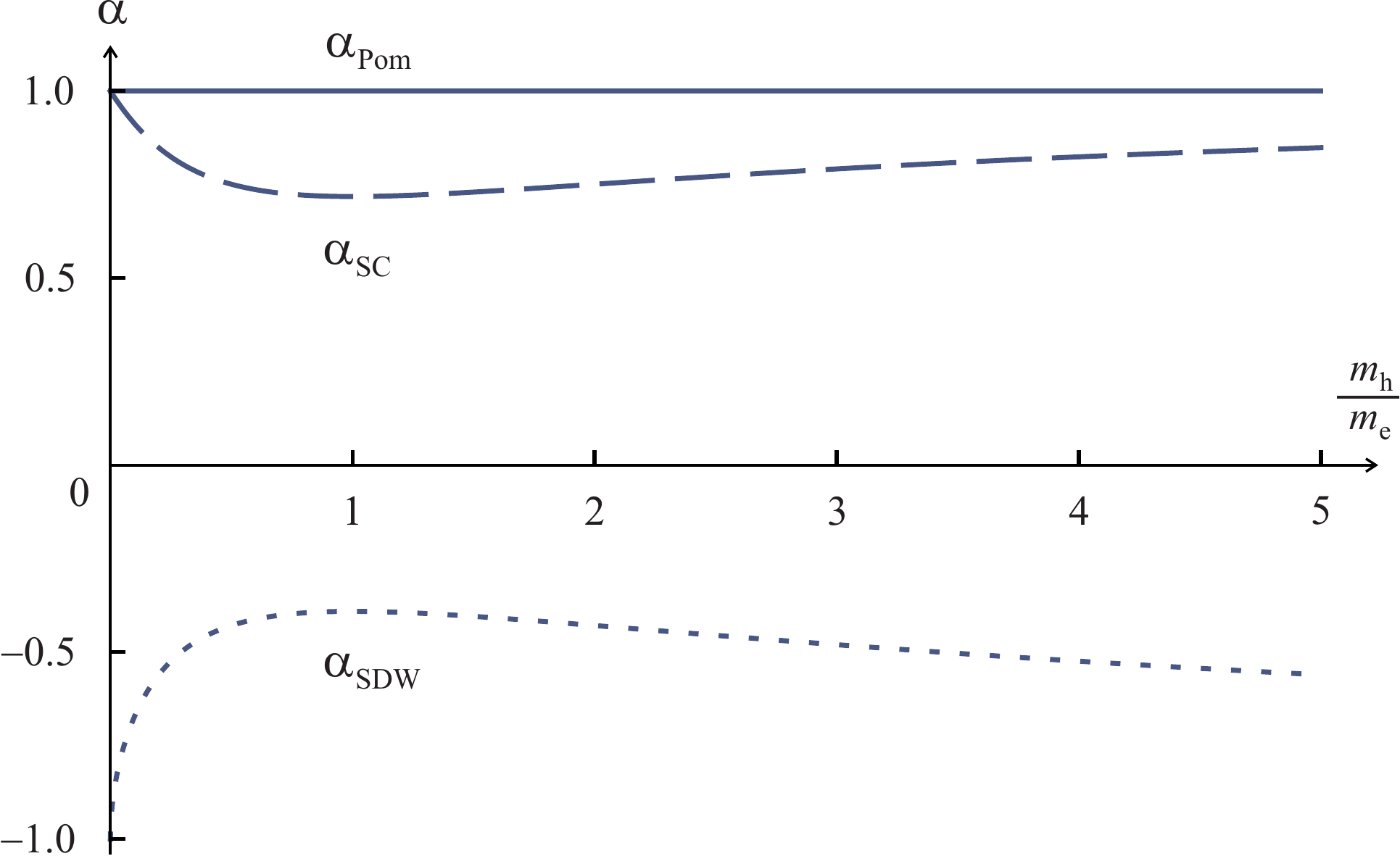} 
\par\end{centering}

\protect\protect\protect\protect\protect\protect\caption{RG results from Ref. \cite{RMF_Chubukov_Khodas} for the orbital-projected
band model showing the behavior of the susceptibility exponents defined
in Eq. (\ref{suscept}) as function of the ratio between the hole
pocket mass $m_{h}$ and the electron pocket mass $m_{e}$. \label{fig_RG_hybrid}}
\end{figure}

In Fig. \ref{fig_RG_hybrid}, we show the behavior of the exponents
$\alpha_{j}$ as functions of the ratio between the electron- and
hole-pocket masses. Across the entire parameter space $\alpha_{\mathrm{POM}}>\alpha_{\mathrm{SC}}>0>\alpha_{\mathrm{SDW}}$,
implying that at the energy scale $E_{c}$ the leading instability
is in the $d$-wave Pomeranchuk channel, towards a spontaneous orbital
order. The SC susceptibility also diverges, albeit with a smaller
exponent, implying that the instability in this channel is the subleading
one. Interestingly, because $\alpha_{\mathrm{SDW}}<0$, the SDW susceptibility
saturates and does not diverge at $E_{c}$, despite the fact that
this susceptibility is the largest at the beginning of the RG flow.

A natural question that arises from these results is why the leading
instability is towards orbital order despite the facts that the free-electron
susceptibility $\chi_{\mathrm{POM},0}$ has no logarithmic divergence
and the bare interaction $\Gamma_{\mathrm{POM}}$ is generally not
attractive. The short answer is that the attractive interaction in
this channel can be viewed as mediated by magnetic fluctuations, like
the attraction in $s^{+-}$ SC channel. In other words, magnetic fluctuations
develop first in the process of the RG flow, and mediate an attractive
interaction, which grows logarithmically as high-energy fluctuations
get progressively integrated out and completely overcomes the bare
interactions in both $s^{+-}$ SC and $d$-wave Pomeranchuk channels.
We see that the mechanisms for attraction in the Pomeranchuk and in
the $s^{+-}$ SC channel are quite similar.

The magnetically-mediated attractive interaction in the pairing channel
also develops within RPA, and in this respect RG and RPA approaches
describe the same physics. However, within RPA, one would always find
the leading instability to be in the SC channel because the bare SC
susceptibility grows logarithmically, while the bare Pomeranchuk susceptibility
is just a constant. In contrast, the RG treatment goes farther than
RPA and shows that, once the SC channel becomes attractive, it starts
competing with the SDW channel, and, as a result of the competition,
the tendency towards instabilities in both channels is reduced. This
in practice implies that the exponents $\alpha_{\mathrm{SC}}$ and
$\alpha_{\mathrm{SDW}}$ become smaller than one (which is their mean-field
values), and that $\alpha_{\mathrm{SDW}}$ even changes sign and becomes
negative. Because the susceptibility in the Pomeranchuk channel is
non-logarithmic, this channel competes much less with the other two
channels. As a consequence, the exponent $\alpha_{\mathrm{POM}}$
remains equal to one. Such an intricate interplay between different
channels illustrates the usefulness of unbiased methods such as RG.

An important point to note is that this result does not imply that
in all cases the leading instability of the system is the Pomeranchuk
one. As we explained previously in Section \ref{sec:Band-basis-models},
once $E$ reaches the scale of the largest Fermi energy, i.e. $L$
reaches $L_{F}\equiv\log\left(\frac{\Lambda}{E_{F}}\right)$, different
instability channels decouple and the RG scheme breaks down. The most
important point for our discussion is that $\chi_{\mathrm{POM}}$
freezes out at $L=L_{F}$, while the susceptibilities in the SC and
SDW channels continue to grow (the SDW susceptibility eventually also
freezes out due to non-perfect nesting, but at a much larger $L$).
Because $\chi_{\mathrm{POM}}$ is only enhanced very close to $L_{c}$
\cite{RMF_Chubukov_Khodas}, in systems where the ratio $E_{F}/\Lambda$
is moderate, such as the 122, 1111, and 111 FeSC compounds, the RG
flow is likely to stop before the Pomeranchuk channel becomes relevant.
As a result, one basically recovers the results of the band-basis
models of Section \ref{sec:Band-basis-models}, in that only SC and
SDW channels are relevant. In this case, a nematic phase can only
arise via a partial melting of the SDW stripe phase, as we discussed
in Section \ref{sub:Three-band-model}. On the other hand, in systems
where $E_{F}/\Lambda$ is small, and $E_{F}$ and $E_{c}$ are comparable,
the leading instability of the system is in the $d$-wave Pomeranchuk
channel, the SC instability is the subleading one, and the SDW instability
does not develop. In this case, nematicity is a result of spontaneous
orbital order.

This general behavior agrees with the phase diagram of FeSe, where
nematic order arises in the presence of weak magnetic fluctuations,
and in the absence of long-range magnetic order \cite{Coldea_FeSe1,Baek15,Bohmer_FeSe}.
Once pressure is applied and $E_{F}/\Lambda$ necessarily increases
for at least one pocket, the system crosses over to a typical iron-pnictide
like behavior, with nematic order preempting a stripe SDW phase \cite{Kothapalli16,Matsuda_FeSe_pressure}.

Besides the SDW, SC, and $d$-wave Pomeranchuk instabilities, another
susceptibility of the system that diverges at $L_{c}$ within the
one-loop RG analysis is in the $s^{+-}$-wave Pomeranchuk channel
(see also \cite{DHLee_09}). For the model of Eq. (\ref{Hint_hybrid}),
a more accurate analysis \cite{RMF_Chubukov_Khodas} shows that this
susceptibility actually diverges at a larger energy (equivalent to
a higher temperature) than the one in the $d$-wave channel. As we
already said, the divergence of the susceptibility in the $s^{+-}$
Pomeranchuk channel is an artifact of the one-loop RG, since in reality
the $s^{+-}$ Pomeranchuk order parameter is non-zero at all temperatures.
Yet, the RG analysis shows that the magnitude of the $s^{+-}$ order
parameter strongly increases around the temperature at which the corresponding
susceptibility diverges in RG. The analysis in Ref. \cite{Ortenzi09}
reveals a self-energy contribution that favors a shift between the
top of the hole band and the bottom of the electron band such that
the areas of both Fermi pockets decrease. Combined with the RG result,
this implies that as temperature decreases, the system should show
a significant temperature-dependent shrinking of both hole-like and
electron-like Fermi pockets.

\subsubsection{Inclusion of the $d_{xy}$ orbital contribution and 5-pocket model
\label{sub:Three-orbitals,-five-band-model}}

To incorporate the $d_{xy}$ orbital into the previous analysis, we
assume first that the $M$-point hole pocket is absent (for instance,
it is sunk below the Fermi level, as in the 111 and 11 materials).
Then the only difference with respect to the model analyzed above
is the presence of $d_{xy}$ spectral weight on the electron pockets.
In the hypothetical case in which these electron pockets are entirely
of $d_{xy}$ character, i.e. $c_{e_{X},\mathbf{k}+\mathbf{Q}_{X}\sigma}\equiv d_{xy,\mathbf{k}+\mathbf{Q}_{X}\sigma}$
and $c_{e_{Y},\mathbf{k}+\mathbf{Q}_{Y}\sigma}\equiv d_{xy,\mathbf{k}+\mathbf{Q}_{Y}\sigma}$,
the number of interactions remains 14, and the RG analysis yields
the same results as for $d_{xz}/d_{yz}$ electron pockets~ \cite{RMF_Chubukov_Khodas}.
Because the results of the RG study are identical in the two limits,
we expect them to hold in a generic situation in which electron pockets
have both $d_{xy}$ and $d_{yz}/d_{xy}$ spectral weight.

The only additional effect introduced by the $d_{xy}$ orbital is
that the nematic order now has two components -- one is the orbital
order component $n_{xz}-n_{yz}$, and the other is $n_{xy}^{X}-n_{xy}^{Y}$,
which is the difference between the $d_{xy}$-orbital charge densities
at the $X$ and $Y$ electron pockets. The latter is not associated
with any type of orbital order, but rather with the fact that the
two electron pockets are located at non-diagonal $X$ and $Y$ points
in the Brillouin zone. This second component can be interpreted as
a $C_{4}$-symmetry breaking anisotropy of the hoppings between nearest-neighbor
$d_{xy}$ orbitals. It is closely related to the $d$-wave Pomeranchuk
order in the pure 3-band model (see the discussion in Section \ref{sub:Three-band-model}).
While the orbital order component of the nematic order parameter splits
the onsite energies of the $d_{xz}$ and $d_{yz}$ orbitals at the
$\Gamma$, $X$, and $Y$ points, the hopping anisotropy component
splits the equivalence between the energy levels of the $d_{xy}$
orbitals at the $X$ and $Y$ points. In general, both components
are present, and their ratio depends on the details of the RG flow
\cite{Borisenko16_FeSe,Laura}.

We now include the fifth Fermi pocket, namely, the $d_{xy}$ hole-pocket
at $M$. An interesting issue is whether this leads to qualitatively
new behavior. A recent analysis argues that the main results remain
the same \cite{Laura}. Specifically, there are several stable and
``almost stable'' fixed trajectories, each with its own basin of
attraction in the parameters space. If the system parameters are such
that the RG flow extends down to the lowest energy, the leading instability
for each fixed trajectory is towards orbital order, the SC instability
is the subleading one, and the SDW susceptibility does not diverge.
If the system parameters are such that the RG flow is halted at higher
energies, the system develops either SDW or SC order. The nematic
order parameter generally has two components, one describing orbital
order and another one the breaking of $C_{4}$ symmetry within the
subset of $d_{xy}$ orbitals.

Nevertheless, the analysis of the RG flow for the orbital-projected
5-pocket model shows a new feature. Depending on the initial parameters,
the system flows at low energies either into the ``phase $A$'',
where the largest interactions are within the subset of the two $\Gamma$
hole pockets and the two $X$, $Y$ electron pockets, or into the
``phase $B$'', where the largest interactions are within the subset
of the $M$ hole pocket and the two $X$, $Y$ electron pockets. Such
a separation has been proposed earlier for LiFeAs \cite{Chubukov_Borisenko},
but for a different reason, related to the topology of the Fermi surfaces.
This separation opens up the possibility for novel $s^{+-}$ superconducting
states, such as the orbital anti-phase state \cite{Kotliar_antiphase},
in which the gap function on the $M$ hole pocket has opposite sign
with respect to the gaps on the $\Gamma$ hole pockets.

The separation between the $A$ and $B$ phases can also provide interesting
insight into the selection of magnetic order -- i.e. whether it is
stripe-like (single-\textbf{Q}) or double-\textbf{Q}. If we consider
only intra-orbital magnetism, we can generally define two magnetic
order parameters for each set (the hermitian conjugate in each expression
is left implicit for simplicity of notation): 
\begin{align}
\boldsymbol{\Delta}_{\mathrm{SDW},X}^{A}(\mathbf{k})\equiv\boldsymbol{\Delta}_{A,X} & \propto d_{yz,\mathbf{k}\alpha}^{\dagger}\boldsymbol{\sigma}_{\alpha\beta}d_{yz,\mathbf{k}+\mathbf{Q}_{X}\beta}\nonumber \\
\boldsymbol{\Delta}_{\mathrm{SDW},Y}^{A}(\mathbf{k})\equiv\boldsymbol{\Delta}_{A,Y} & \propto d_{xz,\mathbf{k}\alpha}^{\dagger}\boldsymbol{\sigma}_{\alpha\beta}d_{xz,\mathbf{k}+\mathbf{Q}_{Y}\beta}\nonumber \\
\boldsymbol{\Delta}_{\mathrm{SDW},X}^{B}(\mathbf{k})\equiv\boldsymbol{\Delta}_{B,X} & d_{xy,\mathbf{k}+\mathbf{Q}_{X}+\mathbf{Q}_{Y}\alpha}^{\dagger}\boldsymbol{\sigma}_{\alpha\beta}d_{xy,\mathbf{k}+\mathbf{Q}_{Y}\beta}\nonumber \\
\boldsymbol{\Delta}_{\mathrm{SDW},Y}^{B}(\mathbf{k})\equiv\boldsymbol{\Delta}_{B,Y} & \propto d_{xy,\mathbf{k}+\mathbf{Q}_{X}+\mathbf{Q}_{Y}\alpha}^{\dagger}\boldsymbol{\sigma}_{\alpha\beta}d_{xy,\mathbf{k}+\mathbf{Q}_{X}\beta}\label{SDW_hybrid}
\end{align}

The total free energy can then be written as 
\begin{equation}
F=F_{A}+F_{B}+F_{AB}\label{F_total_hybrid}
\end{equation}

The terms $F_{A}$ and $F_{B}$ are given by the same expression as
in Eq. (\ref{3band_Fmag}): 
\begin{align}
F_{j} & =\frac{a_{j}}{2}\left(\Delta_{j,X}^{2}+\Delta_{j,Y}^{2}\right)+\frac{u_{j}}{4}\left(\Delta_{j,X}^{2}+\Delta_{j,Y}^{2}\right)^{2}\nonumber \\
 & -\frac{g_{j}}{4}\left(\Delta_{j,X}^{2}-\Delta_{j,Y}^{2}\right)^{2}+w_{j}\left(\boldsymbol{\Delta}_{j,X}\cdot\boldsymbol{\Delta}_{j,Y}\right)^{2}\label{FA_hybrid}
\end{align}
with $j=A,B$. The sign of $g_{j}$ determines whether the ground
state is single-\textbf{Q, $g_{j}>0$} (and therefore orthorhombic),
or double-\textbf{Q}, $g_{j}<0$ (and therefore tetragonal). Expansions
near the perfect nesting limit show that $g_{A}<0$ whereas $g_{B}>0$.
That $g_{B}>0$ can be understood within the 3-band only model of
Sec. \ref{sub:Three-band-model} (see also Ref. \cite{Fanfarillo15},
which includes the orbital content of the Fermi surface). To see that
$g_{A}<0$, one has to include explicitly the matrix elements associated
with the change from orbital to band basis \cite{RMF_Chubukov_Khodas2}.
Because $g_{A}$ and $g_{B}$ have different signs, the two ``phases''
favor different magnetic states: the phase $A$ favors a double-\textbf{Q
}SDW phase and the phase $B$ favors a single-\textbf{Q }phase. A
similar observation was put forward by numerical evaluation of the
elements of the rank-4 nematic tensor in the full five-orbital model
\cite{Christensen16}.

Which of the two types of SDW order is developed by the system depends
on the strength of the biquadratic coupling in the mixed term: 
\begin{equation}
F_{AB}=\lambda\left(\Delta_{A,X}^{2}-\Delta_{A,Y}^{2}\right)\left(\Delta_{B,X}^{2}-\Delta_{B,Y}^{2}\right)+\left(\cdots\right)\label{FAB_hybrid}
\end{equation}

This term in generally renormalizes $g_{A}$ and $g_{B}$. In particular,
if nematic fluctuations arising from $B$ are strong enough, they
change the sign of $g_{A}$ and stabilize the single-\textbf{Q} phase,
even if the $M$ hole pocket rests below the Fermi level. While the
complete analysis is more involved, this simple reasoning already
reveals the key role played by the $d_{xy}$ orbitals in promoting
the experimentally observed stripe SDW phase.

\subsection{Ising-nematic order vs orbital order}

In the previous subsections we identified two possible microscopic
mechanisms for nematic order -- a spontaneous Pomeranchuk instability
for small $E_{F}/\Lambda$ and a partial melting of stripe SDW (a
spin-driven Ising-nematic order) for larger $E_{F}/\Lambda$. Although
these two scenarios may appear completely different, this is actually
not the case because both orders develop due to magnetic fluctuations.

\begin{figure}
\begin{centering}
\includegraphics[width=0.9\columnwidth]{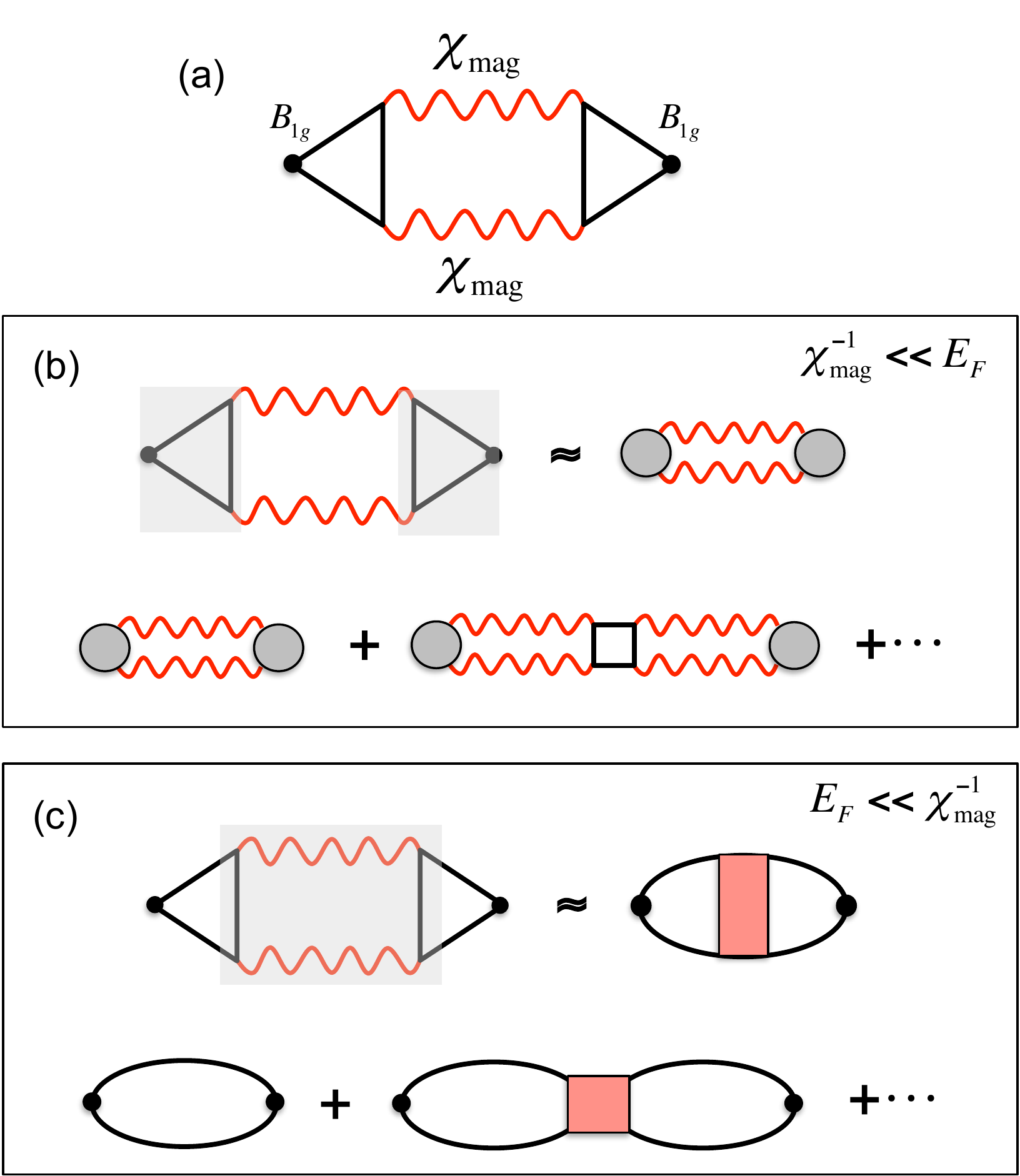} 
\par\end{centering}

\protect\protect\protect\protect\protect\protect\caption{(a) Schematic Aslamazov-Larkin diagram representing the attraction
in the Pomeranchuk channel promoted by the exchange of magnetic fluctuations.
Solid lines represent electronic propagators, wavy lines denote the
magnetic propagator, and the dots in the vertices refer to $B_{1g}$
form factors. (b) In the case where the energy scale of the magnetic
fluctuations is much smaller than the energy scale of the electronic
states, the triangular diagrams involving the electron propagators
can be replaced by an effective vertex. The nematic susceptibility
is obtained by summing the ladder series in which magnetic fluctuations
interact via square diagrams formed by higher-energy electronic propagators.
(c) In the case where the energy scale of the electronic states is
much smaller than the energy scale of the magnetic fluctuations, the
square diagram involving the two magnetic propagators can be replaced
by an effective attractive interaction. The nematic susceptibility
is obtained by summing the corresponding ladder series. \label{fig_Aslamzov_Larkin}}
\end{figure}

We illustrate this point in Fig. \ref{fig_Aslamzov_Larkin}. The fundamental
mechanism by which the exchange of magnetic fluctuations promotes
attraction in the $d$-wave Pomeranchuk channel is via the Aslamazov-Larkin
diagram of Fig. \ref{fig_Aslamzov_Larkin}a \cite{RMF10_nematic,Kontani12,RMF14,Una15}.
This is one of the diagrams that determine the RG flow of the susceptibility
in the $d-$wave Pomeranchuk channel. A ladder series of these diagrams
yields a nematic instability. The composition of the ladder series,
however, depends on how we interpret the fundamental diagram in Fig.
\ref{fig_Aslamzov_Larkin}a \cite{Hinojosa16}.

Near a magnetic instability, the energy scale associated with the
magnetic propagator (wavy lines in the diagram) is much smaller than
the energy scale associated with the electronic degrees of freedom.
In this case, the triangular diagrams in Fig. \ref{fig_Aslamzov_Larkin}a,
which involve only electronic propagators, can be replaced by a constant.
By the same reason, the electronic propagators in higher-order diagrams
can be assembled into effective interactions between low-energy magnetic
fluctuations (Fig. \ref{fig_Aslamzov_Larkin}b). An infinite ladder
series resulting from the interactions between magnetic fluctuations
can then be summed up, yielding a nematic susceptibility of the form
of Eq. (\ref{chi_nem}). When the SDW ground state is stripe-like
($g>0$ in Eq. (\ref{chi_nem})), the nematic susceptibility diverges
before the bare magnetic susceptibility. This is the mechanism in
which nematic order appears as an Ising-nematic order.

Far from a magnetic instability, however, the energy scale associated
with magnetic fluctuations can become larger than $E_{F}$. If this
is the case, then the electronic degrees of freedom should be viewed
as the lowest-energy excitations. As a result, magnetic fluctuations
can be integrated out, what in practice implies that the internal
part of the diagram in Fig. \ref{fig_Aslamzov_Larkin}a, which involves
the two magnetic propagators, can be replaced by an effective attractive
4-fermion interaction in the $d$-wave Pomeranchuk channel ( Fig.
\ref{fig_Aslamzov_Larkin}c). An infinite ladder series of such terms
then gives rise to an instability, which can be naturally identified
as the development of a spontaneous Pomeranchuk instability arising
from this effective attractive interaction. This is the mechanism
by which nematic order arises via a spontaneous orbital order.

\section{1-Fe versus 2-Fe unit cells \label{sec:1-Fe-versus-2-Fe}}

\begin{figure}
\begin{centering}
\includegraphics[width=0.9\columnwidth]{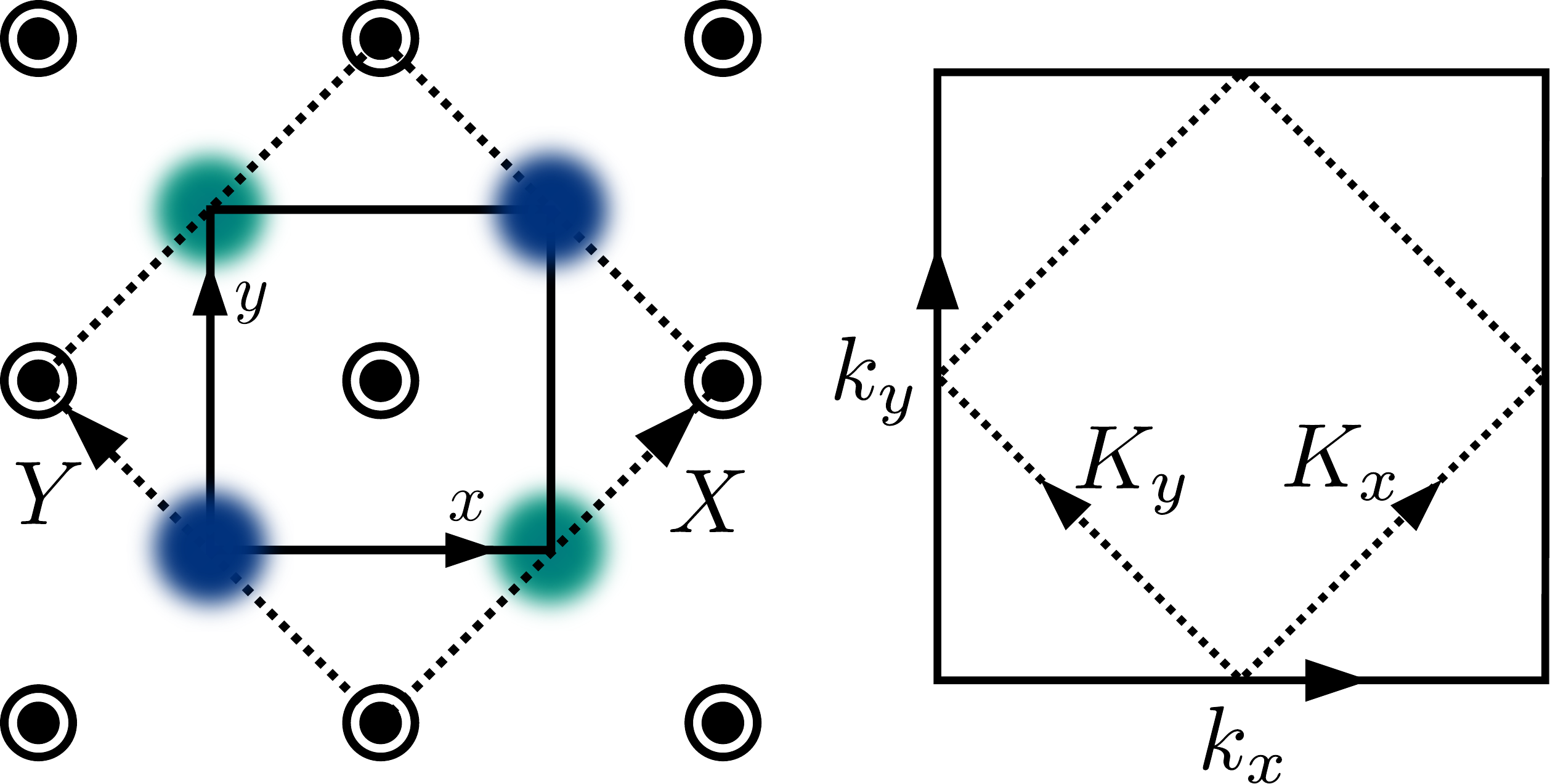} 
\par\end{centering}

\protect\protect\protect\protect\protect\protect\protect\caption{(left panel) The puckering of the As atoms above (green dots) and
below (blue dots) the plane containing the Fe atoms (black dots) increase
the size of the unit cell from 1 Fe atom (solid lines, $x,\: y$ coordinates)
to 2 Fe atoms (dashed lines, $X,\: Y$ coordinates). (right panel)
The unfolded $\left(k_{x},\, k_{y}\right)$ BZ referring to the 1-Fe
unit cell (solid lines) and the folded $\left(K_{x},\, K_{y}\right)$
BZ referring to the 2-Fe unit cell (dashed lines). Figure from Ref.
\cite{Christensen15}. \label{fig_2Fe_unit_cell}}
\end{figure}

Up to this point our analysis of the low-energy microscopic model
for the FeAs plane focused on the BZ formed by the in-plane Fe square
lattice -- the so-called 1-Fe BZ. The puckering of the As atoms, whose
positions at the center of the Fe plaquettes alternate between above
and below the Fe plane, changes the situation significantly. As we
mentioned in the Introduction, one of the effects of the As puckering
is to suppress the crystal field splittings between different orbitals
and to promote a strong hybridization between them \cite{Tesanovic09}.
More importantly, however, the existence of two inequivalent sites
for the As atoms enhances the size of the FeAs crystallographic unit
cell to that containing 2 Fe atoms, see Fig. \ref{fig_2Fe_unit_cell}
\cite{Boeri11,Eschrig09}.

The first effect of the doubling of the unit cell is that one has
to half the BZ and, consequently, fold the Fermi surface accordingly.
Let the unfolded 1-Fe BZ be described by the coordinate system $\left(k_{x},k_{y}\right)$,
and the folded 2-Fe BZ by $\left(K_{x},K_{y}\right)$. The momenta
of each zone are then related by a trivial $45^{\circ}$ rotation:

\begin{align}
K_{x} & =k_{x}-k_{y}\nonumber \\
K_{y} & =k_{x}+k_{y}\label{coordinate_change}
\end{align}
where the momentum in the unfolded zone is measured in units of the
inverse lattice constant of the 1-Fe unit cell, $1/a$, whereas the
momentum in the folded zone is measured in units of the the inverse
lattice constant of the 2-Fe unit cell, $1/\left(\sqrt{2}a\right)$.
Hereafter we denote with symbols with a bar high-symmetry points of
the folded BZ. Using Eq. (\ref{coordinate_change}), we find $\bar{M}=X=Y$
and $\bar{\Gamma}=\Gamma=M$.

\begin{figure}
\begin{centering}
\includegraphics[width=0.8\columnwidth]{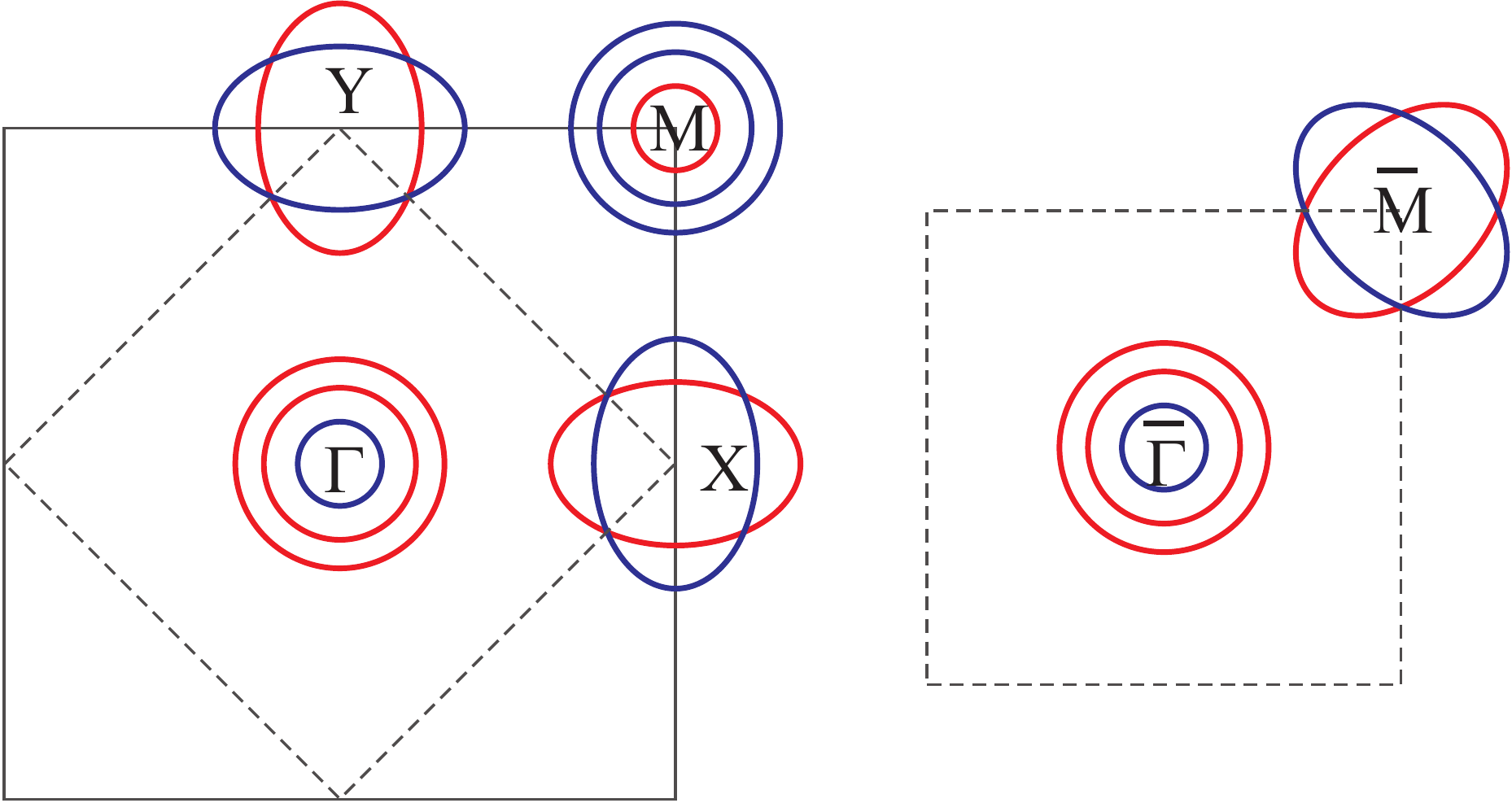} 
\par\end{centering}

\protect\protect\protect\protect\caption{Schematics for the folding of the 1-Fe BZ (solid line) onto the 2-Fe
BZ (dashed line). The red Fermi pockets correspond to the original
ones in the 1-Fe BZ, whereas the blue Fermi pockets correspond to
the original ones translated by the folding vector $\mathbf{Q}_{\mathrm{fold}}=\left(\pi,\pi\right)$.
The folded zone, rotated by $45^{\circ}$ in the right panel for better
visualization, contains both the original and translated pockets.
\label{fig_folding}}
\end{figure}

The band-structure folding resulting from the halving of the BZ is
shown schematically in Fig. \ref{fig_folding}. To obtain the folded
Fermi surface, one makes a copy of the original Fermi surface (in
red in Fig. \ref{fig_folding}) and translates it by the folding vector
$\mathbf{Q}_{\mathrm{fold}}=\left(\pi,\pi\right)$ (in blue in Fig.
\ref{fig_folding}). Besides the $45^{\circ}$ degree rotation, the
main effect of the folding is to move the two electron pockets to
$\bar{M}$ and the third hole pocket to $\bar{\Gamma}$.

\begin{figure}
\begin{centering}
\includegraphics[width=0.9\columnwidth]{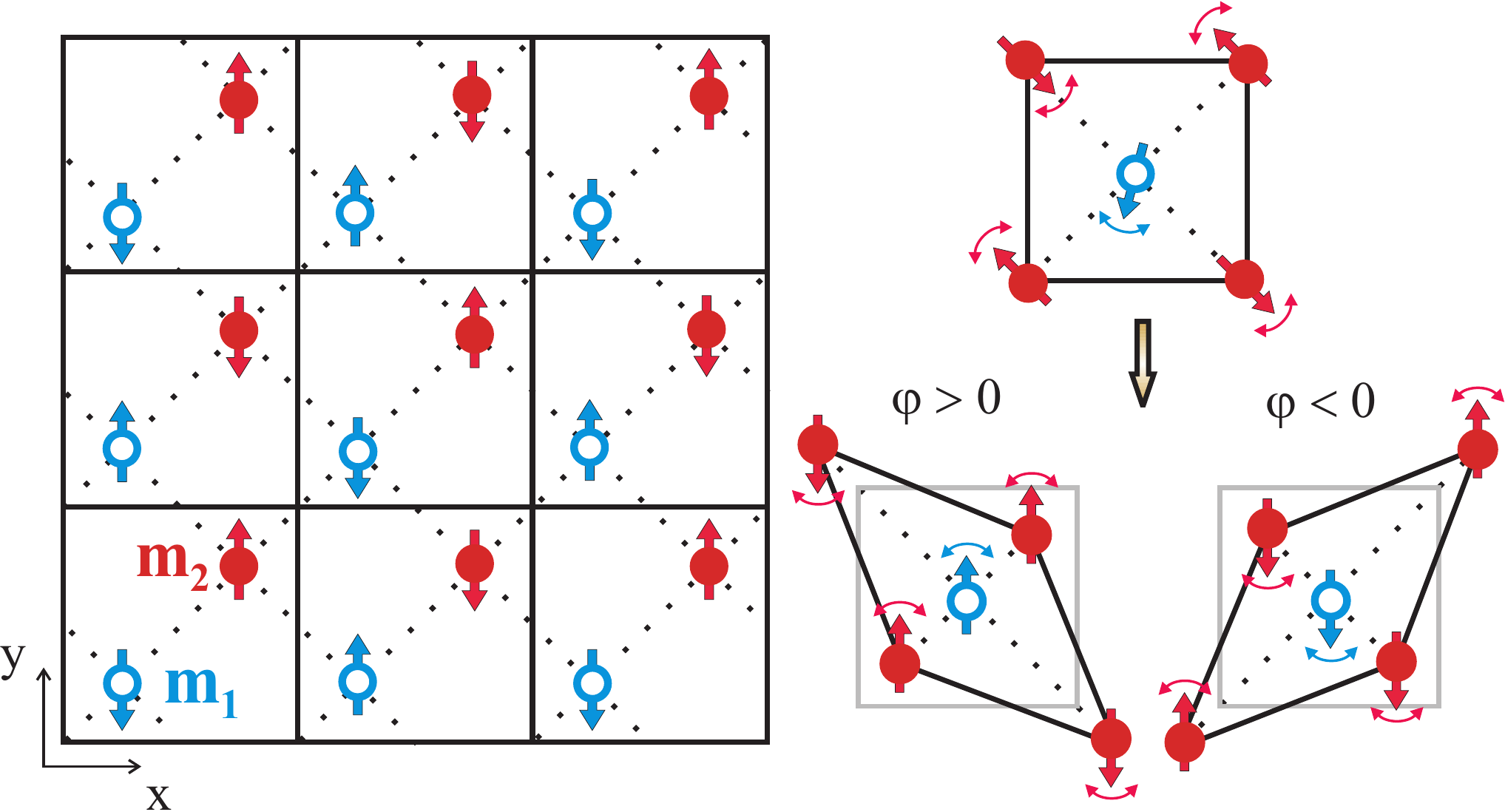} 
\par\end{centering}

\protect\protect\protect\protect\protect\protect\protect\caption{Nematic order in the 2-Fe unit cell: different signs of the nematic
order parameter $\varphi$ correspond to different relative orientations
of the spins of the two Fe atoms (red and blue) in the same unit cell.
Figure from Ref. \cite{RMF10_nematic}. \label{fig_2Fe_nematic}}
\end{figure}

Because $\bar{M}=X=Y$, another consequence of the doubling of the
unit cell is that the two magnetic ordering vectors $\mathbf{Q}_{X}=\left(\pi,0\right)$
and $\mathbf{Q}_{Y}=\left(0,\pi\right)$ in the unfolded zone are
mapped onto the same ordering vector $\mathbf{Q}_{\bar{M}}=\left(\pi,\pi\right)$
of the folded zone. Therefore, nematic order, which in the 1-Fe unit
cell is related to the competition between $\mathbf{Q}_{X}$ and $\mathbf{Q}_{Y}$
SDW orders, is more conveniently associated, in the 2-Fe unit cell,
with the relative orientation of the spins of the two Fe atoms inside
the same unit cell (see Fig. \ref{fig_2Fe_nematic}) \cite{Kivelson08,RMF10_nematic}.
Similarly, the fact that $\bar{\Gamma}=\Gamma=M$ implies that any
instability involving $\mathbf{Q}_{M}=\left(\pi,\pi\right)$ ordering
in the unfolded zone becomes an intra-unit cell order, without additional
translational symmetry breaking. As a result, Neel-type SDW order
becomes more difficult to be observed experimentally since the ordering
vector coincides with a lattice Bragg peak.

The band folding is also accompanied by important effects that affect
both the electronic dispersion as well as the instabilities of the
system. These effects arise from terms in the Hamiltonian that couple
electronic states separated by momentum $\mathbf{Q}_{\mathrm{fold}}=\left(\pi,\pi\right)$.
In the non-interacting level, two terms in the Hamiltonian become
particularly important in the 2-Fe folded zone: the first one corresponds
to the hybridization between states at the $X$ and $Y$ pockets:

\begin{equation}
\mathcal{H}_{\mathrm{hyb}}=\sum_{\mathbf{k}}f_{\mathrm{hyb}}\left(\mathbf{k}\right)c_{e_{X},\mathbf{k}+\mathbf{Q}_{X}\sigma}^{\dagger}c_{e_{Y},\mathbf{k}+\mathbf{Q}_{Y}\sigma}+h.c.\label{hyb}
\end{equation}

As shown in Refs. \cite{Khodas_Chubukov1,Khodas_Chubukov2}, this
term arises from the hybridization between Fe $3d$ states and As
$2p$ states. The momentum dependence of $f_{\mathrm{hyb}}\left(\mathbf{k}\right)$
is a consequence of the orbital content of the Fermi surface, and
vanishes along the diagonals of the folded BZ. The $(\pi,\pi)$ terms
also appear in the interacting part of the Hamiltonian.

The second non-interacting term corresponds to the atomic spin-orbit
coupling (SOC), which connects states at the $X$ and $Y$ pockets
according to $\mathcal{H}_{\mathrm{SOC}}=\lambda\mathbf{L}\cdot\mathbf{S}$.
In terms of the orbital operators, it corresponds to \cite{Vafek13}

\begin{align}
\mathcal{H}_{\mathrm{SOC}} & =\frac{i}{2}\lambda\sum_{\mathbf{k}}d_{xz,\mathbf{k}+\mathbf{Q}_{Y}\alpha}^{\dagger}\sigma_{\alpha\beta}^{x}d_{xy,\mathbf{k}+\mathbf{Q}_{X}\beta}+h.c.\nonumber \\
 & +\frac{i}{2}\lambda\sum_{\mathbf{k}}d_{xy,\mathbf{k}+\mathbf{Q}_{Y}\alpha}^{\dagger}\sigma_{\alpha\beta}^{y}d_{yz,\mathbf{k}+\mathbf{Q}_{X}}\beta+h.c.\label{SOC}
\end{align}

In contrast to the hybridization term in Eq. (\ref{hyb}), the SOC
splits the folded electron pockets into two separate electron pockets
-- an inner one, of mostly $d_{xz}$ and $d_{yz}$ character, and
an outer one, of mostly $d_{xy}$ character. Besides these two non-interacting
terms, interactions involving momentum transfer $\mathbf{Q}_{\mathrm{fold}}=\left(\pi,\pi\right)$
also couple the states at the $X$ and $Y$ pockets.

Below, we discuss how the models presented in the previous sections
need to be modified to account for the doubling of the Fe unit cell.

\subsection{Orbital-basis models}

We start with the models defined in the orbital basis only (Section
\ref{sec:Orbital-basis-models}): in the 2-Fe BZ, one has to consider
ten Fe $3d$ orbitals (assuming that the six As $2p$ orbitals can
be integrated out). The general structure of the non-interacting Hamiltonian,
in the folded zone, can be expressed by introducing the operator \cite{Eschrig09}:

\begin{equation}
\phi_{\mathbf{K}}=\left(\begin{array}{c}
\phi_{1,\mathbf{K}}\\
\phi_{2,\mathbf{K}}
\end{array}\right)\label{phi_operators}
\end{equation}
where $\phi_{i,\mathbf{K}}$ is a 5-component operator consisting
of the orbital-basis operators $d_{j,\mathbf{k}\sigma}^{(i)}$, with
$j=xz,\, yz,\, x^{2}-y^{2},\, xy,\, z^{2}$ (the orbitals remain labeled
with respect to the coordinate system of the 1-Fe BZ). In this notation,
the non-interacting Hamiltonian assumes the form:

\begin{equation}
\mathcal{H}_{0}=\sum_{\mathbf{k}}\phi_{\mathbf{K}}^{\dagger}\left(\begin{array}{cc}
\hat{H}_{11}\left(\mathbf{K}\right) & \hat{H}_{12}\left(\mathbf{K}\right)\\
\hat{H}_{12}^{*}\left(\mathbf{K}\right) & \hat{H}_{11}^{*}\left(\mathbf{K}\right)
\end{array}\right)\phi_{\mathbf{K}}\label{H0_2Fe}
\end{equation}
where $\hat{H}_{i_{1}i_{2}}$ are $5\times5$ matrices. We refrain
here from giving the full expressions for these tight-binding dispersions,
which can be found in Ref. \cite{Eschrig09}.

If terms that couple $\phi_{1,\mathbf{K}}$ and $\phi_{2,\mathbf{K}}$
are present, like those in Eqs. (\ref{hyb}) and (\ref{SOC}), then
one has no choice but to work with the full ten-orbital model. However,
if these specific interactions are absent, it is possible to ``unfold''
the 2-Fe BZ using a glide-plane symmetry of the FeAs plane. Indeed,
the space group of a single FeAs plane is the non-symmorphic $P4/nmm$
group, which contains a glide-plane symmetry corresponding to a translation
by $T=\left(\frac{1}{2},\,\frac{1}{2}\right)$ in the 2-Fe unit cell
followed by a reflection $\sigma_{z}$ with respect the $xy$ plane.
Inspection of Fig. \ref{fig_2Fe_unit_cell} shows that indeed under
this sequence of operations the lattice is mapped back onto itself.

The key point is that the $d_{x^{2}-y^{2}}$, $d_{xy}$, and $d_{z^{2}}$
orbitals are even under the reflection $\sigma_{z}$, while the orbitals
$d_{xz}$ and $d_{yz}$ are odd. Consequently, because the two Fe
sites in the same unit cell are related by a $T=\left(\frac{1}{2},\,\frac{1}{2}\right)$
translation, the $d_{xz}$ and $d_{yz}$ orbitals change sign from
one of these Fe sites to the other. As a result, one can use the eigenvalues
of the operator $T\sigma_{z}$ to diagonalize the Hamiltonian and
express the electronic states in terms of a pseudocrystal momentum
$\tilde{\mathbf{k}}$. The orbital states $\tilde{d}_{\mu,\tilde{\mathbf{k}}\sigma}$
with pseudocrystal momentum $\tilde{\mathbf{k}}$ are related to the
orbital states $d_{\mu,\mathbf{k}\sigma}$ with momentum $\mathbf{k}$
in the unfolded BZ according to \cite{Lee_Wen08}:

\begin{equation}
\tilde{d}_{\mu,\tilde{\mathbf{k}}\sigma}=\left\{ \begin{array}{ccc}
d_{\mu,\mathbf{k}\sigma} & , & \mu\:\mathrm{even}\\
d_{\mu,\mathbf{k}+\mathbf{Q}_{\mathrm{fold}}\sigma} & , & \mu\:\mathrm{odd}
\end{array}\right.\label{folding}
\end{equation}
where $\mathbf{Q}_{\mathrm{fold}}=\left(\pi,\pi\right)$. Therefore,
most of the results obtained in the studies of the orbital models
defined in the unfolded zone can be directly translated to results
in the actual crystallographic zone by means of the pseudocrystal
momentum. Such a procedure has been implemented in different works
\cite{Lee_Wen08,WKu_2Fe,JPHu_antiphase,Valenti_2Fe,WKu_antiphase,Maier_2Fe},
highlighting the importance of the glide-plane symmetry in the properties
of the electronic spectrum (particularly the spectral weight of the
electron pockets observed by ARPES) and of the SC state (such as the
role of the so-called $\eta$-pairing).

We emphasize that this analysis is restricted to a single FeAs plane.
The real materials, however, consist of many coupled layers. In the
materials whose unit cells contain a single FeAs plane, such as the
1111 (e.g. LaFeAsO), the 111 (e.g. NaFeAs), and the 11 (e.g. FeSe)
compounds, the stacking of the FeAs planes is such that the three-dimensional
crystallographic unit cell retains the $P4/nmm$ space group. As a
result, even after including the $k_{z}$ dispersion, this approach
to describe the tight-binding dispersions in the full BZ remains essentially
the same \cite{Vafek13}. The situation is however different in the
122 (e.g. BaFe$_{2}$As$_{2}$) compounds, because their unit cell
becomes body-centered tetragonal, instead of simple tetragonal. As
a result, the space group of the crystallographic unit cell is $I4/mmm$,
which is symmorphic. In this case, the ``folding vector'' changes
from $\mathbf{Q}_{\mathrm{fold}}=\left(\pi,\pi,0\right)$ to $\mathbf{Q}_{\mathrm{fold}}=\left(\pi,\pi,\pi\right)$,
which has important consequences for the $k_{z}$ dispersion of the
different Fermi pockets \cite{Inosov10}. The effect of the inter-layer
coupling to the properties of the FeSC is important \cite{Graser10},
but is beyond the scope of this review in which we consider only the
case of a single FeAs layer effectively uncoupled from the other layers.

\subsection{Orbital-projected band models}

One of the advantages of the orbital-projected band models of Section
\ref{sec:Hybrid-band-orbital-models} is that they can be generalized
in a straightforward way to the 2-Fe BZ, without having to include
additional electronic states. This is in contrast to the orbital-basis
models, in which the number of orbitals double when going from the
1-Fe unit cell to the 2-Fe unit cell.

The reason for this behavior stems from the properties of the $P4/nmm$
space group describing the single FeAs plane. As discussed in details
in Ref. \cite{Vafek13}, the non-symmorphic nature of this group implies
that, while the irreducible representations at the $\bar{\Gamma}$
point are essentially the same as those of the standard $D_{4h}$
group, the irreducible representations at the $\bar{M}$ point must
all be two-dimensional. As a result, all electronic states at the
$\bar{M}$ point must be doubly-degenerate and form doublets, and
the electronic instabilities must be classified according to these
irreducible representations (for details, see Ref. \cite{Vafek13}).

Physically, this double-degeneracy at the $\bar{M}$ point is manifested
in the tight-binding dispersions of the 1-Fe BZ by the fact that $\epsilon_{xx}\left(\mathbf{Q}_{Y}\right)=\epsilon_{yy}\left(\mathbf{Q}_{X}\right)$
and $\epsilon_{xy}\left(\mathbf{Q}_{Y}\right)=\epsilon_{xy}\left(\mathbf{Q}_{X}\right)$.
These doublets can be expressed as spinors $\psi_{\bar{M}_{1}}$ and
$\psi_{\bar{M}_{3}}$ (following the notation of Ref. \cite{Vafek13})
formed by combinations of the spinors $\psi_{X}$ and $\psi_{Y}$
defined in Subsection \ref{sub:Three-orbitals,-five-band-model}:

\begin{align}
\psi_{\bar{M}_{1},\mathbf{k}+\mathbf{Q}_{\bar{M}}} & =\left(\begin{array}{c}
c_{xz,\mathbf{k}+\mathbf{Q}_{2}\sigma}\\
c_{yz,\mathbf{k}+\mathbf{Q}_{1}\sigma}
\end{array}\right)\nonumber \\
\psi_{\bar{M}_{3},\mathbf{k}+\mathbf{Q}_{\bar{M}}} & =\left(\begin{array}{c}
c_{xy,\mathbf{k}+\mathbf{Q}_{2}\sigma}\\
c_{xy,\mathbf{k}+\mathbf{Q}_{1}\sigma}
\end{array}\right)\label{doublets}
\end{align}

\begin{figure}
\begin{centering}
\includegraphics[width=0.9\columnwidth]{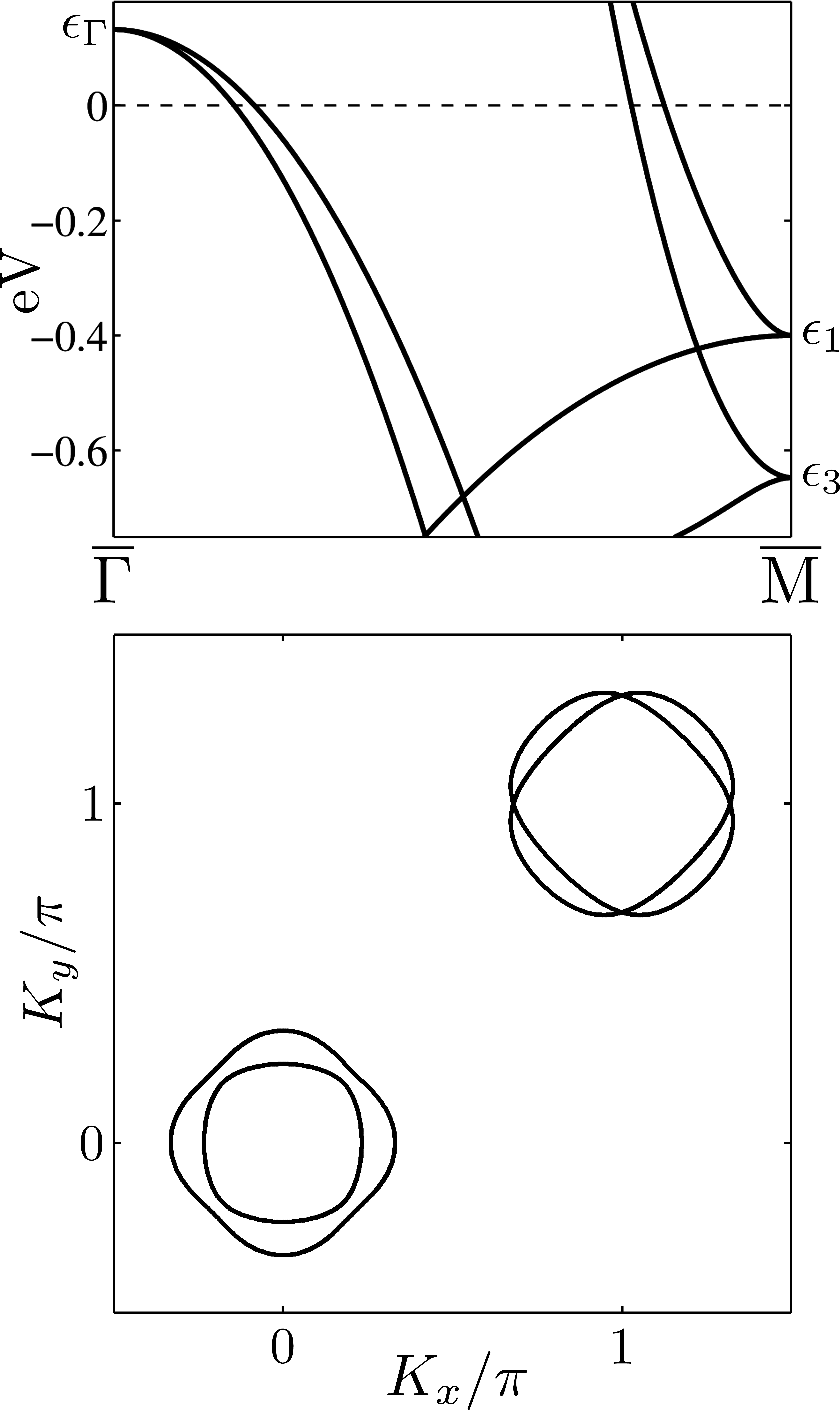} 
\par\end{centering}

\protect\protect\protect\protect\protect\protect\protect\caption{Band dispersion (upper panel) and Fermi surface (lower panel) of the
orbital-projected band model in the folded BZ associated with the
2-Fe unit cell. Figure from Ref. \cite{Christensen15}. \label{fig_2Fe_bands}}
\end{figure}

Note, however, that the block-diagonal non-interacting Hamiltonian
in Eq. (\ref{aux_H0_final}) remains unchanged. To obtain the band
structure and Fermi surfaces in the folded zone, one only needs to
change the coordinates according to Eq. (\ref{coordinate_change}).
Fig. \ref{fig_2Fe_bands} presents both the band dispersions and the
Fermi pockets for this model in the folded zone. The meaning of the
parameters $\epsilon_{1}$ and $\epsilon_{3}$ in Eqs. (\ref{H0_X})
and (\ref{H0_Y}) is now evident: they are nothing but the energies
of the two doublets at the $\bar{M}$ point. Interestingly, these
orbital-projected band models have generally three doublets: two of
them arising from the $\bar{M}_{1}$ and $\bar{M}_{3}$ two-dimensional
irreducible representations at the $\bar{M}$ point and one arising
from the $E_{g}$ two-dimensional irreducible representation at the
$\bar{\Gamma}$ point. These three doublets form the two $\Gamma$
hole pockets and the two $X$, $Y$ electron pockets in the unfolded
zone. On the other hand, the additional hole pocket at the $M$ point
of the unfolded zone does not form a doublet, as it belongs to the
one-dimensional $B_{1g}$ irreducible representation at the $\bar{\Gamma}$
point.

The advantages offered by the orbital-projected band model when dealing
with the 2-Fe BZ become even more clear when one considers the effect
of the spin-orbit coupling (SOC). As we discussed above, the pseudocrystal
approach of the orbital-basis models works well as long as the glide-plane
symmetry is kept intact, i.e. when there are no terms coupling states
of the two different Fe sites of the unit cell. However, the atomic-like
SOC alters this scenario, as it couples the $d_{xy}$ states of one
Fe site with the $d_{xz/yz}$ states of the other Fe site of the unit
cell via the $\sigma_{x}$ and $\sigma_{y}$ spin operators, see Eq.
(\ref{SOC}) above.

To account for SOC in the orbital-basis model, one has to work with
$10\times10$ matrices. On the other hand, in the orbital-projected
band model, the SOC introduces off-diagonal terms into the non-interacting
Hamiltonian (\ref{aux_H0_final}) without increasing the number of
low-energy degrees of freedom. In particular, one finds \cite{Vafek13,RMF_Vafek}:
\begin{equation}
\mathcal{H}_{\mathrm{SOC}}=\sum_{\mbf{k}}\Psi_{\mathbf{k}}^{\dagger}\hat{H}_{\mathrm{SOC}}(\mbf{k})\Psi_{\mathbf{k}}\,,\label{H0_hyrbid_final}
\end{equation}
with: 
\begin{equation}
\hat{H}_{\mathrm{SOC}}(\mbf{k})=\begin{pmatrix}0 & h_{M}^{\text{SOC}}(\mbf{k}) & 0\\
\left(h_{M}^{\text{SOC}}(\mbf{k})\right)^{\dagger} & 0 & 0\\
0 & 0 & h_{\Gamma}^{\text{SOC}}(\mbf{k})
\end{pmatrix}\label{H_SOC_hybrid}
\end{equation}
and $4\times4$ matrices:

\begin{align}
h_{\Gamma}^{\text{SOC}}(\mbf{k}) & =\frac{1}{2}\lambda\left(\tau^{y}\otimes\sigma^{z}\right)\label{aux_H_SOC_hybrid}\\
h_{M}^{\text{SOC}}(\mbf{k}) & =\frac{i}{2}\lambda\left(\tau^{+}\otimes\sigma^{x}+\tau^{-}\otimes\sigma^{y}\right)\nonumber 
\end{align}

Here, $\tau^{\pm}=\frac{1}{2}\left(\tau^{x}\pm i\tau^{y}\right)$
and the Pauli matrices $\sigma$ refer to spin space, whereas $\tau$
refer to spinor space. The SOC has very important consequences for
the electronic properties of the FeSC. While it splits the degeneracy
between the $d_{xz}$ and $d_{yz}$ orbitals at the $\bar{\Gamma}$
point, it preserves the doublets at the $\bar{M}$ point. This feature
allows one to distinguish signatures of nematic order and SOC in the
ARPES spectrum of the FeSC \cite{RMF_Vafek}. Note in this regard
that the typical SOC observed experimentally is $\lambda\sim10$ meV
\cite{Borisenko16}, which is roughly of the same order as the band
splittings due to SDW, SC, and orbital order. Thus, a consistent description
of the normal state of the FeSC must account for the SOC.

The classification of the pairing states also change, as components
identified with singlet and triplet pairing mix (although the Kramers
degeneracy of the electronic states is kept intact by SOC) \cite{Vafek13}.
Finally, the SOC causes a spin anisotropy, which selects different
magnetization directions for the different types of SDW order \cite{Christensen15}.

\section{Concluding remarks \label{sec:Concluding-remarks}}

In this work we reviewed the hierarchy of potential instabilities
in FeSC by analyzing different low-energy models. We focused primarily
on the interplay between superconductivity, SDW order, $\mathbf{Q}=0$
charge Pomeranchuk order (often associated with orbital order), and
Ising-nematic spin order. The last two orders break $C_{4}$ symmetry
and lead to the phase dubbed nematic. We considered three sets of
models: (i) Purely orbital models, in which all computations are performed
within the orbital basis without separation into contributions from
low-energy and high-energy sectors. (ii) Band models, in which the
instabilities are viewed as coming from states near the Fermi surface,
but the orbital composition of the Fermi surfaces is neglected. (iii)
Orbital-projected band models, in which the analysis is restricted
to low energies, but the orbital composition of the Fermi pockets
is fully embraced. In our view, the last class of models are the most
promising ones due to their simplicity and due to the separation between
high-energy and low-energy states.

The orbital-projected band models involve three orbitals ($d_{xz},d_{yz}$,
and $d_{xy}$) from which the low-energy excitations are constructed.
The interactions between low-energy states contain angle-dependent
prefactors that reflect the orbital composition of the Fermi surfaces.
The full five-pocket orbital-projected model is rather involved and
contains 40 distinct coupling constants. The analysis involving the
RG technique, however, yields similar results in different approximated
orbital-projected band models. Namely, at intermediate energies, magnetic
fluctuations are the strongest. These fluctuations give rise to attractive
interactions in $s^{+-}$ and $d$-wave superconducting channels,
as well as in $s^{+-}$ and $d$-wave Pomeranchuk channels. Once interactions
in these two channels become attractive, SC fluctuations compete with
magnetic fluctuations and eventually win over them, while Pomeranchuk
fluctuations develop with little competition with SDW. The final outcome,
i.e. which order develops first, depends on the details of the electronic
dispersion. For certain system parameters, the leading symmetry-breaking
instability is in the $\mathbf{Q}=0$ $d$-wave Pomeranchuk channel,
which gives rise to spontaneous orbital order, the subleading instability
is in the SC channel, and SDW order does not develop. For other system
parameters, however, the leading instability is either SDW or superconductivity,
while spontaneous orbital order does not develop. In this last case,
the nematic order is a vestigial order of the stripe SDW state.

We also discussed the description of the physics in the 1-Fe and 2-Fe
BZ, and the importance of the sizable spin-orbit coupling, which significantly
affects the normal state and superconducting state properties. We
argued that the orbital-projected models are very convenient to study
the problem in the crystallographic 2Fe BZ, as they do not require
the inclusion of additional electronic degrees of freedom. This is
in contrast to orbital-basis models, in which the number of electronic
degrees of freedom doubles.

We believe that the approach we reviewed in this paper is a promising
framework to obtain a unified description of different Fe-based superconductors. 
\begin{acknowledgments}
We thank B. Andersen, E. Bascones, L. Benfatto, E. Berg, L. Classen,
M. Christensen, E. Dagotto, I. Eremin, L. Fanfarillo, M. Gastiasoro,
P. Hirschfeld, C. Honerkamp, J. Kang, S. Kivelson, M. Khodas, H. Kontani,
G. Kotliar, S. Maiti, I. Mazin, A. Millis, A. Moreo, I. Paul, R. Thomale,
J. Schmalian, M. Schuett, O. Vafek, R. Valenti, B. Valenzuela, R.
Xing, X. Wang, and Y. Wang for useful discussions. We would like to
give special thanks to M. Christensen and J. Kang for assistance in
producing some of the figures in this review. This work was supported
by the Office of Basic Energy Sciences, U.S. Department of Energy,
under awards DE-SC0014402 (AVC) and DE-SC0012336 (RMF). 
\end{acknowledgments}

\appendix

\section{Band dispersion parameters}
\begin{widetext}
Here we explicitly present band dispersion parameters for selected
models discussed in the main text.

\subsection{Five-orbital model}

We use the same notation of the Graser \emph{et al }\cite{Graser09}\emph{.}
Note that in Fig. \ref{fig_5orbital_band} of the main text, we used
the parameters of the model of Ikeda \emph{et al}., which contains
many more neighbor hoppings \cite{Ikeda10}. The tight binding parametrization
is given by:

\begin{align}
\epsilon_{xz,xz}\left(\mathbf{k}\right) & =\epsilon_{xz}^{(0)}+2t_{x}^{11}\cos k_{x}+2t_{y}^{11}\cos k_{y}+4t_{xy}^{11}\cos k_{x}\cos k_{y}+2t_{xx}^{11}\left(\cos2k_{x}-\cos2k_{y}\right)\nonumber \\
 & +4t_{xxy}^{11}\cos2k_{x}\cos k_{y}+4t_{xyy}^{11}\cos k_{x}\cos2k_{y}+4t_{xxyy}^{11}\cos2k_{x}\cos2k_{y}\,,\nonumber \\
\epsilon_{yz,yz}\left(\mathbf{k}\right) & =\epsilon_{yz}^{(0)}+2t_{y}^{11}\cos k_{x}+2t_{x}^{11}\cos k_{y}+4t_{xy}^{11}\cos k_{x}\cos k_{y}-2t_{xx}^{11}\left(\cos2k_{x}-\cos2k_{y}\right)\nonumber \\
 & +4t_{xyy}^{11}\cos2k_{x}\cos k_{y}+4t_{xxy}^{11}\cos k_{x}\cos2k_{y}+4t_{xxyy}^{11}\cos2k_{x}\cos2k_{y}\,,\nonumber \\
\epsilon_{x^{2}-y^{2},x^{2}-y^{2}}\left(\mathbf{k}\right) & =\epsilon_{x^{2}-y^{2}}^{(0)}+2t_{x}^{33}\left(\cos k_{x}+\cos k_{y}\right)+4t_{xy}^{33}\cos k_{x}\cos k_{y}+2t_{xx}^{33}\left(\cos2k_{x}+\cos2k_{y}\right)\,,\nonumber \\
\epsilon_{xy,xy}\left(\mathbf{k}\right) & =\epsilon_{xy}^{(0)}+2t_{x}^{44}\left(\cos k_{x}+\cos k_{y}\right)+4t_{xy}^{44}\cos k_{x}\cos k_{y}+2t_{xx}^{44}\left(\cos2k_{x}+\cos2k_{y}\right)\nonumber \\
 & +4t_{xxy}^{44}\left(\cos2k_{x}\cos k_{y}+\cos k_{x}\cos2k_{y}\right)+4t_{xxyy}^{44}\cos2k_{x}\cos2k_{y}\,,\nonumber \\
\epsilon_{z^{2},z^{2}}\left(\mathbf{k}\right) & =\epsilon_{z^{2}}^{(0)}+2t_{x}^{55}\left(\cos k_{x}+\cos k_{y}\right)+2t_{xx}^{55}\left(\cos2k_{x}\cos2k_{y}\right)\nonumber \\
 & +4t_{xxy}^{55}\left(\cos2k_{x}\cos k_{y}+\cos k_{x}\cos2k_{y}\right)+4t_{xxyy}^{55}\cos2k_{x}\cos2k_{y}\,,\nonumber \\
\epsilon_{xz,yz}\left(\mathbf{k}\right) & =-4t_{xy}^{12}\sin k_{x}\sin k_{y}-4t_{xxy}^{12}\left(\sin2k_{x}\sin k_{y}+\sin k_{x}\sin2k_{y}\right)-4t_{xxyy}^{12}\sin2k_{x}\sin2k_{y}\,,\nonumber \\
\epsilon_{xz,x^{2}-y^{2}}\left(\mathbf{k}\right) & =i2t_{x}^{13}\sin k_{y}+i4t_{xy}^{13}\cos k_{x}\sin k_{y}-i4t_{xxy}^{13}\left(\cos k_{x}\sin2k_{y}-\cos2k_{x}\sin k_{y}\right)\,,\nonumber \\
\epsilon_{xz,xy}\left(\mathbf{k}\right) & =i2t_{x}^{14}\sin k_{x}+i4t_{xy}^{14}\sin k_{x}\cos k_{y}+i4t_{xxy}^{14}\sin2k_{x}\cos k_{y}\,,\nonumber \\
\epsilon_{xz,z^{2}}\left(\mathbf{k}\right) & =i2t_{x}^{15}\sin k_{y}-i4t_{xy}^{15}\cos k_{x}\sin k_{y}-i4t_{xxyy}^{15}\cos2k_{x}\sin2k_{y}\,,\nonumber \\
\epsilon_{yz,x^{2}-y^{2}}\left(\mathbf{k}\right) & =-i2t_{x}^{13}\sin k_{x}-i4t_{xy}^{13}\sin k_{x}\cos k_{y}+i4t_{xxy}^{13}\left(\sin2k_{x}\cos k_{y}-\sin k_{x}\cos2k_{y}\right)\,,\nonumber \\
\epsilon_{yz,xy}\left(\mathbf{k}\right) & =i2t_{x}^{14}\sin k_{y}+i4t_{xy}^{14}\cos k_{x}\sin k_{y}+i4t_{xxy}^{14}\cos k_{x}\sin2k_{y}\,,\nonumber \\
\epsilon_{yz,z^{2}}\left(\mathbf{k}\right) & =i2t_{x}^{15}\sin k_{x}-i4t_{xy}^{15}\sin k_{x}\cos k_{y}-i4t_{xxyy}^{15}\sin2k_{x}\cos2k_{y}\,,\nonumber \\
\epsilon_{x^{2}-y^{2},xy}\left(\mathbf{k}\right) & =4t_{xxy}^{34}\left(\sin k_{x}\sin2k_{y}-\sin2k_{x}\sin k_{y}\right)\,,\nonumber \\
\epsilon_{x^{2}-y^{2},z^{2}}\left(\mathbf{k}\right) & =2t_{x}^{35}\left(\cos k_{x}-\cos k_{y}\right)+4t_{xxy}^{35}\left(\cos2k_{x}\cos k_{y}-\cos k_{x}\cos2k_{y}\right)\,,\nonumber \\
\epsilon_{xy,z^{2}}\left(\mathbf{k}\right) & =4t_{xy}^{45}\sin k_{x}\sin k_{y}+4t_{xxyy}^{45}\sin2k_{x}\sin2k_{y}\label{S_matrix_5orb}
\end{align}

The tight-binding hopping parameters from Graser \emph{et al.} are
given in Table \ref{tab_5orb}. For an occupation number of $6$,
the onsite energies are given by: $\epsilon_{xz}^{(0)}=\epsilon_{yz}^{(0)}=130$
meV, $\epsilon_{x^{2}-y^{2}}^{(0)}=-220$ meV, $\epsilon_{xy}^{(0)}=300$
meV, and $\epsilon_{z^{2}}^{(0)}=-211$ meV.

\begin{table}[h]
\begin{centering}
\begin{tabular}{|l|c|c|c|c|c|c|c|}
\hline 
$t_{\alpha}^{\mu\nu}$  & $\alpha=x$  & $\alpha=y$  & $\alpha=xy$  & $\alpha=xx$  & $\alpha=xxy$  & $\alpha=xyy$  & $\alpha=xxyy$\tabularnewline
\hline 
\hline 
$\left(\mu,\nu\right)=\left(xz,xz\right)$  & $-140$  & $-400$  & $280$  & $20$  & $-35$  & $5$  & $35$\tabularnewline
\hline 
$\left(\mu,\nu\right)=\left(x^{2}-y^{2},x^{2}-y^{2}\right)$  & $350$  & $\vartimes$  & $-105$  & $-20$  & $\vartimes$  & $\vartimes$  & $\vartimes$\tabularnewline
\hline 
$\left(\mu,\nu\right)=\left(xy,xy\right)$  & $\vartimes$  & $\vartimes$  & $150$  & $-30$  & $-30$  & $\vartimes$  & $-30$\tabularnewline
\hline 
$\left(\mu,\nu\right)=\left(z^{2},z^{2}\right)$  & $\vartimes$  & $\vartimes$  & $\vartimes$  & $-40$  & $20$  & $\vartimes$  & $-10$\tabularnewline
\hline 
$\left(\mu,\nu\right)=\left(xz,yz\right)$  & $\vartimes$  & $\vartimes$  & $50$  & $\vartimes$  & $-15$  & $\vartimes$  & $35$\tabularnewline
\hline 
$\left(\mu,\nu\right)=\left(xz,x^{2}-y^{2}\right)$  & $-354$  & $\vartimes$  & $99$  & $\vartimes$  & $21$  & $\vartimes$  & $\vartimes$\tabularnewline
\hline 
$\left(\mu,\nu\right)=\left(xz,xy\right)$  & $339$  & $\vartimes$  & $14$  & $\vartimes$  & $28$  & $\vartimes$  & $\vartimes$\tabularnewline
\hline 
$\left(\mu,\nu\right)=\left(xz,z^{2}\right)$  & $-198$  & $\vartimes$  & $-85$  & $\vartimes$  & $\vartimes$  & $\vartimes$  & $-14$\tabularnewline
\hline 
$\left(\mu,\nu\right)=\left(x^{2}-y^{2},xy\right)$  & $\vartimes$  & $\vartimes$  & $\vartimes$  & $\vartimes$  & $-10$  & $\vartimes$  & $\vartimes$\tabularnewline
\hline 
$\left(\mu,\nu\right)=\left(x^{2}-y^{2},z^{2}\right)$  & $-300$  & $\vartimes$  & $\vartimes$  & $\vartimes$  & $-20$  & $\vartimes$  & $\vartimes$\tabularnewline
\hline 
$\left(\mu,\nu\right)=\left(xy,z^{2}\right)$  & $\vartimes$  & $\vartimes$  & $-150$  & $\vartimes$  & $\vartimes$  & $\vartimes$  & $10$\tabularnewline
\hline 
\end{tabular}
\par\end{centering}

\protect\protect\protect\protect\protect\protect\caption{Tight-binding hopping parameters (in meV) for the 5-orbital of Eq.
(\ref{S_matrix_5orb}).\label{tab_5orb}}
\end{table}

\subsection{Two-orbital model}

The band dispersion in the two orbital model by Raghu \emph{et al.}
is \cite{Raghu08}:

\begin{align}
\epsilon_{xx}\left(\mathbf{k}\right) & =-2t_{1}\cos k_{x}-2t_{2}\cos k_{y}-4t_{3}\cos k_{x}\cos k_{y}\nonumber \\
\epsilon_{yy}\left(\mathbf{k}\right) & =-2t_{2}\cos k_{x}-2t_{1}\cos k_{y}-4t_{3}\cos k_{x}\cos k_{y}\nonumber \\
\epsilon_{xy}\left(\mathbf{k}\right) & =-4t_{4}\sin k_{x}\sin k_{y}\label{S_aux_2_orbitals}
\end{align}

The tight-binding parameters used in Fig. \ref{fig_2orbital} are
taken from Ref. \cite{Sknepnek09} and shown in Table \ref{tab_2orb}.
For an occupation number of $2$, the chemical potential is $\mu=550$
meV.

\begin{table}[h]
\begin{centering}
\begin{tabular}{|c|c|c|c|}
\hline 
$t_{1}$  & $t_{2}$  & $t_{3}$  & $t_{4}$\tabularnewline
\hline 
\hline 
$-330$  & $385$  & $-234$  & $-260$\tabularnewline
\hline 
\end{tabular}
\par\end{centering}

\protect\protect\protect\protect\protect\protect\caption{Tight-binding hopping parameters (in meV) for the 2-orbital model
of Eq. (\ref{S_aux_2_orbitals}). \label{tab_2orb}}
\end{table}

\subsection{Three-orbital model}

The band dispersion in the three orbital model by Daghofer\emph{ et
al.} is \cite{Daghofer10}:

\begin{align}
\epsilon_{xz,xz}\left(\mathbf{k}\right) & =-2t_{1}\cos k_{x}-2t_{2}\cos k_{y}-4t_{3}\cos k_{x}\cos k_{y}\nonumber \\
\epsilon_{yz,yz}\left(\mathbf{k}\right) & =-2t_{2}\cos k_{x}-2t_{1}\cos k_{y}-4t_{3}\cos k_{x}\cos k_{y}\nonumber \\
\epsilon_{xy,xy}\left(\mathbf{k}\right) & =-2t_{5}\left(\cos k_{x}+\cos k_{y}\right)-4t_{6}\cos k_{x}\cos k_{y}+\Delta_{\mathrm{CF}}\nonumber \\
\epsilon_{xz,yz}\left(\mathbf{k}\right) & =-4t_{4}\sin k_{x}\sin k_{y}\nonumber \\
\epsilon_{xz,xy}\left(\mathbf{k}\right) & =-2it_{7}\sin k_{x}-4it_{8}\sin k_{x}\cos k_{y}\nonumber \\
\epsilon_{yz,xy}\left(\mathbf{k}\right) & =-2it_{7}\sin k_{y}-4it_{8}\sin k_{y}\cos k_{x}\label{S_3_orb}
\end{align}

The tight-binding parameters are shown in Table \ref{tab_2orb}. For
an occupation number of $4$, the chemical potential is $\mu=212$
meV and the crystal field splitting is $\Delta_{\mathrm{CF}}=400$
meV.

\begin{table}[h]
\begin{centering}
\begin{tabular}{|c|c|c|c|c|c|c|c|}
\hline 
$t_{1}$  & $t_{2}$  & $t_{3}$  & $t_{4}$  & $t_{5}$  & $t_{6}$  & $t_{7}$  & $t_{8}$\tabularnewline
\hline 
\hline 
$-60$  & $-20$  & $-30$  & $10$  & $-200$  & $-300$  & $200$  & $-100$\tabularnewline
\hline 
\end{tabular}
\par\end{centering}

\protect\protect\protect\protect\protect\protect\caption{Tight-binding hopping parameters (in meV) for the 3-orbital model
of Eq. (\ref{S_aux_2_orbitals}). \label{tab_3orb}}
\end{table}

\subsection{Orbital-projected band model}

The band dispersion in the model by Vafek \emph{et al.} is described
in terms of the non-interacting Hamiltonian \cite{Vafek13}:

\begin{equation}
\hat{H}_{0}(\mbf{k})=\begin{pmatrix}h_{Y}(\mbf{k}) & 0 & 0\\
0 & h_{X}(\mbf{k}) & 0\\
0 & 0 & h_{\Gamma}(\mbf{k})
\end{pmatrix}
\end{equation}
with:

\begin{align}
h_{Y}(\mathbf{k})= & \begin{pmatrix}\epsilon_{1}+\frac{k^{2}}{2m_{1}}+a_{1}k^{2}\cos2\theta & -iv_{Y}(\mbf{k})\\
iv_{Y}(\mbf{k}) & \epsilon_{3}+\frac{k^{2}}{2m_{3}}+a_{3}k^{2}\cos2\theta
\end{pmatrix}\otimes\sigma^{0}\nonumber \\
h_{X}(\mathbf{k})= & \begin{pmatrix}\epsilon_{1}+\frac{k^{2}}{2m_{1}}-a_{1}k^{2}\cos2\theta & -iv_{X}(\mbf{k})\\
iv_{X}(\mbf{k}) & \epsilon_{3}+\frac{k^{2}}{2m_{3}}-a_{3}k^{2}\cos2\theta
\end{pmatrix}\otimes\sigma^{0}\nonumber \\
h_{\Gamma}(\mbf{k})= & \begin{pmatrix}\epsilon_{\Gamma}+\frac{k^{2}}{2m_{\Gamma}}+bk^{2}\cos2\theta & ck^{2}\sin2\theta\\
ck^{2}\sin2\theta & \epsilon_{\Gamma}+\frac{k^{2}}{2m_{\Gamma}}-bk^{2}\cos2\theta
\end{pmatrix}\otimes\sigma^{0}\label{S_hybrid}
\end{align}
and:

\begin{align}
v_{X}(\mbf{k})= & 2k\sin\theta\left[v+p_{1}k^{2}(2+\cos2\theta)-p_{2}k^{2}\cos2\theta\right]\nonumber \\
v_{Y}(\mbf{k})= & 2k\cos\theta\left[v+p_{1}k^{2}(2-\cos2\theta)+p_{2}k^{2}\cos2\theta\right]
\end{align}

Here, $k$ is given in units of the inverse lattice constant of the
1-Fe unit cell. To obtain a better description of the Fermi surface,
cubic terms are included in $v_{X}$ and $v_{Y}$, while in the discussion
in the main text we considered only linear terms. All the figures
in the main text refer to the dispersions with the cubic terms present.
The dispersion parameters are presented in Table \ref{tab_hybrid}.
The chemical potential is set to $\mu=0$.

\begin{table}[h]
\begin{centering}
\begin{tabular}{|c|c|c|c|c|c|c|c|c|c|c|c|c|}
\hline 
$\epsilon_{\Gamma}$  & $\epsilon_{1}$  & $\epsilon_{3}$  & $\frac{1}{2m_{\Gamma}}$  & $\frac{1}{2m_{1}}$  & $\frac{1}{2m_{3}}$  & $a_{1}$  & $a_{3}$  & $b$  & $c$  & $v$  & $p_{1}$  & $p_{2}$\tabularnewline
\hline 
\hline 
$132$  & $-400$  & $-647$  & $-368$  & $298$  & $634$  & $419$  & $-533$  & $56.5$  & $124.6$  & $-243$  & $-40$  & $10$\tabularnewline
\hline 
\end{tabular}
\par\end{centering}

\protect\protect\protect\protect\protect\protect\caption{Band dispersion parameters (in meV) for the band-orbital model of
Eq. (\ref{S_hybrid}). \label{tab_hybrid}}
\end{table}

\end{widetext}

 \bibliographystyle{apsrev4-1}
\bibliography{bibliography_review}

\end{document}